\documentclass[10pt]{article}

\usepackage[english]{babel}
\usepackage{authblk}

\usepackage[letterpaper,top=2cm,bottom=2cm,left=3cm,right=3cm,marginparwidth=1.75cm]{geometry}

\usepackage{amsmath}
\usepackage{graphicx}
\usepackage{hyperref}
\hypersetup{colorlinks=true, allcolors=blue}
\usepackage{cancel}

 \usepackage{amssymb}
 \usepackage{amsfonts}
 \usepackage{cases}
 \usepackage{cancel}
\usepackage{ulem}
 \usepackage{here}
\usepackage{dsfont}
\usepackage{amsmath}
\usepackage{mathtools}
 \usepackage{subfigure}
 \usepackage{makecell}
 \usepackage{array}
\usepackage{graphicx}
\usepackage{tabularray}

\usepackage{autobreak}
\usepackage{caption}
\usepackage{enumerate}
\usepackage{svg}
\usepackage{diagbox}
\graphicspath{{./}{figures/}}
\usepackage{amssymb}
\usepackage{ulem} 
\usepackage{adjustbox}

\usepackage{epstopdf}
\usepackage{cellspace}
\begin{document}

\thispagestyle{empty}
\vspace*{1cm}
\begin{center}

{\bf {\LARGE Splitting and gluing in sine-dilaton gravity: matter correlators and the wormhole Hilbert space}}\\

\begin{center}

\vspace{1cm}

{\bf Chuanxin Cui and Moshe Rozali}\\
 \bigskip \rm

\bigskip 

Department of Physics and Astronomy, \\
University of British Columbia, Vancouver, V6T 1Z1,
Canada

\rm
\bigskip 
    \vspace{0.1in}
    
    {\texttt{ccxhans@student.ubc.ca, rozali@phas.ubc.ca}}
  \end{center}

\vspace{2.5cm}
{\bf Abstract}
\end{center}
\begin{quotation}
\noindent

Sine-dilaton gravity has been proposed as the holographic dual of the double scaled SYK (DSSYK) model.  In this work, we examine this duality by deriving general matter correlation functions directly from the bulk perspective. A novel technique we develop is to treat matter lines in the bulk as end-of-the-world (EOW) branes, allowing us to implement a splitting and gluing procedure on the bulk spacetime. This geometric procedure gives rise to the structure of two-sided multi-particle wormhole Hilbert space. In contrast to previous Hilbert space constructions in DSSYK, the length basis in our case factorizes across the subregions, whereas the energy basis acquires a non-local, state-dependent structure determined by the EOW brane quantization in each subregion. Different choices of splitting correspond to distinct representations of the same Hilbert space, all of which are equivalent for physical observables.
Based on this framework, we compute general correlation functions, including the OTOC, and show that they exactly reproduce the DSSYK results obtained from chord diagrams. By calculating the OTOC in different representations, we uncover a new identity for the 6j-symbol of the quantum group $\mathcal{U}_q(su(1,1))$. Finally, the resulting wormhole Hilbert space  enables us to go beyond the disk to compute matter correlation functions on the double trumpet and to include bulk matter loops.

\end{quotation}

\setcounter{page}{0}
\setcounter{tocdepth}{2}
\setcounter{footnote}{0}
\newpage

\parskip 0.1in
 
\setcounter{page}{2}
\tableofcontents

\vspace{10pt}
\section{Introduction}
  The Sachdev-Ye-Kitaev (SYK) model \cite{Sachdev:1992fk,Kitaev:2015KITP,Polchinski:2016xgd,Maldacena:2016hyu} has been widely studied over the past ten years as a toy model for testing many aspects of quantum gravity, owing to its exact solvability in both the IR and UV regimes. Especially, the IR limit of SYK model is described by Schwarzian modes and is holographically dual to  Jackiw–Teitelboim (JT) gravity in  $\mathrm{AdS}_2$ \cite{Maldacena:2016upp,Engelsoy:2016xyb,Jensen:2016pah}. However, beyond the IR limit, the precise holographic dual of the full SYK model is currently unknown \cite{Gross_2017,Das_2017, Das_2018,Goel:2021wim}. Nevertheless, in an interesting double-scaling limit, known as double-scaled SYK (DSSYK) \cite{Cotler:2016fpe}, solving the dynamics of the system reduces to a purely combinatorial problem, which is analytically tractable at all energy scales using "chord diagrams" \cite{Berkooz:2018jqr, Berkooz:2018qkz}. In this sense, finding a gravitational dual of the DSSYK model appears to be a more reachable goal.

 As a first step towards understanding bulk physics, in \cite{Berkooz:2018qkz, Lin:2022rbf} the chord diagrams were argued to be the right language for the emergence of the bulk geometry. Indeed, \cite{Lin:2022rbf} constructed a two-sided chord Hilbert space defined on the two-sided "wormhole" Cauchy slice, with extensions to include matter operators, and \cite{Lin:2023trc} further analyzed the symmetry algebra acting on this Hilbert space. Later, a series of works \cite{Blommaert:2023opb,Blommaert:2024ydx,Blommaert:2024whf,  Blommaert:2025avl} proposed a concrete holographic dual of DSSYK, which is called the "sine-dilaton gravity". The sine-dilaton gravity was shown to reproduce the dynamics of DSSYK, also referred to as q-Schwarzian quantum mechanics \cite{Blommaert:2023opb, Blommaert:2023wad, Bossi:2024ffa}, and its canonical quantization  reproduces the two-sided chord Hilbert space of \cite{Lin:2023trc} . Crucially in this construction, the length in the bulk becomes positive and discrete after gauging the shift symmetry of the conjugate momentum \cite{Blommaert:2024whf}. Besides the connection to DSSYK, the sine-dilaton gravity itself has some very unusual features for a gravitational theory, such as bounded spectrum and the resulting good UV properties, and therefore merits investigation beyond its connection to the SYK model.
 
 In this work, we take the next step in exploring this duality, by including bulk matter fields and reproducing general correlation functions of boundary-inserted operators from the gravitational perspective.  A nontrivial example is the out-of-time ordered correlator (OTOC). In DSSYK, the OTOC has been calculated using the Wick contractions in the chord diagram \cite{Berkooz:2018jqr}, where it gives the 6j-symbol of the quantum group $\mathcal{U}_q(su(1,1))$. Notably, this 6j-symbol leads to sub-maximal chaos, which is related to the "fake thermal circle" of DSSYK \cite{Lin:2023trc}. On the gravity side, a direct calculation of OTOC is more subtle. For example, previous OTOC calculations in JT gravity rely on boundary-bulk mappings or bootstrap techniques \cite{Mertens:2017mtv, Blommaert:2018oro, Iliesiu_2019, Suh:2020lco, Xu:2024gfm}. However, in sine-dilaton gravity, a more straightforward bulk calculation can be expected -- since the bulk geometry is discrete, the separation of different bulk regions in the OTOC contour is less ambiguous, as we will see below.

 Studying general correlation functions also provides a direct access to the structure of bulk Hilbert space with matter. In this work, we are led to a convenient representation of the bulk Hilbert space, whose details are different from existing proposals in the literature. This Hilbert space is expected to be isomorphic to various matter Hilbert spaces proposed in DSSYK \cite{Lin:2023trc,Okuyama:2024yya,Okuyama:2024gsn}. However, the direct identification between them may not be immediately transparent, as the isomorphisms can be quite complicated \cite{xu2025vonneumannalgebrasdoublescaled}. Nevertheless, they should all give the same results for any correlation functions.

\vspace{5pt}
\subsection{Summary}
In this paper, we give a bulk derivation of general correlation functions in sine-dilaton gravity with matter, and identify the correct structure of the multi-particle Hilbert space implied by these correlation functions. Here the multi-particle Hilbert space refers to the Hilbert space on a bulk Cauchy slice intersected by multiple matter particles. Below we briefly summarize the main results, leaving  the details to the main text.

\paragraph{Splitting and gluing procedure}
The motivation to investigate this procedure originates from a natural bulk perspective: a boundary-to-boundary particle worldline on the  Euclidean $\mathrm{AdS}_2$ disk divides the disk into two regions and can be considered as an end-of-the-world (EOW) brane for each region. When the time direction is chosen such that the Cauchy slices are orthogonal to the EOW brane, this brane acts as a defect operator, leading to a twisted Hilbert space, which we denote by $\mathcal{H}_{\mathrm{\Delta}}$, with $\Delta$ being one of the brane parameter corresponding to the scaling dimension of the inserted boundary matter operator. The canonical quantization result of this twisted Hilbert space was established in \cite{Blommaert:2025avl}, where the resolution of the identity operator takes the form
\begin{align}
   \textbf{Twisted Hilbert space {$\mathcal{H}_{\mathrm{\Delta}}$}}: \quad   \mathds{1}=\int_{0}^{\pi} d\theta ~\rho_{\Delta,\bar\alpha}(\theta)|\theta \rangle_{\Delta,\bar{\alpha}}   {_{\Delta,\bar{\alpha}}}\langle \theta| =\sum_{n=0}^{\infty}|n\rangle \langle n|.
\end{align}
Here, $\Delta$ and $\bar\alpha$ are two EOW brane parameters, while $|\theta\rangle_{\Delta,\bar\alpha}$ and $|n\rangle$ denote energy and length eigenstates in $\mathcal{H}_{\Delta}$, respectively. We have normalized the states so that the length basis is independent of the brane parameters, whereas the energy basis corresponds to eigenstates of a family of Hamiltonians that do depend on the brane parameters.  The corresponding density of states $\rho_{\Delta,\bar\alpha} (\theta)$ is given in (\ref{dos_so}). In the absence of the brane parameters, this Hilbert space reduces to the standard sine-dilaton Hilbert space $\mathcal{H}_{\mathrm{SD}}$.

The two-point function should be obtained by appropriately gluing the two regions, with each involving an EOW brane
\begin{align}
    \langle\mathcal{O}(\tau)\mathcal{O}(0)\rangle_\beta =\adjincludegraphics[width=5.2cm,valign=c]{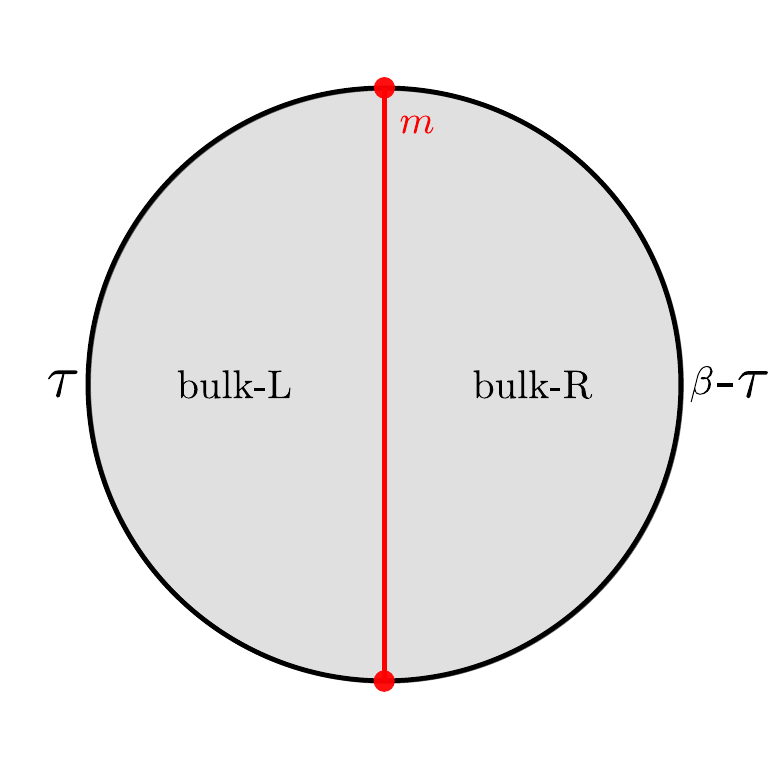}\quad=\adjincludegraphics[width=6cm,valign=c]{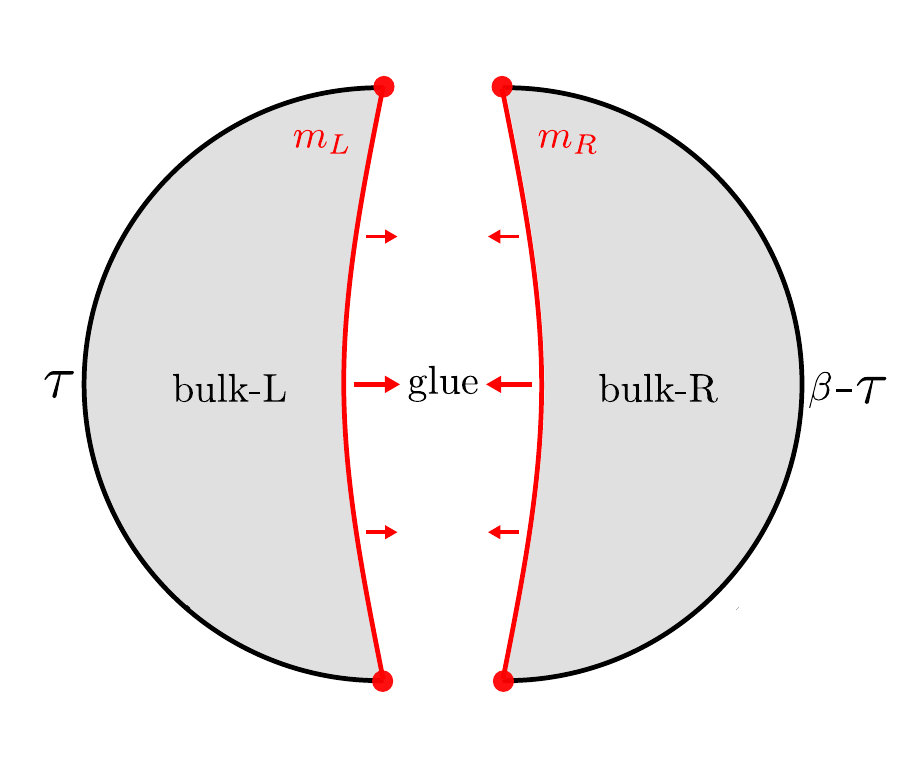}.
\end{align}
This procedure can be interpreted as "splitting" the matter geodesic into two, treating each as an EOW brane, and then gluing them back. In Section \ref{matter_sec}, we derive the splitting and gluing rules explicitly starting at the Lagrangian level.

\paragraph{One-particle Hilbert space} In Section \ref{1-parti_HS}, we discuss the structure of the one-particle Hilbert space implied by the splitting and gluing procedure. We denote the energy and the length bases of this Hilbert space as $|\Delta;\theta_L, \theta_R\rangle$ and $|\Delta;n_L, n_R\rangle$, respectively.  Here $\Delta$ labels the boundary matter operator, and $\theta_L/n_L$, $\theta_R/n_R$ are energy/length parameters of the left and right side regions relative to the particle.
The twisted Hilbert space introduced above does not directly yield the full one-particle Hilbert space. Rather, the one-particle state is constructed by appropriately patching up the states on the left and right sides of the particle's worldline using the gluing rules. There are two natural ways to carry out this patching, depending on how the EOW brane is associated with each side. Each choice leads to a different realization of the twisted Hilbert spaces, and corresponds to different representations of the one-particle state:
\begin{itemize}
    \item \textbf{0-split representation} One may associate the EOW brane to only one side of the particle worldline - say, the right side. In this case, the right Hilbert space is twisted by the EOW brane, while the left Hilbert space remains untwisted, but the brane parameter $\bar\alpha$ is fixed to the left energy parameter $\theta_L$. The other brane parameter $\Delta$ labels the scaling dimension of the boundary matter operator. The energy basis $|\Delta;\theta_L, \theta_R\rangle$  can then be decomposed as
    \begin{align}
 |\Delta;\theta_L,\theta_R\rangle_0=\adjincludegraphics[width=4cm,valign=c]{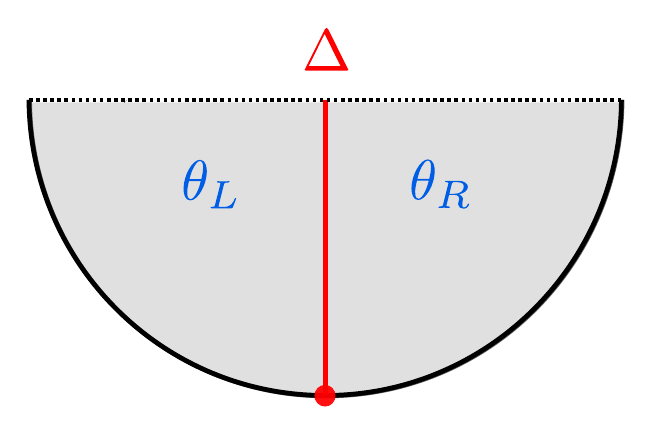} =  |{\theta_L}\rangle \otimes^{\theta_L} |{\theta_R}\rangle_{\Delta,\theta_L},
    \end{align}
   where $|\theta_L\rangle \in \mathcal{H}_{\mathrm{SD}}$ and $|\theta_R\rangle_{\Delta,\theta_L}\in \mathcal{H}_{\Delta}$. Here we introduce the notation $\otimes^{\theta_L}$ to indicate the non-factorization property of this basis: the second brane parameter of the right twisted state depends on the choice of energy in the left region. This reflects a non-local "state-dependence" feature of the energy basis.

The length basis $|\Delta;n_L, n_R\rangle$, on the other hand, simply factorizes as 
    \begin{align}
        |\Delta;n_L,n_R\rangle_0=|n_L\rangle \otimes |n_R\rangle.
    \end{align}

As we will show, the transformation between these two bases is a global quasi-unitary operator (i.e. unitary up to scaling), which is non-local with respect to left and right regions individually.

    \item  \textbf{1-split representation}  One can also associate the EOW brane to both sides of the particle worldline, which means that the brane is twisting both regions simultaneously.  The correct bulk picture corresponds to "splitting" the brane into two - each residing in one of the two regions and giving separate twisted Hilbert spaces in each region - and then "gluing" two regions back together. After identifying the correct gluing condition, the energy basis $|\Delta;\theta_L, \theta_R\rangle$ in this case can be decomposed as
     \begin{align}
 |\Delta;\theta_L,\theta_R\rangle_1=\adjincludegraphics[width=4cm,valign=c]{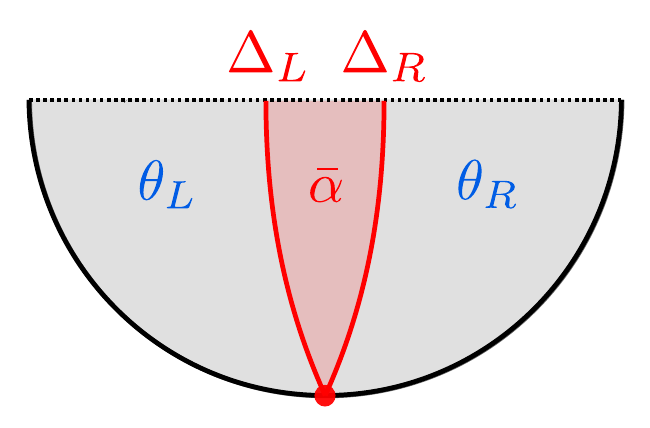} =  \int_0^{\pi} d\bar\alpha \, \rho(\bar\alpha) \, |{\theta_L}\rangle_{\Delta_L,\bar\alpha} \otimes^{\bar\alpha} |\bar\alpha\rangle \otimes^{\bar\alpha} |\theta_R\rangle_{\Delta_R,\bar\alpha}
    \end{align}
    Note that $\Delta_L+\Delta_R=\Delta$, and the presence of the middle "untwisted" region is essential. In this 1-split representation, the left and right regions are completely symmetric, where they are both state-dependent on the energy in the middle region.

The length basis in this representation factorizes into three states in each region
    \begin{align}
        |\Delta;n_L,n_R\rangle_1= |n_L\rangle \otimes |n_M=0\rangle \otimes |n_R\rangle.
    \end{align}
\end{itemize}

\paragraph{Multi-particle Hilbert space} The above splitting procedure is iterative and one can easily write out the $n$-split representation of the one-particle state. Crucially, when viewed locally in the bulk, the $(n-1)$-split representation of the one-particle Hilbert space directly implies the structure of the n-particle Hilbert space, as discussed in Section \ref{multi_parti_HS}. For instance, locally near a bulk Cauchy slice, the energy eigenstate of the n-particle Hilbert space  is illustrated as follows
\begin{align}
  |\Delta_1,...,\Delta_n; \theta_1,...,\theta_{n+1}\rangle=\adjincludegraphics[width=4.2cm,valign=c]{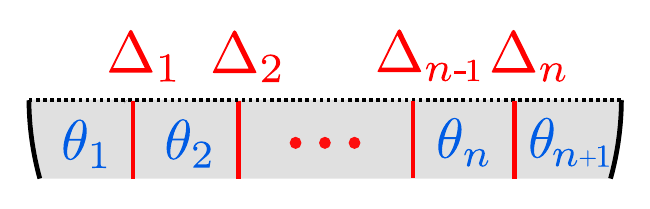},
    \end{align}
where the black lines are asymptotic boundaries, and the dotted line represents the Cauchy slice on which the state is defined. The $n$   matter lines divide the Cauchy slice into $n+1$ regions, and this $n$-particle energy eigenstate can be obtained by patching $n+1$ states in these regions together. The patching rule is very simple: one associates all the matter lines as EOW branes to the adjacent regions, such that each region contains at most one EOW brane. This results in $n$ twisted states and one untwisted state. An example of the decomposition takes the form
\begin{align}
     |\Delta_1,...,\Delta_n; \theta_1,...,\theta_{n+1}\rangle=|{\theta_1}\rangle_{\Delta_{1},\theta_2}\otimes^{\theta_2} |\theta_2\rangle_{\Delta_{2},\theta_3}\otimes^{\theta_3} \cdot\cdot\cdot \otimes |\theta_{n-1}\rangle_{\Delta_{n-1},\theta_{n}}\otimes^{\theta_n} |\theta_n\rangle \otimes^{\theta_n}
|\theta_{n+1}\rangle_{\Delta_{n},\theta_n} . 
\end{align}
 Other decompositions can be obtained simply by placing the untwisted state $|\theta_i\rangle$ at any other regions. All different decompositions will give the same result for any observable. 
 
The length eigenstate still factorizes into $n+1$ length eigenstates in each region
 \begin{align}
     |\Delta_1,...,\Delta_n; n_1,...,n_{n+1}\rangle=|n_1\rangle\otimes  \cdot \cdot \cdot \otimes|n_{n+1}\rangle.
 \end{align}

It is important to emphasize that these rules and structures are not constructed by hand. Rather, they are a direct result of the splitting and gluing procedure in the bulk, which can be derived starting from the Lagrangian of sine-dilaton gravity. The resulting state-dependence in the energy basis seems likely to be related to the representation theory of quantum groups (for example, state dependence did not appear in similar constructions in JT gravity). We leave details of such algebraic construction to future study. It is also fascinating to relate the state dependence here to the discussions of e.g. bulk reconstruction \cite{Papadodimas:2013jku}, as in that context the structure of the combined Hilbert space on a global Cauchy slice is non-linear \cite{Marolf:2015dia}.

 \paragraph{General correlation functions}
General correlation functions can be directly computed using the previous Hilbert spaces resulting from the splitting and gluing procedure. A key example is the OTOC, which is computed in Section \ref{OTOC_sec}, where it can be obtained as a ”two-point” function in the one-particle Hilbert space. Since this Hilbert space admits different representations, the OTOC can be computed in multiple ways
\begin{align}
    \mathrm{OTOC}=\adjincludegraphics[width=5cm,valign=c]{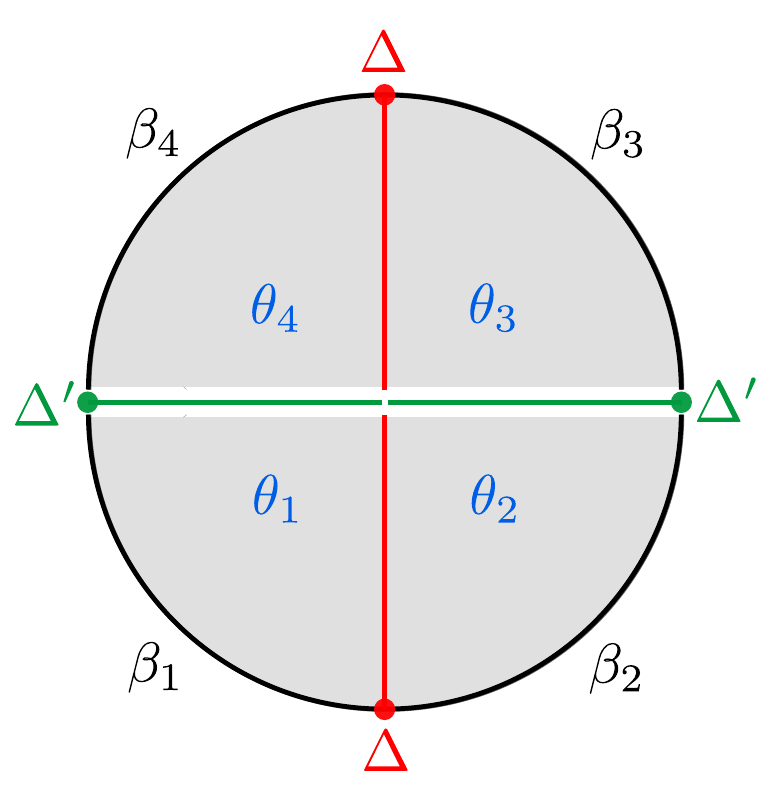}=\adjincludegraphics[width=5cm,valign=c]{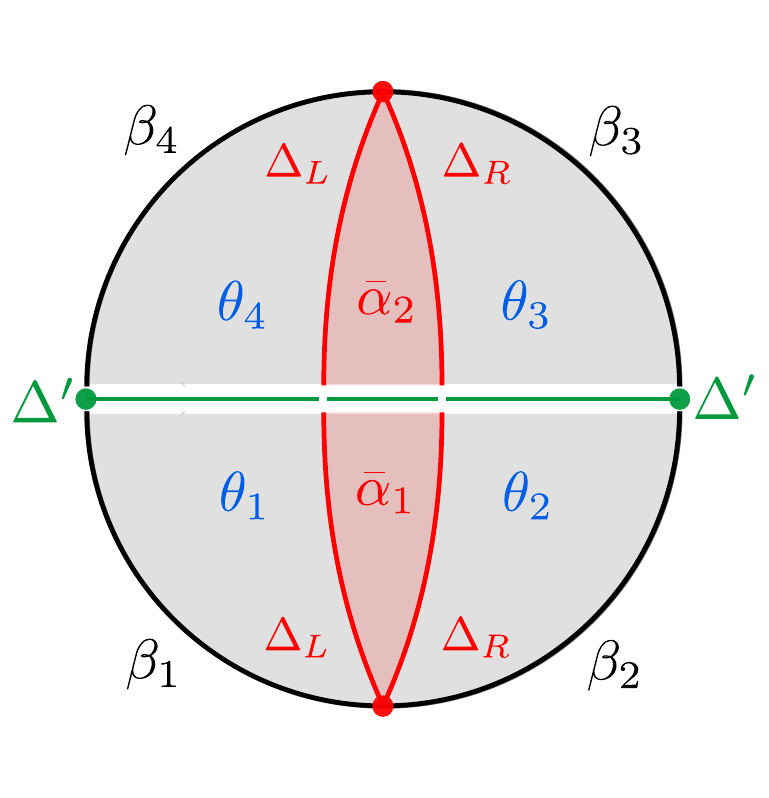}.
\end{align}
The 0-split representation exactly reproduces the DSSYK OTOC result, with the presence of the 6j-symbol of the quantum group $\mathcal{U}_q(su(1,1))$. In the 1-split representation, the result involves two 6j-symbols. Interestingly, we find a new identity relating the convolution of two 6j-symbols to a single one. This identity is given in (\ref{R_relation}) and we illustrate it pictorially below
\begin{align}
    \int_0^{\pi}\int_0^{\pi}d\bar\alpha_1 d\bar\alpha_2 \, \rho(\bar\alpha_1) \, \rho(\bar\alpha_2) \quad  \adjincludegraphics[width=4.5cm,valign=c]{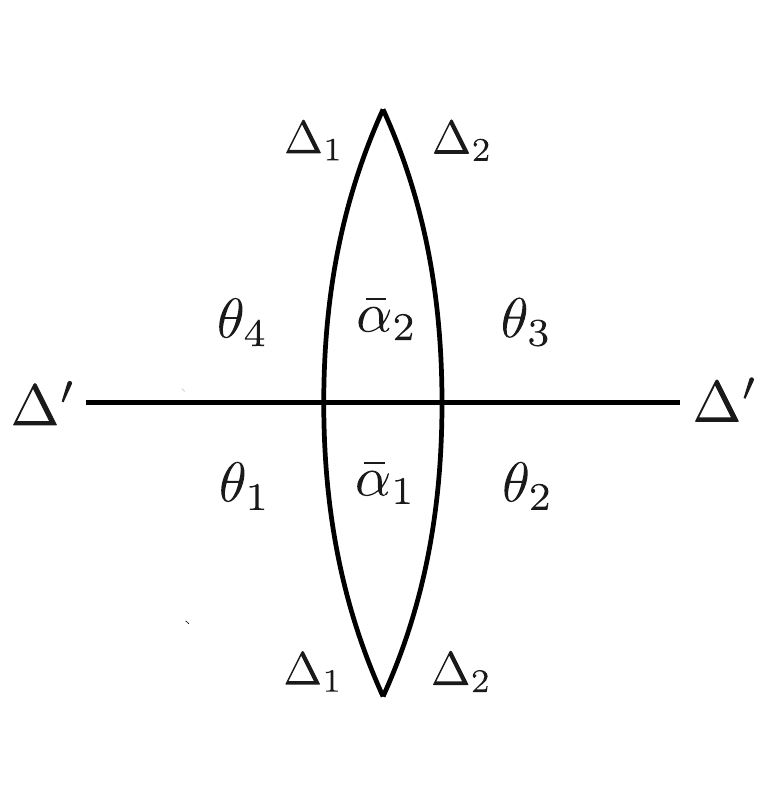}= \adjincludegraphics[width=4.5cm,valign=c]{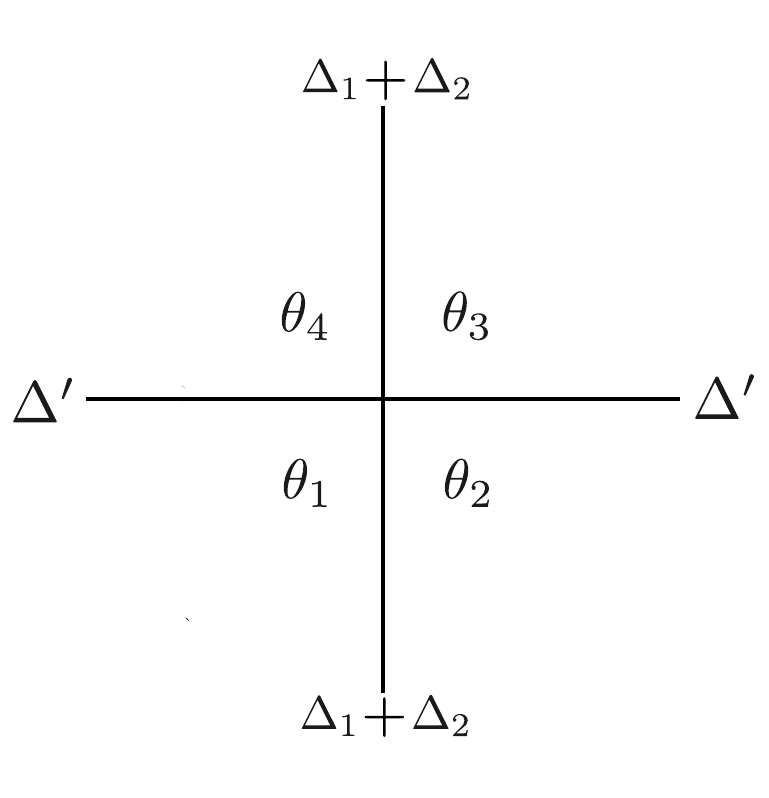}.
\end{align}
To our knowledge this is a new identity for 6j-symbols, and we have verified it numerically. With this identity all different expressions for the OTOC are shown to be equivalent.

We also compute several higher point correlation functions in Section \ref{n-pt_sec}, including the fully-crossed 8-point function, which involves a middle region that is disconnected from the boundary, and two-point functions on double trumpet with or without matter loops. Based on  these results, we formulate a set of bulk Feynman rules for computing general correlation functions. These rules are similar to those given in DSSYK \cite{Berkooz:2018jqr}, but extended to include bulk correlators and the double trumpet.

Finally in Section \ref{JT limit} we discuss the JT limit of matter correlators -- while the methods we use do not seem to have a direct analogue in JT, the formulas we obtain have a  good JT limit.  We conclude with a summary and open questions for future work.

\vspace{5pt}
\subsection{Sine-dilaton gravity setup}
Before proceeding to the main text we briefly summarize the basics of sine-dilaton gravity and explain our notations. The sine-dilaton gravity is defined by the two-dimensional gravitational path integral \cite{Blommaert:2024ydx,Blommaert:2024whf}
\begin{align}
    \int \mathcal{D}g \mathcal{D}\Phi ~\mathrm{exp}\left( \frac{1}{2|\log q|}\left\{\frac{1}{2} \int d^2 x \sqrt{g} \left( \Phi R+2 \sin(\Phi)\right)+\int_{\mathrm{bdy}} du \sqrt{h}~\Phi K -i\int_{\mathrm{bdy}} du \sqrt{h} ~e^{-i\Phi/2} \right\} \right),
    \label{full_action}
\end{align}
with the boundary conditions at the asymptotic boundaries,
\begin{align}
    \sqrt{h}~e^{i\Phi/2}\Big{|}_{\Phi=\Phi_{\mathrm{bdy}}}=\frac{i}{2|\log q|}, \quad \Phi_{\mathrm{bdy}}=\frac{\pi}{2}+i\infty.
    \label{asy_bdy}
\end{align}
Note that $0<q<1$ and $2|\log q|\equiv \hbar$ defines the semiclassical regime to be $q \rightarrow 1$. The classical solutions are parametrized by $\theta\in [0,\pi]$
\begin{align}
    ds^2=F(r) d\tau^2+\frac{1}{F(r)}dr^2, \quad F(r)=-2\cos(r)+2\cos(\theta), \quad \Phi=r.
\end{align}
The solutions have a black hole horizon at $r_{\mathrm{h}}=\theta$ and an asymptotic boundary which we locate at $r_{\mathrm{bdy}}=\Phi_{\mathrm{bdy}}=\frac{\pi}{2}+i\infty$. We discuss a path integral defined on a patch consisting of a complex contour from $r_{\mathrm{h}}$ to $r_{\mathrm{bdy}}$\footnote{There are also other patches corresponding for example to cosmological solutions.}. When we apply a Weyl rescaling we obtain from this patch a Euclidean $\mathrm{AdS}_2$ space, with
\begin{align}
    ds^2_{\mathrm{AdS}}=e^{-i\Phi}ds^2=F_{\mathrm{AdS}}(\rho)d\tau^2+\frac{1}{F_{\mathrm{AdS}}(\rho)}d\rho^2, 
    \label{AdS_metric}
\end{align}
where
\begin{align}
 F_{\mathrm{AdS}}(\rho)=\rho^2-\sin(\theta)^2, \quad r=\frac{\pi}{2}+i\log(\rho+i\cos \theta).
\end{align}
We can define a length variable for Cauchy slices on this  $\mathrm{AdS}_2$ patch
\begin{align}
    L=\int_{\mathrm{slice}} ds ~e^{-i\Phi/2}=\int_{\mathrm{slice}} du \sqrt{h} ~e^{-i\Phi/2}.
\end{align}
Canonical quantization of Sine-dilaton gravity was discussed in \cite{Blommaert:2024ydx}. The Hamiltonian is
\begin{align}
    \hat{H}_{\mathrm{SD}}=-\frac{1}{2\hbar}\left(e^{i\hat{P}}\sqrt{1-e^{-\hat{L}}}+ \sqrt{1-e^{-\hat{L}}} e^{-i\hat{P}}\right),
    \label{hal_SD}
\end{align}
where the canonical commutation relations are $[\hat{L},\hat{P}]= i \hbar$. As emphasized in \cite{Blommaert:2024whf}, the correct way to quantize this Hamiltonian is to gauge the shift symmetry of $\hat{P}$. This gauge symmetry discretizes the length and also projects out negative length states
\begin{align}
    \hat{P}\sim \hat{P}+2\pi \quad \rightarrow \quad \hat{L}=\hbar \, \hat{n}, \quad n\in \mathbb{N}_{0}.
\end{align}
Eigenvalues of the Hamiltonian can be obtained as a scattering problem at $L\rightarrow \infty$ as 
\begin{align}
    E=-\frac{\cos (\theta)}{\hbar}, \quad \theta\in [0,\pi].
\end{align}
Different bases of the Hilbert space are characterized by the resolutions  of the identity
\begin{align}
    \mathds{1}=\int_{0}^{\pi} d\theta ~\rho(\theta) |\theta\rangle \langle\theta | =\sum_{n=0}^{\infty}|n\rangle \langle n|,
    \label{resolution_1}
\end{align}
and the density of state $\rho(\theta)$ is given as
\begin{align}
    \rho(\theta)=\frac{(q^2;q^2)_{\infty}}{2\pi}(e^{\pm2i\theta};q^2 )_{\infty},
\end{align}
where $(e^{\pm 2i\theta};q^2)_{\infty}=(e^{ 2i\theta};q^2)_{\infty}(e^{- 2i\theta};q^2)_{\infty}$ is the q-Pochhammer symbol defined by
\begin{align}
    (a;q)_{n}=\prod_{k=1}^{n}(1-aq^{k-1}), \quad (a;q)_{0}=1.
\end{align}
We also use the shorthand notation $(a,b;q^2)_{\infty}=(a;q^2)_{\infty}(b;q^2)_{\infty}$. The overlap between $\theta$ and $n$ bases is given by the q-Hermite polynomials (which also appear as the DSSYK wavefunctions)
\begin{align}
    \langle n|\theta\rangle=\frac{H_n(\cos(\theta)|q^2)}{(q^2;q^2)_{n}^{1/2}}.
\end{align}
The partition function of sine-dilaton gravity can be obtained from the following "transition matrix element"
\begin{align}
    Z_{\mathrm{SD}}(\beta) &=\langle n=0| e^{-\beta \hat{H}_{\mathrm{SD}}}|n=0\rangle 
    =\int_{0}^{\pi}d\theta ~ \rho(\theta)~ e^{\beta \frac{\cos(\theta)}{\hbar}},
\end{align}
where we used the initial conditions $H_0(\cos(\theta)|q^2)=1$ and $(a;b)_0=1$.

\vspace{10pt}
\section{Matter fields in sine-dilaton gravity}
\label{matter_sec}

We now add massive matter fields to sine-dilaton gravity. Importantly this coupling is "non-minimal", meaning that matter field also couples to the dilaton field \cite{Blommaert:2024ydx}. Rather, the matter field couples minimally to the Weyl rescaled $AdS_2$ metric
\begin{align}
    S_{\phi}=\frac{1}{2}\int d^2 x \sqrt{g_{\mathrm{AdS}}}\left( g_{\mathrm{AdS}}^{\mu \nu} \partial_{\mu} \phi \partial_{\nu}\phi + m^2 \phi^2 \right).
\end{align}
In this way, the matter boundary-to-boundary propagator in  $\mathrm{AdS}_2$ can be obtained from the worldline formalism of a particle in $AdS_2$,
\begin{align}
    \langle\mathcal{O}(x_1) \mathcal{O}(x_2)\rangle=\int \mathcal{D}x \exp\left( -\frac{m}{\hbar}\int_{x(u_i
    )=x_1}^{x(u_f
    )=x_2}  du \sqrt{h} ~e^{-i\Phi/2}\, + \mathrm{c.t.}\right)= \ \exp\left(-\Delta L (x_1,x_2)\right),
    \label{bdy-to-bdy}
\end{align}
where we emphasize that the action $-\frac{m}{\hbar}\int_{x(u
    )}  du \sqrt{h} ~e^{-i\Phi/2}$ is formally divergent as the bulk operator approaches the boundary, and $L(x_1,x_2)$ is a suitably renormalized geodesic length between the boundary points $x_1$ and $x_2$ in $\mathrm{AdS}_2$, obtained after adding the appropriate counter term (c.t.)\footnote{The precise form of the counter term, as well as the procedure for obtaining the effective action on the geodesic, is currently unknown. The difficulty in identifying it arises from the fact that the boundary of sine-dilaton gravity is essentially non-geometric and singular \cite{Belaey:2025ijg}, as $\mathrm{AdS}_2$ patch of sine-dilaton gravity is defined along a complex contour that extends to $i\infty$ and the classical curvature diverges. In this paper, we take the conjectured dictionary (\ref{dic_mass}), inspired by the quantum group structure and the JT gravity result \cite{Yang:2018gdb}, where the effective action $e^{-\Delta L(x_1,x_2)}$ is in fact the exact result for any $\Delta$,  as the definition of this counter term. It would be interesting and important to investigate this point in the future. For recent explorations of the dictionary, see \cite{Bossi:2024ffa, Aguilar-Gutierrez:2025pqp}}, and $\Delta$ is the scaling dimension of the operator inserted at the boundary. The relation of the scaling dimension to the particle mass is conjectured to be \cite{Blommaert:2025avl,Blommaert:2023opb} 
\begin{align}
    m=\cosh(\hbar\Delta).
    \label{dic_mass}
\end{align}
The Euclidean two-point function is easily calculated as
\begin{align}
    \langle \mathcal{O}(\tau) \mathcal{O}(0)\rangle_{\beta} &=\langle n=0|e^{-(\beta-\tau) \hat{H}_{\mathrm{SD}}} e^{-\Delta \hat{L}} e^{-\tau \hat{H}_{\mathrm{SD}}} |n=0\rangle
    \nonumber\\
    &= \int_{0}^{\pi} \int_{0}^{\pi}d\theta_1 d\theta_2 \rho(\theta_1) \rho(\theta_2) e^{(\beta-\tau) \frac{\cos(\theta_1)}{2|\log q|}} e^{\tau \frac{\cos(\theta_2)}{2|\log q|}} \langle\theta_1| q^{2\Delta \hat{n}} |\theta_2\rangle,
    \label{2-pt_correlator1}
\end{align}
The matrix element $ \langle\theta_1| q^{2\Delta \hat{n}} |\theta_2\rangle$ can be calculated by using properties of the q-Hermite polynomials. The result is
\begin{align}
    \langle\theta_1| q^{2\Delta \hat{n}} |\theta_2\rangle=\frac{(q^{4\Delta};q^2)_{\infty}}{(q^{2\Delta}e^{\pm i\theta_1\pm i \theta_2};q^2)_{\infty}}.
    \label{2-pt_correlator2}
\end{align}

The result for the two-point function is well-known in both DSSYK and sine-dilaton gravity. In the current context it is most easily derived in the quantization channel where the particle is represented as an operator. If we draw the time running vertically and the Cauchy slices as horizontal, the particle is an operator inserted at a fixed Cauchy slice with matrix elements given by (\ref{2-pt_correlator2}). Our initial goal in this section is to re-derive this result  by  in the "dual'' channel where the particle is represented as a ("vertical'') defect or a twist operator. In other words, we treat the particle as a thin-shell whose trajectory divides the space of sine-dilaton gravity into left and right regions.  

In this dual channel, we proceed by splitting the thin-shell into two parts and performing independent quantizations in each region with the corresponding boundary conditions. It turns out that the quantization in each region corresponds to the EOW brane quantization developed in \cite{Blommaert:2025avl} and the two-point function can then be obtained by appropriately gluing back the two regions (and their Hilbert spaces). Although for the purpose of reproducing the two-point function this may appear to be a detour, the splitting and gluing procedure we develop plays a crucial role in identifying the correct wormhole Hilbert space with matter and the bulk derivation of general matter correlation functions, including crucially the crossed 4-point function. 

\vspace {5pt}
\subsection{Splitting a particle worldline in $\mathrm{AdS}_2$}
\begin{figure}[htbp]
    \centering
    \subfigure[]{\includegraphics[width=0.4\textwidth]{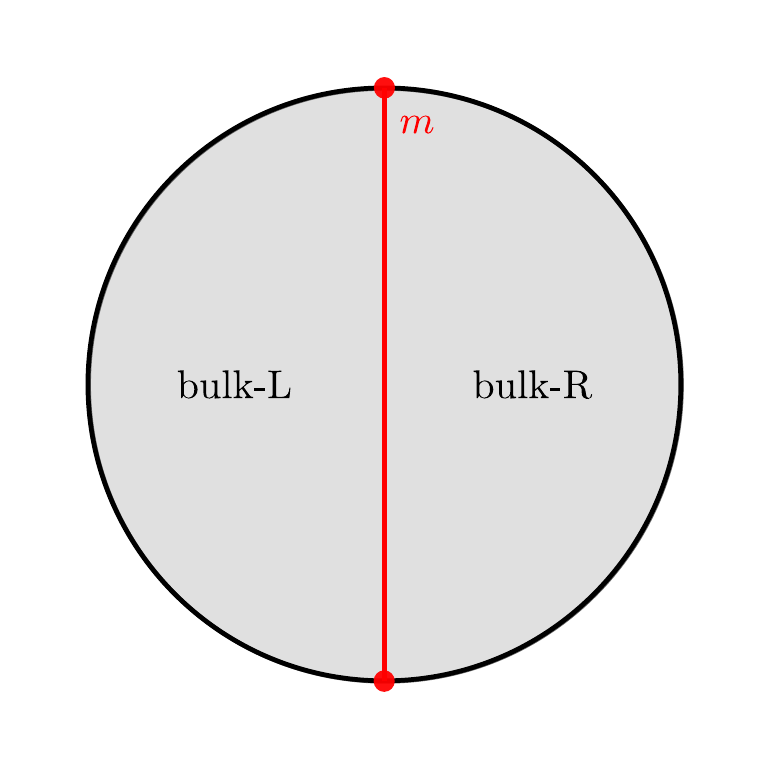}} 
    \subfigure[]{\includegraphics[width=0.46\textwidth]{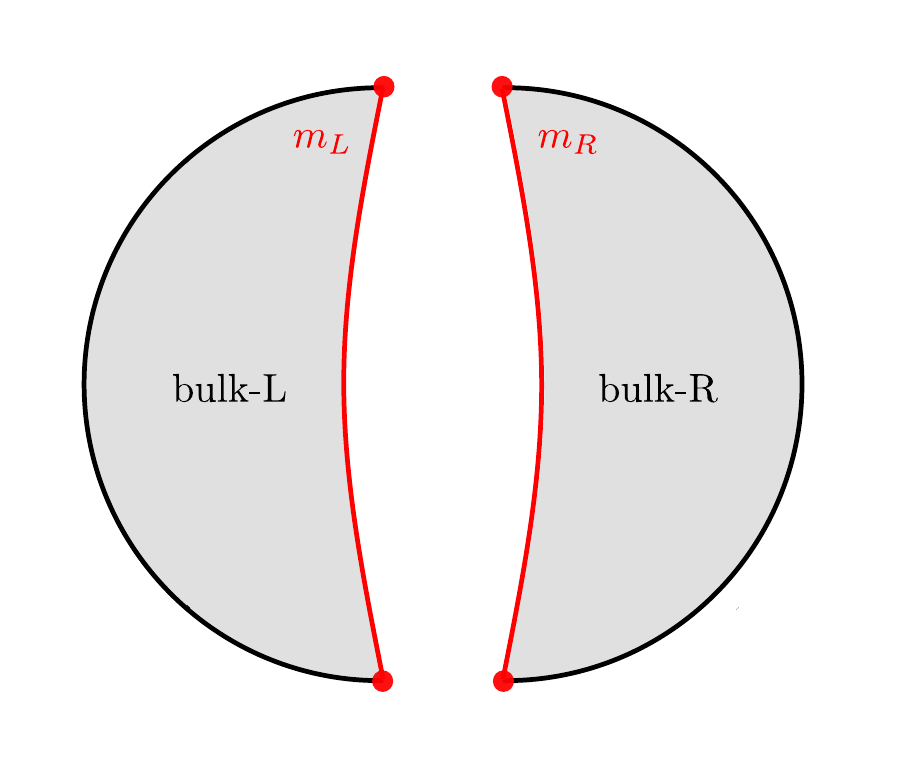}} 
    \caption{(a) Matter geodesic as a thin shell in the bulk. (b) Splitting the shell into two parts.}
    \label{shell_split}
\end{figure}

A matter geodesic in the bulk can be treated as a thin shell that divides the $\mathrm{AdS}_2$  spacetime into left and right regions as in Fig.\ref{shell_split}(a).

To ensure a valid gravity solution, the metric and dilaton fields must be continuous everywhere in the bulk. However, the extrinsic curvature and the normal derivative of the dilaton field  can have a jump across the shell \cite{Israel:1966rt}. We denote these discontinuous quantities as $K_L, K_R$ and $\partial_{n_L}\Phi=n_{L}^{\mu}\partial_{\mu}\Phi, \partial_{n_R}\Phi=n_{R}^{\mu}\partial_{\mu}\Phi$, respectively. The action in the presence of the shell is\footnote{We suppress asymptotic boundary terms for conciseness. These terms are the same as in (\ref{full_action}).  }
\begin{align}
    I=&\frac{1}{2}\int_{\mathrm{bulk-L}} d^2 x \sqrt{g}\left[\Phi R+2 \sin(\Phi) \right] +\frac{1}{2}\int_{\mathrm{bulk-R}} d^2 x \sqrt{g}\left[\Phi R+2 \sin(\Phi) \right]
    \nonumber\\
    &-im\int_{\mathrm{shell}}du \sqrt{h}~e^{-i\Phi/2} +\int_{\mathrm{shell}}du \sqrt{h}~\Phi K_L + \int_{\mathrm{shell}}du \sqrt{h}~\Phi K_R.
    \label{TS_action}
\end{align}
where we added the worldline of matter field as a thin shell  and included the GHY boundary term on the shell separately for left and right regions\footnote{Note that, subtly, an extra factor of $i$ has been added in front of the mass term. This is required to ensure a consistent boundary value problem on the thin shell in sine-dilaton gravity, as the variation of dilaton field introduces  an additional factor of $i$. As we will see later, the jump of extrinsic curvature in AdS metric (\ref{AdS_metric}) across the shell is equal to $m$, which is a real number, only when this extra $i$ factor is included. This provides one way to correctly identify the mass in the boundary action.}.
Note that we treat the metric and dilaton as continuous across the interface, even off-shell, thus continuity is automatically enforced by our choice of variables.

We now split the shell into two parts, treating each part as the boundary of the corresponding region, as illustrated in Fig.\ref{shell_split}(b). Naively, one may attempt to directly split the worldline action of the shell into two contributions, each associated with its own mass parameter. However, this splitting does not make sense, as the worldline action describes boundary-to-boundary propagator and is formally divergent. Rather, we need to perform holographic renormalization to obtain finite expression prior to the splitting. The splitting procedure is then,
\begin{align}
    &\exp\left( -\frac{m}{\hbar}\int_{x(u
    )}  du \sqrt{h} ~e^{-i\Phi/2}\right) 
   \rightarrow \exp\left( -\Delta L\right)= \exp\left( -(\Delta_L+\Delta_R) L\right) 
    \nonumber\\
    &\rightarrow  \exp\left( -\frac{m_L}{\hbar}\int_{x(u
    )}  du \sqrt{h} ~e^{-i\Phi/2}-\frac{m_R}{\hbar}\int_{x(u
    )}  du \sqrt{h} ~e^{-i\Phi/2}\right)
\end{align}
where in the first line, the arrow indicates the use of holographic renormalization and summing over all possible shell locations to obtain the finite action $e^{ -\Delta L}$, as in (\ref{bdy-to-bdy}).  At this stage, we are able to correctly split the action by writing $\Delta=\Delta_L+\Delta_R$.\footnote{One can choose any decomposition of $\Delta=\Delta_L+\Delta_R$, and any observables only depends on $\Delta_L+\Delta_R$. This provides a non-trivial consistency check of our calculation later.} In the second line, we "undo'' the holographic renormalization to recover the worldline action for each region separately. Crucially, the dictionary (\ref{dic_mass}) implies that $m\neq m_L+m_R$ but instead,
\begin{align}
    m_{L/R}=\cosh(\hbar \Delta_{L/R}), \quad m=\cosh(\hbar \Delta)=\cosh(\hbar (\Delta_L+\Delta_R))\neq m_L+m_R.
    \label{dic_massLR}
\end{align}

Using the bulk equations of motion, the variation of the split action localizes as usual on the boundary, giving
\begin{align}
    \delta I=&\int_{\mathrm{shell}}du \left[\delta \sqrt{h} (\partial_{n_L}\Phi+\partial_{n_R}\Phi-\Phi K_L-\Phi K_R)-\delta K_L  \sqrt{h} \Phi-\delta K_R \sqrt{h} \Phi \right]
    \nonumber\\
    &+\int_{\mathrm{shell}} du \Big{[} \delta \sqrt{h} (\Phi K_L+\Phi K_R-i(m_L+m_R) e^{-i\Phi/2})+\delta K_L \sqrt{h}\Phi+\delta K_R \sqrt{h} \Phi
    \nonumber\\
    &\quad \quad \quad \quad \quad+  \delta \Phi  \sqrt{h}(K_L+K_R-\frac{1}{2}(m_L+m_R) e^{-i\Phi/2})\Big{]}
    \nonumber\\
    =&\int_{\mathrm{shell}}du\left[ \delta \sqrt{h} (\partial_{n_L}\Phi+\partial_{n_R}\Phi-i(m_L+m_R)e^{-i\Phi/2}) +  \delta \Phi \sqrt{h} (K_L+K_R-\frac{1}{2}(m_L+m_R) e^{-i\Phi/2}) \right].
\end{align}
To have a well-defined  variational principle without fixing the metric or dilaton field on-shell, we obtain the following Israel junction condition for sine-dilaton gravity\footnote{Similar junction conditions for general dilaton gravity theories, including JT gravity, were derived in \cite{Shen:2024itl}.}
\begin{align}
    \partial_{n_L}\Phi+\partial_{n_R}\Phi=i(m_L+m_R)e^{-i\Phi/2}, \quad  K_L+K_R=\frac{1}{2}(m_L+m_R) e^{-i\Phi/2}.
    \label{Israel_JC}
\end{align}
The presence of the thin shell gives a jump in the extrinsic curvature and the normal derivative of the dilaton field across the shell, both of which are controlled by the fixed parameter $m$ and the on-shell value of the field $\Phi$.

 We now proceed to calculate the two-point function by splitting the action into two parts and applying quantization separately on each side. To do so, observe that the Israel junction condition (\ref{Israel_JC}) specifies only the jump of $K$ and $\partial_n \Phi$. To fully specify the boundary data for one-sided quantization, one needs to further fix the boundary conditions for $K$ and $\partial_{n}\Phi$ on one side (by adding appropriate new boundary terms). The boundary conditions on the other side are then determined via the junction condition (\ref{Israel_JC}). 
 In particular, one can impose the following boundary conditions for the left and right regions on the thin shell
\begin{align}
    &\partial_{n_R}\Phi=\bar{\mu}e^{i\Phi/2}+im_R e^{-i\Phi/2}, \quad K_R=\frac{i}{2}\bar{\mu}e^{i\Phi/2}+\frac{1}{2}m_R e^{-i\Phi/2}.
    \label{right_bdy_con}\\
    &\partial_{n_L}\Phi=-\bar\mu e^{i\Phi/2}+im_L e^{-i\Phi/2}, \quad K_L=-\frac{i}{2} \bar\mu e^{i\Phi/2}+\frac{1}{2}m_L e^{-i\Phi/2}.
    \label{left_bdy_con}
\end{align}
As we will show in the next subsection, these boundary conditions can be imposed by a seemingly trivial rewriting of the action (\ref{TS_action}) as
\begin{align}
    &I=I_L+I_R.
    \nonumber\\
    &I_L=\frac{1}{2}\int_{\mathrm{bulk-L}} d^2 x \sqrt{g}\left[\Phi R+2 \sin(\Phi) \right]+ \int_{\mathrm{shell}}du \sqrt{h}~\Phi K_L -im_L\int_{\mathrm{shell}}du \sqrt{h}~e^{-i\Phi/2} +\bar{\mu} \int_{\mathrm{shell}}du \sqrt{h}e^{i\Phi/2}.
    \label{left_action}
    \\
    &I_R=\frac{1}{2}\int_{\mathrm{bulk-R}} d^2 x \sqrt{g}\left[\Phi R+2 \sin(\Phi) \right]+ \int_{\mathrm{shell}}du \sqrt{h}~\Phi K_R-im_R\int_{\mathrm{shell}}du \sqrt{h}~e^{-i\Phi/2}- \bar{\mu} \int_{\mathrm{shell}}du \sqrt{h}e^{i\Phi/2}.
    \label{right_action}
\end{align}
 Here, $I_L$ and $I_R$ correspond to the actions of the left and right regions, respectively. Note that, at the action level, the $\bar\mu$-terms cancel trivially, so the total action $I$ is the same as before. But the boundary conditions imposed on the thin shell make $\bar\mu$ a physical parameter, and consequently, the partition functions for each region depend on $\bar\mu$. This dependence persists even when the two partition functions are multiplied together.

With both thin shell and asymptotic boundary conditions specified, we are able to calculate the partition functions for each region, denoted by $Z_L(\beta_L,\bar\mu,m_L)$ and $Z_R(\beta_R,\bar\mu,m_R)$. One might expect that multiplying these two partition functions together should reproduce the two-point function. However, this is not the case, as can be seen for example by the fact that such a product depends on the arbitrary choice of $\bar\mu$. This mismatch arises because, in the calculation of the two-point function, only the jump of $K$ and $\partial_n\Phi$ is determined, not their individual values on each side. The correct procedure is to integrate over $\bar\mu$, with a possible measure, to include all possible values of $K$ and $\partial_n \Phi$ consistent with the jump condition,
\begin{align}
     \langle \mathcal{O}(\tau) \mathcal{O}(0)\rangle_{\beta} =\int_{D} d\bar\mu~ \mathcal{M}(\bar\mu) ~Z_L^{\star}(\tau,\bar\mu,m_L) ~Z_R(\beta-\tau,\bar{\mu},m_R),
     \label{2pt_measure}
\end{align}
where $D$ is the domain of integration and $\mathcal{M}(\bar\mu)$ is the measure\footnote{This is reminiscent of the gluing procedure in JT gravity, where one integrates over one of the phase space variables while keeping the other one fixed. For example, when gluing open-slice wavefunctions in JT, one integrates over $\ell$ with a unity measure, while keeping the extrinsic curvature $K$ fixed. Similarly, when gluing closed surfaces, one integrates over the baby universe size with the measure given by the twist factor of the Riemann surface. We thank Andreas Blommaert for making relevant comments.}. We refer to (\ref{2pt_measure}) as the "gluing procedure" and argue for it further, including determining the appropriate measure, below. Note that the complex conjugation of the left region partition function has been taken when performing the gluing, as the left and right region should be interpreted as the bra and ket states in the two-point function  \cite{Held:2024rmg,Blommaert:2025avl}. 
In the following subsections, we present the quantization procedure for each region.  The domain $D$ will then be determined by the quantization result, and the measure can be fixed by matching to the known two-point function.

\vspace{5pt}
\subsection{EOW brane quantization}
\label{sec_right_reg}

The left and right region actions, (\ref{left_action}) and (\ref{right_action}), have the same form as sine-dilaton gravity with an EOW brane boundary, whose quantization  was worked out in \cite{Blommaert:2025avl}.  The prescription of EOW brane there utilizes the reformulation of  sine-dilaton gravity as a sum of two Liouville theories, where the EOW brane corresponds to the FZZT brane in each Liouville theory. Therefore, the introduction of an EOW brane leads to a two-parameter space of boundary conditions labeled by $(\mu, \bar\mu)$
\begin{align}
    I_{\mathrm{EOW}}= \int_{\mathrm{brane}}du \sqrt{h} \Phi K -\mu\int_{\mathrm{brane}} du \sqrt{h} e^{-i\Phi/2} -\bar{\mu}\int_{\mathrm{brane}} du \sqrt{h} e^{i\Phi/2} .
    \label{brane_action}
\end{align}
The right region action is thus equivalent to this EOW brane action upon identifying $\mu=im_R$. The left region action is also equivalent to an EOW brane action with $\mu=im_L$, except that the sign of the $\bar\mu$ term is flipped. In what follows, we focus on the right region action; the left region analysis will be straightforwardly obtained  by replacing $\bar \mu$ with $-\bar \mu$. 

Variation of the right region action  gives the following terms localized on the brane boundary
\begin{align}
    \delta I_R=\int_{\mathrm{shell}}du \left[\delta \sqrt{h}(\partial_{n_R}\Phi -im_Re^{-i\Phi/2}-\bar\mu e^{i\Phi/2})+\delta \Phi \sqrt{h}(K_R-\frac{1}{2}m_Re^{-i\Phi/2}-\frac{i}{2}\bar\mu e^{i\Phi/2})\right].
    \label{right_action_vary}
\end{align}
The boundary conditions for the EOW brane are then,
\begin{align} \partial_{n_R}\Phi=\bar{\mu}e^{i\Phi/2}+im_R e^{-i\Phi/2}, \quad K_R=\frac{i}{2}\bar{\mu}e^{i\Phi/2}+\frac{1}{2}m_R e^{-i\Phi/2},
    \label{EOW_bdy_1}
\end{align}
which is exactly the boundary conditions we need in (\ref{right_bdy_con}).
To understand these boundary conditions better, one can express them in terms of parameters in AdS metric (\ref{AdS_metric}), where the extrinsic curvature and the normal derivative of the dilaton field transform as follows
\begin{align}
    K_{\mathrm{AdS}}=e^{i\Phi/2}K-\frac{i}{2}e^{i\Phi/2}\partial_{n}\Phi, \quad \partial_{n_{\mathrm{AdS}}}\Phi_{\mathrm{AdS}}=e^{-i\Phi/2}\partial_n\Phi.
    \label{AdS_SD}
\end{align}
Then the boundary conditions (\ref{EOW_bdy_1}) are translated into the AdS space as
\begin{align}
    K_{AdS}^{R}=m_R, \quad \partial_{n_{\mathrm{AdS}}^{R}}\Phi_{\mathrm{AdS}}^{R}=\bar\mu+im_Re^{-i\Phi}.
    \label{AdS_bdycond}
\end{align}
Note that this confirms the necessity of including the additional factor of $i$ in the thin-shell action (\ref{TS_action}), which ensures that the extrinsic curvature in AdS space correctly matches the physical mass parameter $m$.

In \cite{Blommaert:2025avl}, two quantization channels relevant to our right region setup were developed: the fully-open channel (FC) and the semi-open channel (SC). We now summarize the procedure and provide further details.

\begin{figure}[htbp]
  \begin{center}
   \includegraphics[width=6cm]{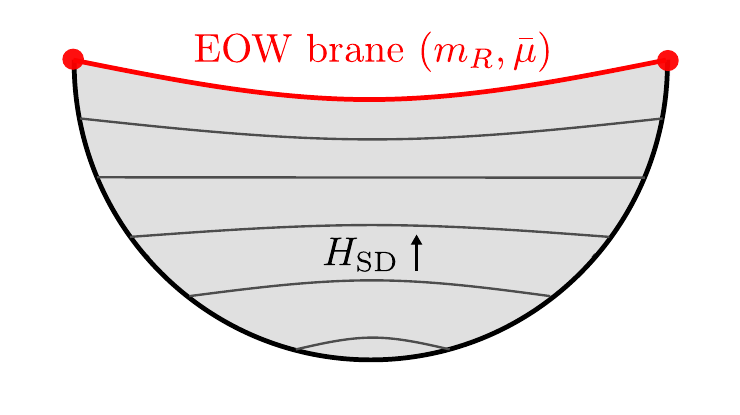}
  \end{center}   
\vspace*{-0.5cm}
\caption{Fully-open channel}
\label{F-O}
\end{figure}

\paragraph{Fully-open channel}
In the fully-open channel, the bulk Cauchy slices are given as in Fig.\ref{F-O}, where the EOW brane coincides with the final Cauchy slice at the red boundary. In this setup, the brane is interpreted as a state $|m_R,\bar\mu\rangle$ in the bulk Hilbert space.
Since the Cauchy slices are unaffected by the presence of the brane, the Hamiltonian that generates the time flow of the Cauchy slices is the same as the ordinary sine-dilaton Hamiltonian (\ref{hal_SD}). Thus, the Hilbert space is characterized by the same resolution of the identity and wavefunctions as before
\begin{align}
    \mathds{1}=\int_{0}^{\pi} d\theta ~\rho(\theta) |\theta\rangle \langle\theta | =\sum_{n=0}^{\infty}|n\rangle \langle n|, \quad   \langle n|\theta\rangle=\frac{H_n(\cos(\theta)|q^2)}{(q^2;q^2)_{n}^{1/2}}.
    \label{FC_R_Hilb}
\end{align}

The only unknown quantity is the wavefunction $\langle n|m_R, \bar\mu\rangle$. As explained in \cite{Blommaert:2025avl}, this wavefunction can be determined by expressing the effective mass of the brane in AdS space 
 in two different ways.\footnote{ The first way is to directly match the parameter of EOW brane here to the JT gravity brane mass $m_{\mathrm{AdS}}$: $m_{\mathrm{AdS}}=\partial_{n_{\mathrm{AdS}}}\Phi_{\mathrm{AdS}}-\Phi_{\mathrm{AdS}}K_{\mathrm{AdS}}$. The second way is to use the Hamiltonian constraint to relate $m_{\mathrm{AdS}}$ to the JT momentum and extrinsic curvature. See \cite{Blommaert:2025avl} for details.}
\begin{align}
    m_{\mathrm{AdS}}=\bar{\mu}+im_R\cos(P)\sqrt{1-e^{-L}}=i\sin(P)\sqrt{1-m_R^2}\sqrt{1-e^{-L}}.
    \label{2_mass}
\end{align}
 Using the relation (\ref{dic_massLR}) and analytically continuing $\bar\mu$ to an angular variable 
\begin{align}
    \bar\mu=-i\cos(\bar\alpha),
    \label{mubar_ang}
\end{align}
one obtains
\begin{align}
    2\cos(\bar{\alpha})=q^{-2\Delta_R}\sqrt{1-e^{-\hat{L}}}e^{i\hat{P}}+q^{2\Delta_R}e^{-i\hat{P}}\sqrt{1-e^{-\hat{L}}},
    \label{FC_wavefunction}
\end{align}
where an operator ordering has been chosen. We can treat the above equation as a Schrodinger eigenvalue equation
\begin{align}
    2\cos(\bar\alpha) \langle m_R,\bar\mu|n\rangle= \langle m_R,\bar\mu|q^{-2\Delta_R}\sqrt{1-e^{-\hat{L}}}e^{i\hat{P}}+q^{2\Delta_R}e^{-i\hat{P}}\sqrt{1-e^{-\hat{L}}}|n\rangle.
\end{align}
Noting that $\hat{L}=\hbar \hat{n}$ and $e^{\pm i \hat{P}}|n\rangle=|n\pm1\rangle$, this can be reduced to the recursion relation for continuous q-Hermite polynomials if the wavefunction takes the form,
\begin{align}
    \langle m_R,\bar\mu|n\rangle=\frac{H_n(\cos(\bar{\alpha})|q^2)}{\sqrt{(q^2;q^2)_n}}q^{2\Delta_Rn}.
\end{align}
This wavefunction provides an identification of the EOW brane state in the fully-open channel as the operator $q^{2\Delta \hat{n}}$ acting on the energy eigenstate $|\bar\alpha\rangle$,
\begin{align}
    |m_R,\bar\mu \rangle= q^{2\Delta_R \hat{n}} |\bar\alpha\rangle.
    \label{Dic_branestate}
\end{align}
This identification is reminiscent of expressing the EOW brane state in DSSYK as a coherent state of the q-deformed oscillator in \cite{Okuyama:2023byh}.

\begin{figure}[htbp]
  \begin{center}
   \includegraphics[width=4.8cm]{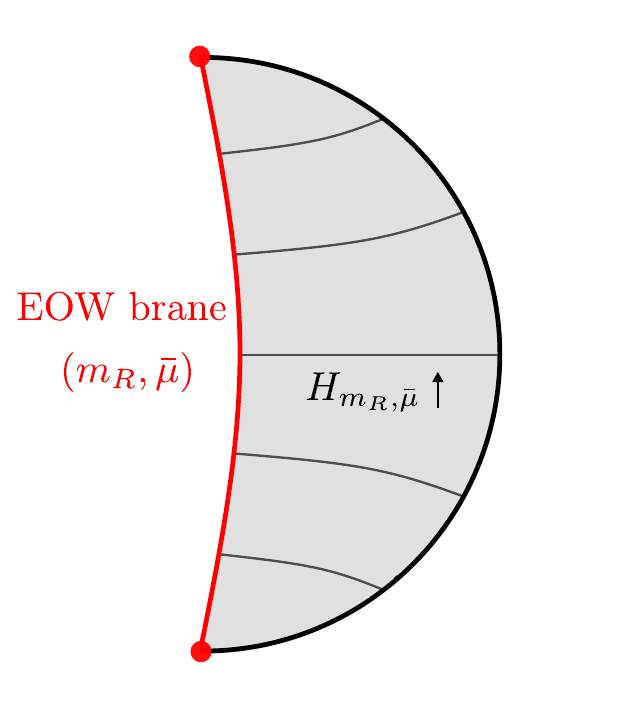}
  \end{center}   
\vspace*{-0.5cm}
\caption{Semi-open channel}
\label{S-O}
\end{figure}

\paragraph{Semi-open channel}
In the semi-open channel, the bulk Cauchy slices are given in Fig.\ref{S-O}, where each bulk time slice ends on the asymptotic boundary and perpendicularly on the EOW brane. 

 The ADM Hamiltonian generating this time flow is affected by the presence of the brane, leading to a "twisted'' Hilbert space denoted by $\mathcal{H}_{\Delta_R
 }$. Here, $\Delta_R$ is the scaling dimension of the boundary inserted operator that twists the Hilbert space, and is related to the first brane parameter by (\ref{dic_mass}). The fact that this twisted Hilbert space depends only on the first, but not the second, brane parameter can be understood as follows. Referring to the brane action (\ref{brane_action}), if one sets $\mu=0$, the EOW brane boundary term reduces to $\bar\mu \int du \sqrt{h} e^{i\Phi/2}$. This term under the asymptotic boundary condition (\ref{asy_bdy}) becomes $\frac{1}{\hbar}\cos(\bar\alpha)\beta$, where $\beta$ is the coordinate length of the brane. However, since EOW boundary condition does not fix the metric and dilaton field,  the contribution of this term on the EOW brane boundary to the path integral is   $\int_0^{\infty} d\beta \, e^{\frac{\beta \cos(\bar\alpha)}{\hbar}}$, which is an inverse Laplace transform. This only changes the ensemble from canonical to  microcanonical, but does not affect the structure of the Hilbert space.  
 
The second brane parameter comes in by labeling a family of Hamiltonians in this twisted Hilbert space, which we denote by $H_{m_R, \bar{\mu}}$.
 One can find these new Hamiltonians by calculating the AdS length of each time slice ending perpendicularly on the brane as a function of boundary time and Hamiltonian, and use the symplectic form to invert the canonical transformation to find this Hamiltonian: $\omega=dL\wedge dP=dT\wedge dH_{m_R, \bar{\mu}}$. The result is\footnote{We refer the reader to Section 3.5 and Appendix A of \cite{Blommaert:2025avl} for the detailed derivation of the Hamiltonian. Note that an extra complex conjugation has been applied, which swaps the meaning of the boundary cosmological constants: $\mu \rightarrow \bar\mu^{\star}, \quad \bar\mu \rightarrow \mu^{\star}$. Whether or not to take this conjugation is ambiguous due to the non-Hermitian nature of the Hamiltonian in the associated eigenvalue problem: $H=-\frac{1}{\hbar}\cos(\theta),$ where the right-hand side is real. Here we choose to apply the conjugation so that the quantization reproduces the correct two-point function. The original calculation in JT gravity can be found in \cite{Gao:2021uro}.}
\begin{align}
    H_{m_R,\bar{\mu}}=-\frac{1}{\hbar}\left( \frac{1}{2}(e^{iP}+e^{-iP})\sqrt{1+2im_Re^{-L}-e^{-2L}}+\bar\mu e^{-L}\right). 
\end{align}
This Hamiltonian retains the $\hat{P}\sim \hat{P}+2\pi$ symmetry as in the fully-open channel. Thus, gauging this shift symmetry still leads to the discreteness of the length: $L=\hbar n$ as before. Note that the operator ordering has not yet been fixed.
By performing a similarity transformation,
\begin{align}
    H_{m_R,\bar{\mu}} \rightarrow g^{-1}(L)H_{m_R,\bar{\mu}} ~g(L), \quad g(L)=\prod_{k=0}^{\infty}\left( 1+ 2im_Re^{-(L+k\hbar)}-e^{-2(L+k\hbar)} \right)^{-1/2},
\end{align}
and using the commutation relation $[\hat{L},\hat{P}]=i\hbar$, which implies
\begin{align}
  e^{\pm i \hat{P}}g(L)=g(L\pm \hbar)   e^{\pm i \hat{P}},
\end{align}
the Hamiltonian can be cast into the form
\begin{align}
     \hat{H}_{m_R,\bar{\mu}}=-\frac{1}{\hbar}\left(\frac{1}{2}e^{i\hat{P}}+\bar\mu e^{-\hat{L}}+\frac{1}{2}\sqrt{1+2im_Re^{-\hat{L}}-e^{-2\hat{L}}}  e^{-i\hat{P}} \sqrt{1+2im_Re^{-L}-e^{-2L}}\right).
     \label{pre-Hal}
\end{align}
where the operator ordering of the last square-root term remains unfixed. Using (\ref{dic_mass}) and (\ref{mubar_ang}),
and choosing an operator ordering for the last square-root term in (\ref{pre-Hal}), the Hamiltonian becomes
\begin{align}
     \hat{H}_{m_R,\bar{\mu}}=&-\frac{1}{2\hbar}\Big(  e^{i\hat{P}}-i2\cos(\bar\alpha) e^{-\hat{L}}+e^{-i\hat{P}}
     \nonumber\\
     &+ ie^{-\hbar \Delta_R} e^{-\hat{L}} e^{-i\hat{P}}+ ie^{\hbar \Delta_R} e^{-i\hat{P}} e^{-\hat{L}} -e^{-\hat{L}}e^{-i\hat{P}} e^{-\hat{L}}
     \Big).
\end{align}
 By shifting $\hat{L}\rightarrow \hat{L}+\hbar \Delta_R -i\frac{\pi}{2}$\footnote{Here the length shift in the quantization depends only on $\Delta_R$, not on $\bar\alpha$.  This is another way to see that the twisted Hilbert space is labeled solely by $\Delta_R$.}, the Hamiltonian takes the final form
 \begin{align}
      \hat{H}_{m_R,\bar{\mu}}=&-\frac{1}{2\hbar}\left(  e^{i\hat{P}}+(e^{-\hbar\Delta_R+i\bar\alpha}+ e^{-\hbar\Delta_R-i\bar\alpha})e^{-\hat{L}}+\left( 1-e^{-2\hbar \Delta_R}e^{-\hat{L}}\right)e^{-i\hat{P}}\left(1-e^{-\hat{L}} \right)\right).
      \label{Hal_final}
 \end{align}
 
 We now consider the eigenvalue equation for this final form of the Hamiltonian. Here, we choose to normalize the new basis such that all the $m_R,\bar\mu$ dependence resides in the energy basis, denoted by $|\theta\rangle_{\Delta_R,\bar{\alpha}} \in \mathcal{H}_{\Delta}$, while the length basis $|n\rangle \in \mathcal{H}_{\Delta}$ remains independent of these brane parameters. A simple reason is that the length basis is discrete, and thus it would not be possible for it to have non-trivial dependence on continuous parameters. The eigenvalue equation now reads
\begin{align}
   _{\Delta_R,\bar{\alpha}}\langle \theta|\hat{H}_{m,\bar{\mu}}|n\rangle=-\frac{\cos(\theta)}{\hbar}  {_{\Delta_R,\bar{\alpha}}}\langle \theta|n\rangle.
   \label{eigen_func}
\end{align}
Using (\ref{Hal_final}) and noting that $e^{\pm i \hat{P}}|n\rangle=|n\pm1\rangle$, the wavefunction satisfies the following recursion relation:
\begin{align}
    2\cos(\theta) {_{\Delta_R,\bar{\alpha}}\langle} \theta|n\rangle=_{\Delta_R,\bar{\alpha}}{\langle} \theta|n+1\rangle +(q^{2\Delta}e^{i\bar{\alpha}}+q^{2\Delta}e^{-i\bar{\alpha}})q^{n} {_{\Delta_R,\bar{\alpha}}\langle} \theta |n\rangle +(1-q^{4\Delta}q^{n-1})(1-q^{n}){_{\Delta_R,\bar{\alpha}}\langle} \theta|n-1\rangle.
\end{align}
This is exactly the recursion relation for the Al Salam-Chihara polynomial $Q_n(\cos(\theta)|q^{2\Delta}e^{i\bar\alpha},q^{2\Delta}e^{-i\bar{\alpha}};q^2)$.\footnote{This wavefunction can also be directly recovered from the chord Hilbert space of the DSSYK by gauging the chord parity symmetry \cite{Aguilar-Gutierrez:2025hty}.} Taking into account the fact that the left and right eigenfunctions differ for a non-Hermitian Hamiltonian, the wavefunctions can be normalized as \cite{Blommaert:2025avl, Yao:2018fsg}
\begin{align}
   \langle n|\theta\rangle_{\Delta_R,\bar{\alpha}}=  {_{\Delta_R,\bar{\alpha}}\langle\theta} |n\rangle=\frac{Q_n(\cos(\theta)|q^{2\Delta_R}e^{i\bar\alpha},q^{2\Delta_R}e^{-i\bar{\alpha}};q^2)}{(q^2,q^{4\Delta_R};q^2)^{1/2}_n}.
   \label{Qn_right}
\end{align}
The Al Salam-Chihara polynomials satisfy the following orthogonality relation
\begin{align}
    \delta_{n_1,n_2}=\int_{0}^{\pi}d\theta \frac{(q^{4\Delta_R},e^{\pm2i\theta};q^2)_{\infty}}{(q^{2\Delta_R}e^{\pm i\theta\pm i \bar\alpha};q^2)_{\infty}}    \langle n_1|\theta \rangle_{\Delta_R,\bar{\alpha}} \,  {_{\Delta_R,\bar{\alpha}}}\langle \theta|n_2\rangle,
\end{align}
and the completeness relation
\begin{align}
    \sum_{n=0}^{\infty}{_{\Delta_R,\bar{\alpha}}}\langle \theta_1|n\rangle  \langle n|\theta_2\rangle_{\Delta_R,\bar{\alpha}}=\frac{1}{\rho_{\Delta_R,\bar\alpha}(\theta_1)}\delta(\theta_1-\theta_2).
    \label{comp_Al}
\end{align}
Consequently, the modified density of states due to the presence of the brane is obtained as
\begin{align}
    \rho_{\Delta_R,\bar\alpha}(\theta)=\frac{(q^{4\Delta_R};q^2)_{\infty}}{(q^{2\Delta_R}e^{\pm i\theta\pm i\bar{\alpha}};q^2)_{\infty}}\rho(\theta).
    \label{dos_so}
\end{align}
Thus, we have the new resolution of the identity operator in semi-open channel as follows
\begin{align}
    \mathds{1}=\int_{0}^{\pi} d\theta ~\rho_{\Delta_R,\bar\alpha}(\theta) \, |\theta \rangle_{\Delta_R,\bar{\alpha}} \,  {_{\Delta_R,\bar{\alpha}}}\langle \theta| =\sum_{n=0}^{\infty}|n\rangle \langle n|.
    \label{resolution_2}
\end{align}

\paragraph{Partition functions of the left and right regions}

The partition function for the right region can be calculated in either fully-open channel or the semi-open channel, and the results are identical
\begin{align}
    Z_{R}(\beta_R,\bar{\mu},m_R)&=\langle n=0|e^{-\beta_R \hat{H}_{m_R,\bar\mu}}|n=0\rangle_{\mathrm{SC}}= \langle m_R,\bar\mu|e^{-\beta_R \hat{H}_{\mathrm{SD}}}|n=0\rangle_{\mathrm{FC}}
    \nonumber\\
    &=\int_0^{\pi}d\theta \, \rho(\theta) \, e^{\beta_R \frac{\cos(\theta)}{\hbar}} \frac{(q^{4\Delta_R};q^2)_{\infty}}{(q^{2\Delta_R}e^{\pm i\bar\alpha\pm i \theta};q^2)_{\infty}},
\end{align}
where  we have used the property of Al Salam-Chihara plynomials: $Q_0(x|A,B;q)=1$. 

For the left region, the changes we need to make are simply to change all the subscripts $R$ into $L$ and to replace $\bar\mu$ with $-\bar\mu$, which corresponds to shifting $\bar\alpha \rightarrow \bar\alpha+\pi$. The partition function for the left region can therefore be calculated as
\begin{align}
     Z_{L}(\beta_L,\bar{\mu},m_L)=Z_R(\beta_L,-\bar\mu,m_L)
     =\int_{0}^{\pi}d\theta \, \rho(\theta) \, e^{\beta_L \frac{\cos(\theta)}{\hbar}} \frac{(q^{4\Delta_L};q^2)_{\infty}}{(-q^{2\Delta_L}e^{\pm i\bar\alpha\pm i \theta};q^2)_{\infty}}.
\end{align}
Importantly, the complex conjugation of the left region partition function should be taken by treating $Z_L$ as a function of the real variable $\bar\mu$ rather than $\bar\alpha$. The analytic continuation to $\bar\mu=-i\cos(\bar\alpha)$ is performed only after taking the complex conjugation.\footnote{Simply speaking, the analytic continuation $\bar\mu=-i\cos(\bar\alpha)$ does not commute with the complex conjugation. The reason why we should treat $\bar\mu$ as the real variable in $Z_L$ can be seen from (\ref{AdS_bdycond}) and (\ref{2_mass}) by setting $m_R=0$, where the boundary conditions in AdS space reduce to the usual EOW brane boundary conditions in JT gravity, and $m_{\mathrm{AdS}}$ in this case should be real. } As a result, the conjugation gives
\begin{align}
     Z_{L}^{\star}(\beta_L,\bar{\mu},m_L)=Z_L(\beta_L,-\bar\mu,m_L)=Z_R(\beta_L,\bar\mu,m_L)
     =\int_{0}^{\pi}d\theta \, \rho(\theta) \, e^{\beta_L \frac{\cos(\theta)}{\hbar}} \frac{(q^{4\Delta_L};q^2)_{\infty}}{(q^{2\Delta_L}e^{\pm i\bar\alpha\pm i \theta};q^2)_{\infty}}.
\end{align}
We emphasize again the following dictionary
\begin{align}
    m=\cosh (\hbar \Delta), \quad \bar{\mu}=-i\cos(\bar\alpha).
\end{align}

\vspace{5pt}
\subsection{Gluing condition and the two-point function}
We now connect the two-point function to the partition functions of the left and right regions and identify the gluing condition. We compare our gluing procedure (\ref{2pt_measure}) to the calculation in the fully closed (FC) channel, in which the matter particle is represented as an operator, as discussed above. By inserting an identity operator expressed in the energy basis, the two-point function can be cast into the form
\begin{align}
    \langle \mathcal{O}(\tau) \mathcal{O}(0)\rangle_{\beta} &= \langle n=0|e^{-(\beta-\tau) \hat{H}_{\mathrm{SD}}} e^{-\Delta \hat{L}} e^{-\tau \hat{H}_{\mathrm{SD}}} |n=0\rangle_{\mathrm{FC}},
     \nonumber\\
    &= \int_{0}^{\pi} d \bar\alpha ~\rho(\bar\alpha) ~
    \langle n=0|e^{-(\beta-\tau) \hat{H}_{\mathrm{SD}}}e^{-\Delta_L \hat{L}}|\bar{\alpha}\rangle_{\mathrm{FC}}  ~\langle\bar\alpha| e^{-\Delta_R \hat{L}} e^{-\tau \hat{H}_{\mathrm{SD}}} |n=0\rangle_{\mathrm{FC}},
    \label{2pt_alpha}
    \\
    &=\int_{0}^{\pi}  d \bar\alpha~ \rho(\bar\alpha)~\Psi_L(\beta-\tau,\bar\alpha,\Delta_L) \Psi_R(\tau,\bar{\alpha},\Delta_R),
     \nonumber\\
    &=\int_{0}^{\pi}  d \bar\alpha~ \rho(\bar\alpha) ~Z_{L}^{\star}(\beta-\tau,\bar\mu,m_L) Z_{R}(\tau,\bar\mu,m_R),
    \label{2pt_parti}
\end{align}
where in the second line we have split $e^{-\Delta \hat{L}}=e^{-\Delta_L \hat{L}}e^{-\Delta_R \hat{L}}$ with $\Delta=\Delta_L+\Delta_R$ and inserted a resolution of identity operator in the energy basis. This yields a convolution of two wavefunctions in the third line. In the last line we have identified the wavefunctions $\Psi_L(\beta-\tau,\bar\alpha,\Delta_L) $ and $\Psi_R(\tau,\bar{\alpha},\Delta_R)$ appearing in two-point function calculation with the partition function of the left  and right regions written in the fully-open channel. For example,
\begin{align}
    \langle\bar\alpha| e^{-\Delta_R \hat{L}} e^{-\tau \hat{H}_{\mathrm{SD}}} |n=0\rangle=\langle\bar\alpha| q^{2\Delta_R \hat{n}} e^{-\tau \hat{H}_{\mathrm{SD}}} |n=0\rangle=\langle m_R,\bar\mu|  e^{-\tau \hat{H}_{\mathrm{SD}}} |n=0\rangle=Z_{R}(\tau,\bar\mu,m_R).
\end{align}
Note that the use of (\ref{Dic_branestate}) and the relation (\ref{dic_massLR}) are crucial here.
By comparing with (\ref{2pt_measure}),  the integration domain $D$ and the measure $\mathcal{M}(\bar\mu)$ for patching the left and right region partition functions can be expressed using the $\bar\alpha$ variable as
\begin{align}
    \int_D d\bar\mu \mathcal{M}(\bar\mu)=\int_0^{\pi}d\bar\alpha \, \rho(\alpha).
    \label{glue_cond}
\end{align}
This determines the desired gluing condition.


We now have multiple ways to calculate the two-point function, as each partition function can be computed in either fully-open or semi-open channel. At this stage, these approaches are trivially equivalent. For example, by placing both sides in the semi-open channel, the two-point function can be obtained by
\begin{align}
     \langle \mathcal{O}(\tau) \mathcal{O}(0)\rangle_{\beta} &= \int_{0}^{\pi}d\bar{\alpha}~ \rho(\bar{\alpha}) \langle n_{L}=0|e^{-\tau \hat{H}_{m_L,\bar{\mu}}}|n_{L}=0\rangle_{\mathrm{SC}} \, \langle n_{R}=0|e^{-\tau \hat{H}_{m_R,\bar{\mu}}}|n_{R}=0\rangle_{\mathrm{SC}},
     \nonumber\\
     &=
     \int_{0}^{\pi} d\bar\alpha \rho(\bar\alpha) \int_0^{\pi} \int_0^{\pi} d\theta_L d\theta_R ~\rho_{\Delta_L,\bar\alpha}(\theta_L) \rho_{\Delta_R,\bar\alpha}(\theta_R) \, e^{(\beta-\tau) \frac{\cos(\theta_L)}{\hbar}} e^{\tau \frac{\cos(\theta_R)}{\hbar}} 
     \label{2pt_sdp}
     \\
     &= \int_0^{\pi} \int_0^{\pi} d\theta_L d\theta_R ~\rho(\theta_L) \rho(\theta_R)\, e^{(\beta-\tau) \frac{\cos(\theta_L)}{\hbar}} e^{\tau \frac{\cos(\theta_R)}{\hbar}} \nonumber
     \\
     &\quad \times \int_{0}^{\pi} d\bar\alpha \rho(\bar\alpha) \frac{(q^{4\Delta_L};q^2)_{\infty}}{(q^{2\Delta_L}e^{\pm i\theta_L}e^{\pm i \bar\alpha};q^2)_{\infty}}\frac{(q^{4\Delta_R};q^2)_{\infty}}{(q^{2\Delta_R}e^{\pm i\theta_R\pm i \bar\alpha};q^2)_{\infty}}
     \nonumber
     \\
     &=\int_{0}^{\pi} \int_{0}^{\pi}d\theta_L d\theta_R \, \rho(\theta_L) \rho(\theta_R) e^{(\beta-\tau) \frac{\cos(\theta_1)}{\hbar}} e^{\tau \frac{\cos(\theta_2)}{\hbar}}  \frac{(q^{4\Delta};q^2)_{\infty}}{(q^{2\Delta}e^{\pm i\theta_L\pm i \theta_R};q^2)_{\infty}}, 
\end{align}
which indeed matches (\ref{2-pt_correlator1}). To get to the last line, we used the remarkable Askey–Wilson integral result \cite{askey1985some}
\begin{align}
    \int_0^{\pi} \frac{d\theta}{2\pi}  \frac{(q^2;q^2)_{\infty}(e^{\pm2i\theta};q^2)_{\infty}}{(Ae^{\pm i \theta};q^2)_{\infty}(Be^{\pm i \theta};q^2)_{\infty}(Ce^{\pm i \theta};q^2)_{\infty}(De^{\pm i \theta};q^2)_{\infty}}=\frac{(ABCD;q^2)_{\infty}}{(AB,AC,AD,BC,BD,CD;q^2)_{\infty}}.
    \label{AW-integral}
\end{align}
Indeed the above splitting and gluing procedure can be thought of as geometrization of this identity.

\vspace{5pt}
\section{Wormhole Hilbert space with matter}

In this section, we clarify how the splitting and gluing procedure developed above leads to a correct way of patching the left and right states on a global Cauchy slice that crosses the shell. We refer to the Hilbert space defined on this global Cauchy slice as the "wormhole Hilbert space with matter", following the terminology of \cite{Lin:2022rbf}. In our conventions, the wormhole Hilbert spaces discussed in this section are labeled by the number of matter particles inserted on the boundary\footnote{For example, the one-particle Hilbert space describes a single thin shell dividing the Cauchy slice into two parts.}. In particular, we start by discussing the one-particle Hilbert space, which will be essential for computing the crossed four-point function in the next section. Moreover, since the splitting and gluing procedure can be used iteratively, there exist different representations of this one-particle Hilbert space, labeled by the number of splittings. Surprisingly, we also find that these different representations, when viewed locally in the bulk, automatically give rise to the multi-particle Hilbert space, which we will use to compute general correlation functions.


\vspace{5pt}
\subsection{One-particle Hilbert space}
\label{1-parti_HS}
\begin{figure}[htbp]
    \centering
    \subfigure[]{\includegraphics[width=0.4\textwidth]{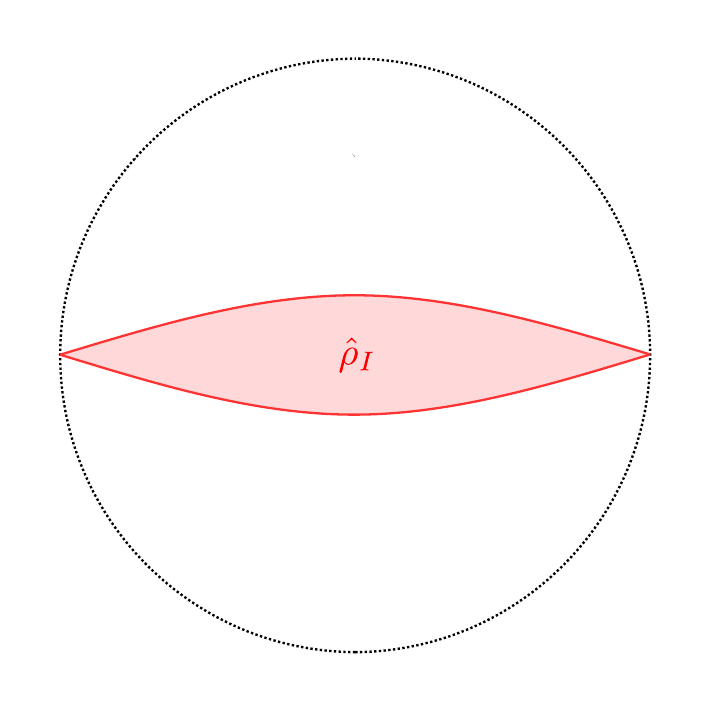}} 
    \subfigure[]{\includegraphics[width=0.385\textwidth]{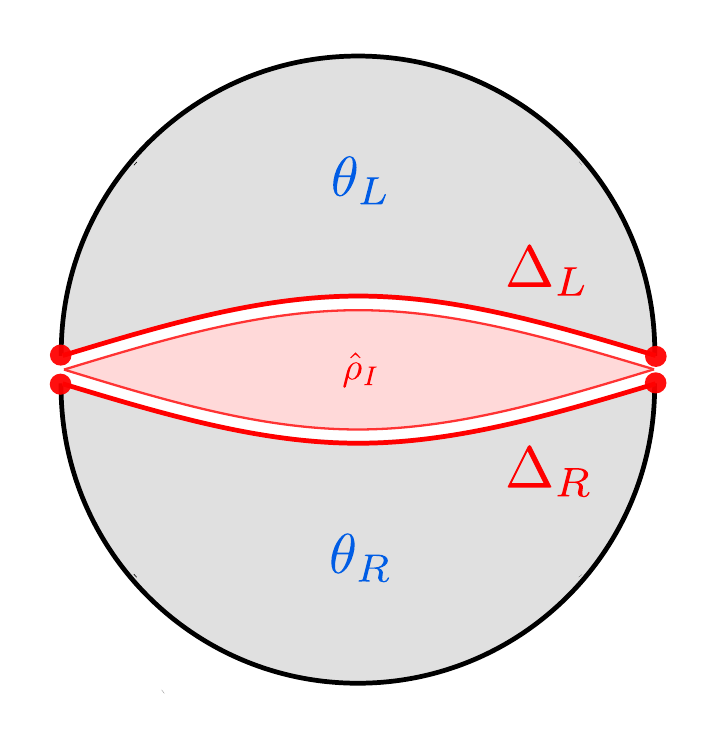}} 
    \caption{Bulk pictures in the fully-open channel: (a) Identity operator as a density matrix; (b) Gluing of two regions with each  having an EOW brane boundary.}
    \label{patching_grav}
\end{figure}

Let us discuss the derivation of the gluing condition more geometrically. In the fully open channel we insert a resolution of the identity operator expressed in the energy basis in the two-point function. The key component of this process in (\ref{2pt_alpha}) is
\begin{align}
    \langle\theta_L|e^{-\Delta \hat{L}}|\theta_R\rangle_{\mathrm{FC}}=\int_0^{\pi} d\bar\alpha ~\rho(\bar\alpha) \langle\theta_L|e^{-\Delta_L \hat{L}}|\bar\alpha\rangle _{\mathrm{FC}} \langle\bar\alpha|e^{-\Delta_R \hat{L}}|\theta_R\rangle_{\mathrm{FC}}.
    \label{1-split}
\end{align}
In gravity, one should think of this identity operator as a "density matrix"
\begin{align}
    \hat\rho_{I}=\int d\bar\alpha \rho(\bar\alpha) |\bar\alpha \rangle \langle\bar\alpha|.
\end{align}
This density matrix corresponds to a portion of the $\mathrm{AdS}_2$ disk with two open cuts, shown as the red region in Fig.\ref{patching_grav}(a). The two open cuts end at the same points on the asymptotic boundary because this density matrix involves no boundary evolution\footnote{In this sense, the two cuts are related by bulk diffeomorphisms.}. The gluing procedure can therefore be visualized as in Fig.\ref{patching_grav}(b).
Note that in this channel the splitting of a single particle geodesic into two and then gluing them back is trivial.

Now we use this geometrical picture in  the semi-open channel, where the matter worldline is now vertical. To obtain a Hilbert space we cut open along an horizontal Cauchy slice which crosses the matter worldline (i.e., the wormhole slice). This gives rise to the wormhole Hilbert space with a single matter particle on that Cauchy slice. Let us now discuss the construction in detail.  

\begin{figure}[htbp]
    \centering
    \subfigure[]{\includegraphics[width=0.4\textwidth]{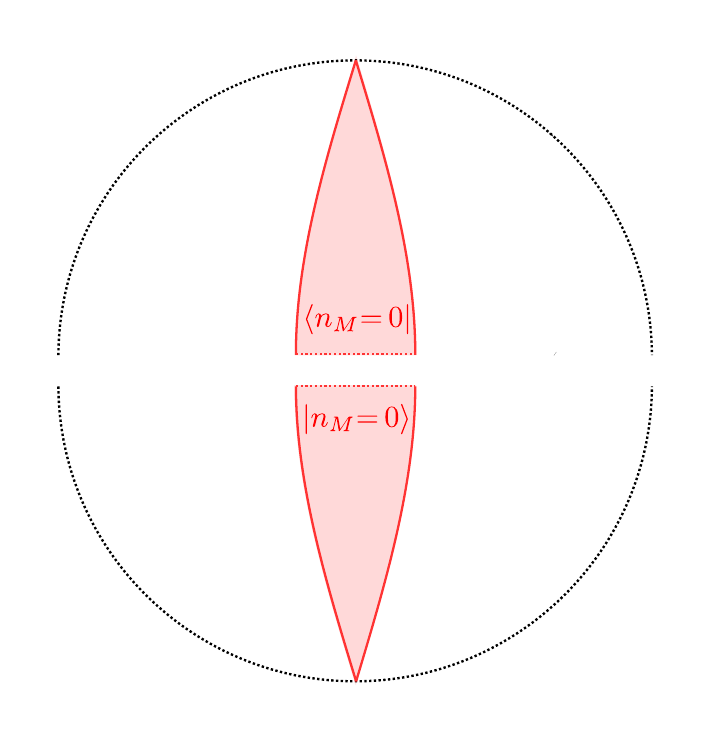}} 
    \subfigure[]{\includegraphics[width=0.4\textwidth]{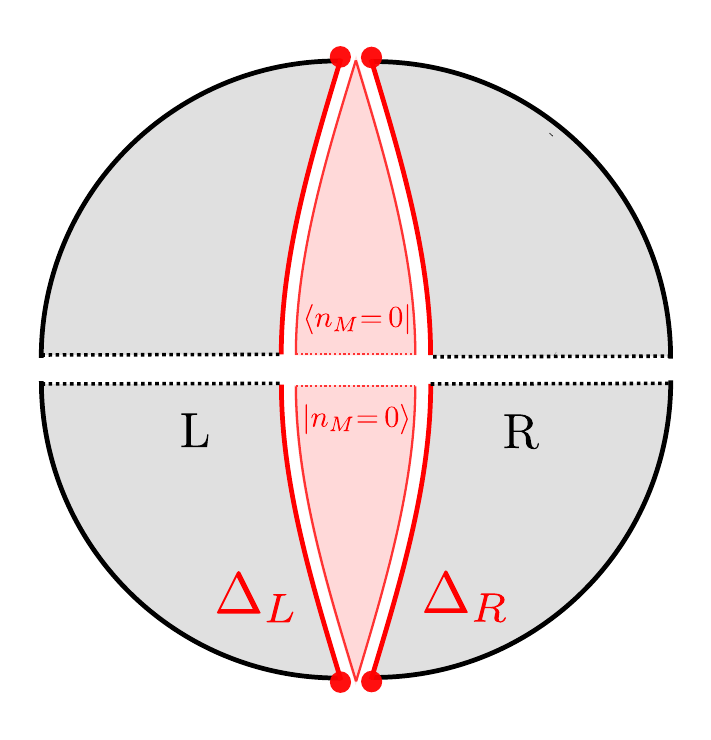}} 
    \caption{Bulk pictures in the semi-open channel: (a) $|n_M=0\rangle$ state; (b) State on the wormhole slice.}
    \label{patching_grav_SC}
\end{figure}

For the identity operator in semi-open channel, as illustrated in Fig. \ref{patching_grav_SC}(a), there is no time evolution along the boundary. Thus, a cut in the middle just prepares a state with $n=0$, reflecting the "absence of structure" \cite{Blommaert:2024whf,Blommaert:2023wad}\footnote{ As discussed in \cite{Blommaert:2024whf}, setting $n_M=0$ implements a smooth gluing of the left and right boundaries. In the bulk, this can be understood as ensuring that the two split geodesics coincide, leaving no intermediate space between them.}
\begin{align}
   |n_M=0\rangle=\int \, d \bar\alpha \, \rho(\bar\alpha) |\bar\alpha\rangle,
   \label{n=0}
\end{align}

To define the wormhole Hilbert space with one particle, we rotate Fig.\ref{patching_grav}(b) to obtain the corresponding picture in the semi-open channel, and cut it open as showed in Fig.\ref{patching_grav_SC}(b).  The resulting wormhole slice consists of three regions: the left and right regions, which involve states in the twisted Hilbert spaces $\mathcal{H}_{\Delta_L}$ and $\mathcal{H}_{\Delta_R}$ due to the presence of EOW brane boundaries, and a middle region represented by the state $|n_M=0\rangle \in \mathcal{H}_{\mathrm{SD}}$. 

We denote the energy and length bases of the global states on this wormhole slice (after the gluing procedure) by $|\Delta;\theta_L,\theta_R\rangle$ and $|\Delta;n_L,n_R\rangle$, where  $\theta_L/n_L$ and $\theta_R/n_R$ label the energy/length of the left and right regions, respectively, and $\Delta=\Delta_L+\Delta_R$. We now discuss how each of these bases is obtained by patching together the corresponding bases in the three regions, as implied by the gluing condition.

For the length basis $|\Delta;n_L,n_R\rangle$, note that the gluing condition (\ref{glue_cond}) only involves the integration over the brane parameter, which the length basis in the twisted Hilbert space is independent of, as mentioned above (\ref{eigen_func}). Therefore,  $|\Delta;n_L,n_R\rangle$ is simply a product state:
\begin{align}
    |\Delta;n_L,n_R\rangle=|n_L\rangle\otimes|n_M=0\rangle \otimes |n_R\rangle.
\end{align}

For the energy basis  $|\Delta;\theta_L,\theta_R\rangle$, the gluing condition (\ref{glue_cond}) and the two-point function in semi-open channel (\ref{2pt_sdp}) shows that,  the correct way to patch the left and right energy bases is to first decompose the middle region state $|n_M=0\rangle$ in the energy basis as in (\ref{n=0}), and  then associate the energy label $\bar{\alpha}$ in the middle region with EOW branes for the left and right regions. The Hilbert space on either side is now a twisted Hilbert space with brane parameters $(\Delta_{L/R}, \bar{\alpha})$.
In this way, the left and right Hilbert spaces are "state-dependent'', in the sense that the Hamiltonian in each region is determined  with respect to the choice of energy eigenstate in the middle region. It follows that, $|\Delta;\theta_L,\theta_R\rangle$ can be decomposed  as
\begin{align}
    |\Delta;\theta_L,\theta_R\rangle=\adjincludegraphics[width=4cm,valign=c]{figure_pdf/1split_SC.pdf} =\int_0^{\pi} d\bar\alpha \rho(\bar\alpha)\, |{\theta_L}\rangle_{\Delta_L,\bar\alpha} \otimes^{\bar\alpha} |\bar\alpha\rangle \otimes^{\bar\alpha} |\theta_R\rangle_{\Delta_R,\bar\alpha},
    \label{1-split_rep}
\end{align}
where we used the notation $\otimes^{\bar\alpha}$ to indicate that the structure is not a tensor product of three states: the patching induces an explicit dependence of the left and right states  on the state in the middle region. This state-dependence breaks the factorization. However, since the length bases in each region do factorize, this non-factorized energy basis can be viewed as an entangled state once the energy basis in each region is decomposed into the corresponding length basis
\begin{align}
     |\Delta;\theta_L,\theta_R\rangle=\sum_{n_L=0}^{\infty} \sum_{n_M=0}^{\infty} \sum_{n_R=0}^{\infty}  \, \int_0^{\pi} d\bar\alpha \, \rho(\bar\alpha)  \,U_{\theta_L,\bar\alpha,\theta_R}^{n_L,n_M,n_R} \, |n_L\rangle\otimes|n_M\rangle \otimes |n_R\rangle, 
\end{align}
where the change-of-basis operator $\hat U$ is unitary up to normalization (as the energy bases are not normalized), which we refer to as "quasi-unitary". Its matrix elements are given by
\begin{align}
U_{\theta_L,\bar\alpha,\theta_R}^{n_L,n_M,n_R}=\frac{Q_{n_L}(\cos(\theta_L)|q^{2\Delta_L}e^{i\bar\alpha},q^{2\Delta_L}e^{-i\bar{\alpha}};q^2)}{(q^2,q^{4\Delta_L};q^2)^{1/2}_{n_L}} \frac{H_{n_M}(\cos(\bar\alpha)|q^2)}{(q^2;q^2)_{n_M}^{1/2}}\frac{Q_{n_R}(\cos(\theta_R)|q^{2\Delta_R}e^{i\bar\alpha},q^{2\Delta_R}e^{-i\bar{\alpha}};q^2)}{(q^2,q^{4\Delta_R};q^2)^{1/2}_{n_R}}  .
\label{cob_matrix}
\end{align}
Importantly, this basis transformation is quasi-unitary globally, but not locally: $\hat{U}$ cannot be decomposed into a tensor product of three quasi-unitary operators acting separately in each region, as can be seen from its matrix elements
\begin{align}
    U_{\theta_L,\bar\alpha,\theta_R}^{n_L,n_M,n_R}\neq U_{\theta_L}^{n_L} \times {U^{\prime}}_{\bar\alpha}^{n_M} \times {U^{\prime \prime}}_{\theta_R}^{n_R}.
\end{align}
This is because in (\ref{cob_matrix}), the $\bar\alpha$ index couples non-linearly to both left and right region indices via the Al Salam-Chihara polynomial. This non-linear mixing produces a non-local basis transformation, which is the origin of the state-dependence and the breakdown of factorization in the energy basis.

For later convenience, we call the above decompositions of the one-particle state the "1-split representation", and we indicate this by adding a subscript 1 to the states
\begin{align}
     \textbf{1-split representation:} \quad   &|\Delta;n_L,n_R\rangle_1=|n_L\rangle\otimes|n_M=0\rangle \otimes |n_R\rangle 
     \nonumber
     \\
     &  |\Delta;\theta_L,\theta_R\rangle_1=\int_0^{\pi} d\bar\alpha \rho(\bar\alpha)\, |{\theta_L}\rangle_{\Delta_L,\bar\alpha} \otimes^{\bar\alpha} |\bar\alpha\rangle \otimes^{\bar\alpha} |\theta_R\rangle_{\Delta_R,\bar\alpha}.
\end{align}
The reason for this name will become clear presently.
\footnote{\label{LS_compare}Here we make some more comments on the factorization of the length basis.

In some previous works \cite{Lin:2022rbf, Lin:2023trc, xu2025vonneumannalgebrasdoublescaled}, the length basis of their one-particle Hilbert space is  orthogonal only with respect to the total length: $\langle\Delta;n_L^{\prime},n_R^{\prime}|\Delta;n_L,n_R\rangle \propto \delta_{n_L^{\prime}+n_R^{\prime},n_L+n_R}$, while in our case (and similarly \cite{Okuyama:2024yya,Okuyama:2024gsn}), we have independent orthogonality of the length indices in the left and right regions: ${_1}\langle\Delta;n_L^{\prime},n_R^{\prime}|\Delta;n_L,n_R\rangle_1=\delta_{n_L^{\prime},n_L}\,\delta_{n_R^{\prime},n_R}$. The difference comes from the precise definition of the bra and ket states and their inner products on the bulk wormhole Cauchy slice, which we compare below:

\begin{center}
    \includegraphics[width=0.3\textwidth]{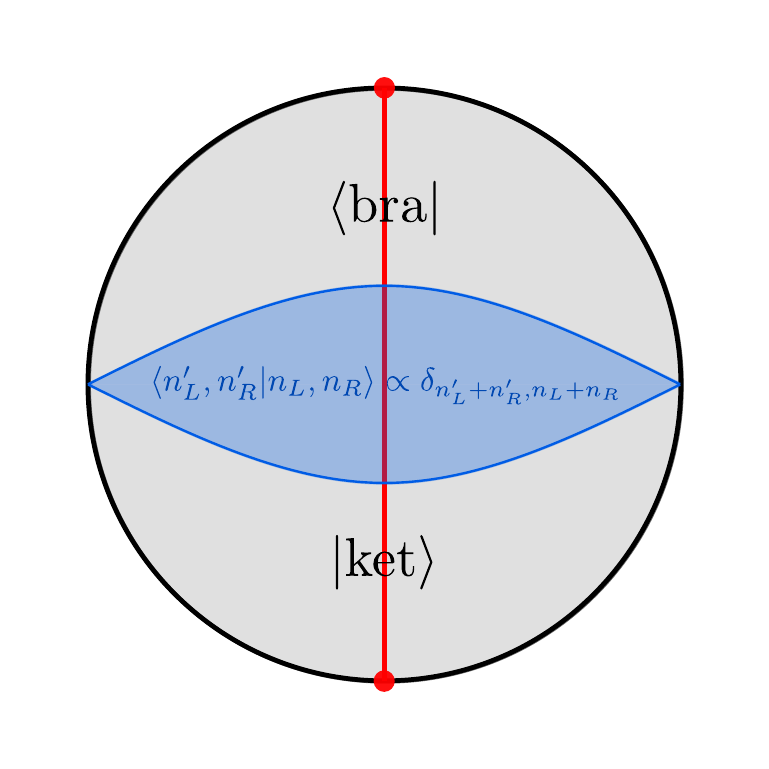} \quad  \quad 
    \includegraphics[width=0.315\textwidth]{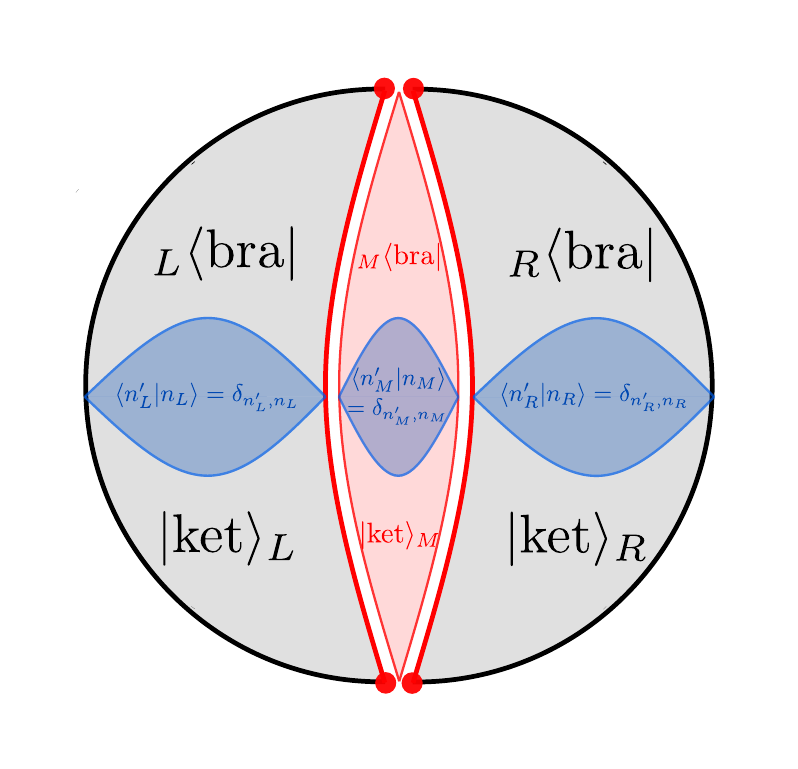}.
\end{center}

In both figures, the two-point function is expressed as the inner product between the top states $\langle\mathrm{bra}|$ and the bottom states $|\mathrm{ket}\rangle$, with  the middle blue region defining the equal Euclidean time inner product.

The left figure illustrates the non-factorized length basis in \cite{Lin:2022rbf, Lin:2023trc, xu2025vonneumannalgebrasdoublescaled}. In DSSYK, this non-factorization comes from the fact that chords on either side can cross the matter chords in the middle. More generally, any geometry dual to DSSYK on this slicing should exhibit this feature. The reason is that the DSSYK has chord rearrangement symmetry, which should become a gauge symmetry in the dual geometry. The combination $n_L+n_R$ is independent of the ordering of boundary chords relative to the matter line, and is thus gauge invariant, while $n_L-n_R$ is not \cite{Lin:2023trc}. Consequently, the orthogonality of $|\Delta;n_L,n_R\rangle$  can only be chosen with respect to the total length $n_L+n_R$. 

Our case corresponds to the right figure, where we split the matter line into two physical EOW brane boundaries and applied EOW brane quantization separately in the left and right regions. The intersecting points on each EOW brane are uniquely determined once the Euclidean time is fixed on the boundary, since EOW brane quantization requires a geodesic shooting from the asymptotic boundary at fixed Euclidean time ends perpendicularly to the EOW brane boundary. This is consistent with the boundary condition that the ADM energy of the EOW brane is zero.  The length bases $|n_L\rangle$ and $|n_R\rangle$ are individually orthonormal and gauge invariant within each region. During the Euclidean time evolution at asymptotic boundary, the states in different regions are dynamically patched together via the non-factorized energy basis.
}

We can also define the Laplace conjugate states of the energy basis by specifying the asymptotic boundary length $\beta_L$ and $\beta_R$, which is also called the Hartle-Hawking state
\begin{align}
     &|\Delta;\beta_L,\beta_R\rangle_1 
     \nonumber\\
     &=  \int_0^{\pi} \, d\bar\alpha \, \rho(\bar\alpha) \left(e^{-\beta_L\hat{H}_{m_L,\bar{\mu}}} |n_L=0\rangle \right) \otimes^{\bar\alpha} |\bar\alpha\rangle \otimes^{\bar\alpha}  \left(e^{-\beta_R\hat{H}_{m_R,\bar{\mu}}} |n_R=0\rangle \right)
     \nonumber\\
     &=\int_0^{\pi} d\bar\alpha \, \rho(\bar\alpha) \int_0^{\pi}  d\theta_L   \, \rho_{\Delta_L,\bar\alpha}(\theta_L) \, e^{\beta_L\frac{\cos(\theta_L)}{\hbar}} \int_0^{\pi} d\theta_R \,
     \rho_{\Delta_R,\bar\alpha}(\theta_R)  \, e^{\beta_R\frac{\cos(\theta_R)}{\hbar}}
     |{\theta_L}\rangle_{\Delta_L,\bar\alpha} \otimes^{\bar\alpha} |\bar\alpha\rangle \otimes^{\bar\alpha} |\theta_R\rangle_{\Delta_R,\bar\alpha}.
     \label{1split_HH}
\end{align}
One can check that the inner product of two such states indeed reproduces the two-point function in (\ref{2pt_sdp}), by using the orthogonality property of the energy basis in the twisted Hilbert space (\ref{comp_Al})
\begin{align}
    {_{\Delta,\bar{\alpha}}}\langle \theta_1|\theta_2\rangle_{\Delta_R,\bar{\alpha}}=\frac{1}{\rho_{\Delta,\bar\alpha}(\theta_1)}\delta(\theta_1-\theta_2).
\end{align}

At this point, one might have the following straightforward questions: What happens if we don't split the geodesic of the particle? And conversely, what happens if we keep splitting the geodesics further? The short answer is that these choices lead to different bases and correspond to different representations of the wormhole Hilbert space with one matter particle. However, they will all yield the same physical results for any observable. In particular, the splitting procedure is iterative, and when viewed locally in the bulk, it gives rise to a transparent construction of the multi-particle Hilbert space, which we discuss in the next subsection. We now discuss these generalizations of the splitting and gluing process in more detail.

First, we can choose not to split the particle geodesic, rather associate it asymmetrically to the left or the right regions. For example, if we associate it to the right region, this corresponds to inserting the identity operator in the fully-open channel as follows
\begin{align}
    \langle\theta_L|e^{-\Delta \hat{L}}|\theta_R\rangle_{\mathrm{FC}}=\int_0^{\pi} d\bar\alpha \, \rho(\bar\alpha) \, \langle\theta_L|\bar\alpha\rangle _{\mathrm{FC}} \langle\bar\alpha|e^{-\Delta \hat{L}}|\theta_R\rangle_{\mathrm{FC}}.
    \label{0-split}
\end{align}
In this case, the right region still has an EOW brane with the brane parameter $(\Delta,\bar\alpha)$, while the left regions is the standard sine-dilaton gravity with the bulk energy parameter fixed to $\bar\alpha$, since $\langle\theta_L|\bar\alpha\rangle=\frac{\delta(\theta_L-\bar\alpha)}{\rho(\bar\alpha)}$. Following the previous analysis, in the semi-open channel, the length and energy bases of the one-particle state on the wormhole slice in this case can be decomposed in the 0-split representation as
\begin{align}
   \textbf{0-split representation:}  &\quad |\Delta;n_L,n_R\rangle_0=|n_L\rangle \otimes |n_R\rangle
   \nonumber
   \\
   &\quad|\Delta;\theta_L,\theta_R\rangle_0=|\theta_L\rangle\otimes^{\theta_L} |\theta_R\rangle_{\Delta,\theta_L}.
    \label{0-split_rep}
\end{align}
The Hartle-Hawking state becomes
\begin{align}
  |\Delta;\beta_L,\beta_R\rangle_0=\int_0^{\pi}d\theta_L \, \rho(\theta_L) \, e^{\beta_L \frac{\cos(\theta_L)}{\hbar}} \int_0^{\pi} d\theta_R \, \rho_{\Delta,\theta_L}(\theta_R) \, e^{\beta_R\frac{\cos(\theta_R)}{\hbar}} |\theta_L\rangle\otimes^{\theta_L} |\theta_R\rangle_{\Delta,\theta_L}.
  \label{0split_HH}
\end{align}
The inner product of two such states again reproduces the two-point function.

Next, we can continue splitting the geodesics by inserting additional identity operators into (\ref{1-split}). For example, one can further split the term  $\langle\bar\alpha|e^{-\Delta_R \hat{L}}|\theta_R\rangle_{\mathrm{FC}}$ in (\ref{1-split}),
\begin{align}
    \langle\bar\alpha|e^{-\Delta_R \hat{L}}|\theta_R\rangle_{\mathrm{FC}}=\int_0^{\pi}d\bar\beta \, \rho(\bar\beta) \, \langle\bar\alpha|e^{-\Delta_{R_1} \hat{L}}|\bar\beta \rangle_{\mathrm{FC}}\langle\bar\beta|e^{-\Delta_{R_2}\hat{L}}|\theta_R\rangle_{\mathrm{FC}},
    \label{2-split}
\end{align}
where $\Delta_R=\Delta_{R_1}+\Delta_{R_2}$. This insertion has exactly the same form as going from (\ref{0-split}) to (\ref{1-split}), where the corresponding energy basis changes from $|\theta_L\rangle\otimes^{\theta_L} |\theta_R\rangle_{\Delta,\theta_L}$ to $\int_0^{\pi} d\bar\alpha \, \rho(\bar\alpha) \, |{\theta_L}\rangle_{\Delta_L,\bar\alpha} \otimes^{\bar\alpha} |\bar\alpha\rangle \otimes^{\bar\alpha} |\theta_R\rangle_{\Delta_R,\bar\alpha}$. Thus, the length and energy bases in the 2-split representation can be written as
\begin{align}
    &\textbf{2-split representation:}  
    \nonumber\\
    &\quad   |\Delta;n_L,n_R\rangle_2= |n_L\rangle \otimes |n_{M_1}=0\rangle \otimes |n_{M_2}=0\rangle \otimes |n_R\rangle
    \nonumber
    \\
    &\quad
    |\Delta;\theta_L,\theta_R\rangle_2=\int_0^{\pi} \int_0^{\pi} d\bar\alpha \, d\bar\beta \, \rho_{\Delta_{R_1},\bar\beta}(\bar\alpha)  ~\rho(\bar\beta)~ |{\theta_L}\rangle_{\Delta_L,\bar\alpha} \otimes^{\bar\alpha} |\bar\alpha\rangle_{\Delta_{R_1},\bar\beta} \otimes^{\bar\beta} |\bar\beta\rangle \otimes^{\bar\beta} |\theta_R\rangle_{\Delta_{R_2},\bar\beta}.
    \label{2-split_rep}
\end{align}
Here, $\Delta=\Delta_L+\Delta_{R_1}+\Delta_{R_2}$. Note that we have also replaced $\rho(\bar\alpha)$ with $\rho(\bar\alpha)_{\Delta_{R_1},\bar\beta}$, since now the $\alpha$-region is twisted by EOW brane state $e^{-\Delta_{R_1}\hat{L}}|\bar\beta\rangle$ in the fully-open channel, as can be seen in (\ref{2-split}). Accordingly, in the semi-open channel, one should expand $|n_{M_1}=0\rangle$ state in the twisted Hilbert space as: $|n_{M_1}=0\rangle=\int_0^{\pi}d\bar\alpha \, \rho(\bar\alpha)_{\Delta_{R_1},\bar\beta} ~|\bar\alpha\rangle_{\Delta_{R_1},\bar\beta}$.

This splitting procedure can be iterated. In the fully-open channel, each splitting step corresponds to a further insertion of an identity operator into the middle of the boundary-to-boundary operator $e^{-\Delta \hat{L}}$. In the semi-open channel, it corresponds to the following change of energy basis
\begin{align}
|\bar\alpha\rangle\otimes|\bar\beta\rangle_{\Delta,\bar\alpha}\rightarrow \int_0^{\pi}d\bar\gamma \, \rho(\bar\gamma) \, |\bar\alpha\rangle_{\Delta_1,\bar\gamma} \otimes^{\bar\gamma}  |\bar\gamma\rangle \otimes^{\bar\gamma} |\bar\beta\rangle_{\Delta_2,\bar\gamma},
\end{align}
with $\Delta=\Delta_1+\Delta_2$, and the density of states changes accordingly.

We summarize the splitting procedure and the resulting representations of the energy basis of the one-particle Hilbert space in Table~\ref{rep_table}. In each case, we have $\sum_{i=1}^{n}\Delta_i=\Delta$. The length basis always simply factorizes as
\begin{align}
    |\Delta;n_L,n_R\rangle_n=|n_L\rangle \otimes |n_{M_1}=0\rangle \otimes \cdot\cdot\cdot \otimes |n_{M_n}=0\rangle \otimes |n_R\rangle.
\end{align}

One can also check that all these representations reproduce the same two-point function. For example, the Hartle-Hawking state in the $n$-split representations is given by
\begin{align}
    |\Delta;\beta_L,\beta_R\rangle_n
    
    &=\prod_{i=1}^{n-1}\left(\int_0^{\pi} d\bar\alpha_i  \rho_{\Delta_{i+1},\bar\alpha_{i+1}}(\bar\alpha_i) \right) \int_0^{\pi} d\bar\alpha_n \rho(\bar\alpha_n) \nonumber 
    \\
    &\quad\cdot \int_0^{\pi}  d\theta_L   \, \rho_{\Delta_1,\bar\alpha_1}(\theta_L) \, e^{\beta_L\frac{\cos(\theta_L)}{\hbar}}\int_0^{\pi}  d\theta_R   \, \rho_{\Delta_{n+1},\bar\alpha_n}(\theta_R) \, e^{\beta_R\frac{\cos(\theta_R)}{\hbar}}  \nonumber \\
&\quad \cdot|{\theta_L}\rangle_{\Delta_1,\bar\alpha_1} \otimes^{\bar\alpha_1} 
|\bar\alpha_1\rangle_{\Delta_{2},\bar\alpha_2}  \otimes ^{\bar\alpha_2} \cdot\cdot\cdot \otimes^{\bar\alpha_{n-1}}  |\bar\alpha_{n-1}\rangle_{\Delta_n,\bar\alpha_n}\otimes^{\bar\alpha_n}  |\bar\alpha_n\rangle\otimes^{\bar\alpha_n} 
|\theta_R\rangle_{\Delta_{n+1},\bar\alpha_n}.
\end{align}
The inner product of two $n$-split Hartle-Hawking states recovers the two-point function
\begin{align}
   &_n\langle\Delta;\beta_L^{\prime},\beta_R^{\prime} |\Delta;\beta_L,\beta_R\rangle_n
   \nonumber\\
   &=  \int_0^{\pi}  d\theta_L   \, \rho_{\Delta_1,\bar\alpha_1}(\theta_L) \, e^{\beta_L\frac{\cos(\theta_L)}{\hbar}}\prod_{i=1}^{n-2}\left(\int_0^{\pi} d\bar\alpha_i  \rho_{\Delta_{i+1},\bar\alpha_{i+1}}(\bar\alpha_i) \right) \int_0^{\pi} d\bar\alpha_{n-1}  \rho(\bar\alpha_{n-1}) \nonumber
   \\
   &\quad \times \left( \int_{0}^{\pi}d\bar\alpha_{n} \, \rho(\bar\alpha_{n}) \frac{(q^{4\Delta_{n}};q^2)_{\infty}}{(q^{2\Delta_n}e^{\pm i\bar\alpha_{n-1}\pm i \bar\alpha_n};q^2)_{\infty}} \frac{(q^{4\Delta_{n+1}};q^2)_{\infty}}{(q^{2\Delta_{n+1}}e^{\pm i\bar\alpha_{n}\pm i \theta_R};q^2)_{\infty}} \right)
   \nonumber\\
   &= \int_{0}^{\pi} \int_{0}^{\pi}d\theta_L d\theta_R \, \rho(\theta_L) \rho(\theta_R) e^{(\beta_L+\beta_L^{\prime}) \frac{\cos(\theta_L)}{\hbar}} e^{(\beta_R+\beta_R^{\prime}) \frac{\cos(\theta_R)}{\hbar}}  \frac{(q^{4\Delta};q^2)_{\infty}}{(q^{2\Delta}e^{\pm i\theta_L\pm i \theta_R};q^2)_{\infty}},
\end{align}
where to reach to the last line, we applied the Askey-Wilson integral (\ref{AW-integral}) iteratively.

It may appear that the untwisted state $|\bar\alpha_i\rangle$ always occupies the second-to-last position in the decomposition. However, this is simply because at each iteration, we have inserted the new identity operator into the last $e^{-\Delta_i\hat{L}}$ in the fully-open channel.  If we have arbitrary insertions, the untwisted state can be placed at any position. For example, the energy basis in the $n$-split representation can also be written as
\begin{align}
 |\Delta;\theta_L,\theta_R\rangle_n= &\int_0^{\pi}d\bar\alpha_1 \rho_{\Delta_1,\theta_L}(\bar\alpha_1)\prod_{i=2}^{n}\left(\int_0^{\pi} d\bar\alpha_i  \rho_{\Delta_i,\bar\alpha_{i-1}}(\bar\alpha_i) \right)\nonumber \\
&\cdot |{\theta_L}\rangle\otimes^{\theta_L}  |\bar\alpha_1\rangle_{\Delta_{1},\theta_L} \otimes^{\bar\alpha_1}  |\bar\alpha_2\rangle_{\Delta_2,\bar\alpha_1} \otimes^{\bar\alpha_2}  \cdot\cdot\cdot \otimes^{\bar\alpha_{n-1}}  |\bar\alpha_n\rangle_{\Delta_{n},\bar\alpha_{n-1}}\otimes^{\bar\alpha_n} 
|\theta_R\rangle_{\Delta_{n+1},\bar\alpha_n},
\end{align}
where the untwisted state now becomes $|\theta_L\rangle$.

\newcolumntype{C}[1]{>{\centering\arraybackslash}m{#1}}
\begin{table}[!htbp]
 \centering
 \renewcommand{\arraystretch}{2}
 \setlength{\tabcolsep}{2mm}
\begin{tabular}{|c|C{6.5cm}|C{7cm}|}
\hline
\textbf{Split} &  \textbf{ Fully-open channel} & 
    \textbf{Semi-open channel state $|\Delta;\theta_L,\theta_R\rangle$} 
 \\   \hline 
0-split & \vspace{-2mm} \makecell{$\begin{aligned} \\
         \langle\theta_L|e^{-\Delta \hat{L}}|\theta_R\rangle  
    \end{aligned}$ \vspace{2mm} \\ \includegraphics[width=2.5cm]{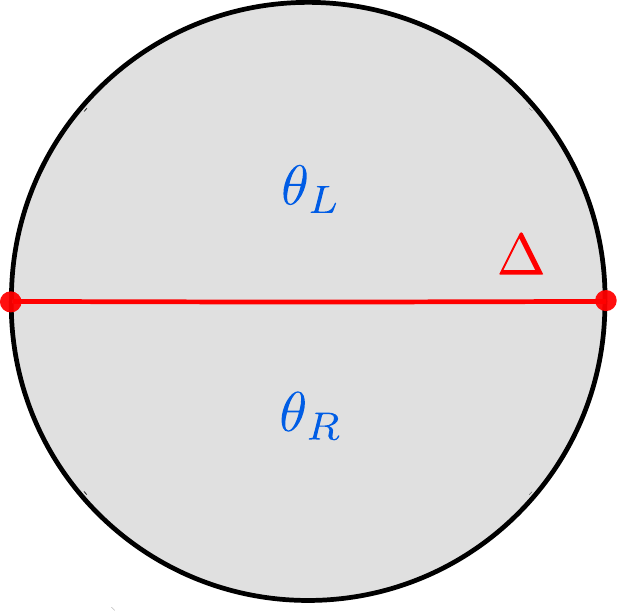}} & \vspace{4mm} \makecell{$\begin{aligned}
    |{\theta_L}\rangle \otimes^{\theta_L} |{\theta_R}\rangle_{\Delta,\theta_L} 
\end{aligned}$\\ \\ \includegraphics[width=2.5cm]{figure_pdf/0split_SC.pdf} } \\ \hline

1-split & \vspace{-2mm} \makecell{ $\begin{aligned}  \\ 
    \int_0^{\pi} d\bar\alpha \rho(\bar\alpha) \langle\theta_L|e^{-\Delta_1 \hat{L}}|\bar\alpha\rangle  \langle\bar\alpha|e^{-\Delta_2 \hat{L}}|\theta_R\rangle 
\end{aligned}$ \\  \includegraphics[width=2.5cm]{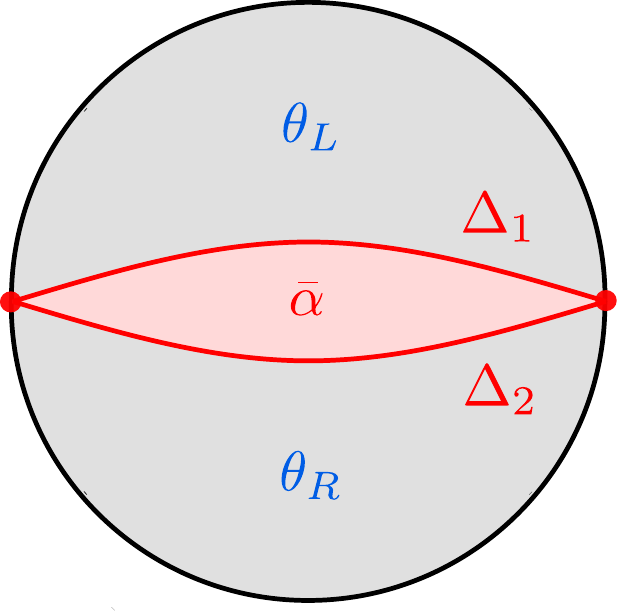}}        & \vspace{4mm}\makecell{$\begin{aligned}
     \int_0^{\pi} d\bar\alpha \, \rho(\bar\alpha) \, |{\theta_L}\rangle_{\Delta_1,\bar\alpha} \otimes^{\bar\alpha} |\bar\alpha\rangle \otimes^{\bar\alpha} |\theta_R\rangle_{\Delta_2,\bar\alpha} 
 \end{aligned}$ \vspace{4mm} \\ \includegraphics[width=2.5cm]{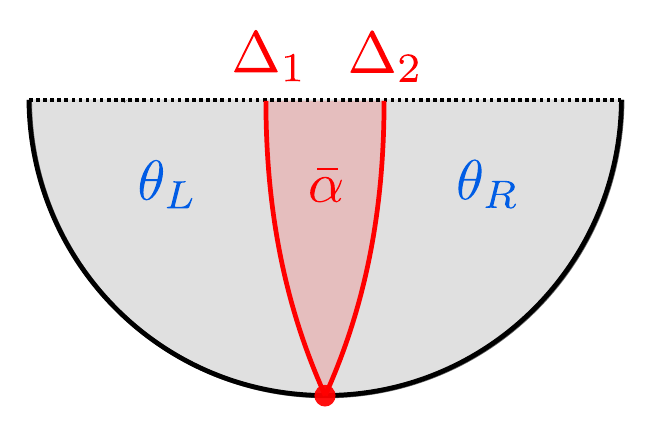}    }       \\ \hline

2-split & \vspace{-8mm}   \makecell{$\begin{aligned}  \\ 
&\int_0^{\pi}\int_0^{\pi} d\bar\alpha_1 d\bar\alpha_2 \rho(\bar\alpha_1) \rho(\bar\alpha_2)  \\
&\langle\theta_L|e^{-\Delta_1 \hat{L}}|\bar\alpha_1\rangle 
  \langle\bar\alpha_1|e^{-\Delta_{2} \hat{L}}|\bar\alpha_2 \rangle\langle\bar\alpha_2|e^{-\Delta_{3}\hat{L}}|\theta_R\rangle   
\end{aligned} $ \vspace{2mm} \\ \hspace{-6mm}\includegraphics[width=2.5cm]{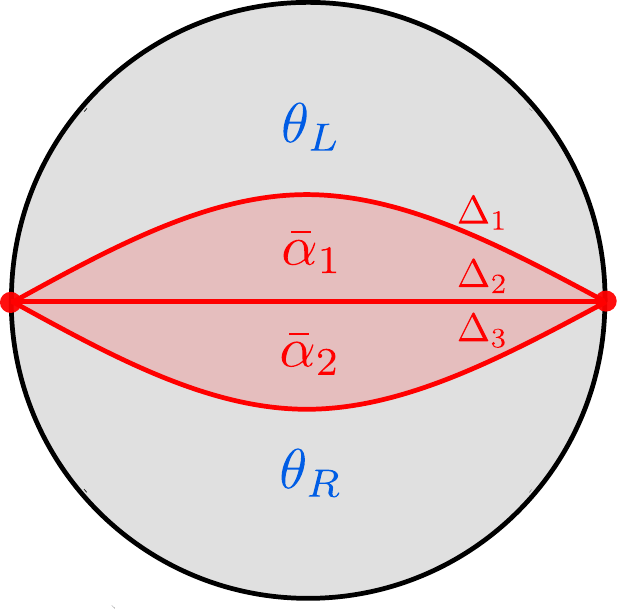} }         & \vspace{-6.5mm} \makecell{
$\begin{aligned}
    &\int_0^{\pi} \int_0^{\pi} d\bar\alpha_1 d\bar\alpha_2 \rho_{\Delta_2,\bar\alpha_2}(\bar\alpha_1)  \rho(\bar\alpha_2)
    \\
    &|{\theta_L}\rangle_{\Delta_1,\bar\alpha_1} \otimes^{\bar\alpha_1} |\bar\alpha_1\rangle_{\Delta_{2},\bar\alpha_2} \otimes^{\bar\alpha_2} |\bar\alpha_2\rangle \otimes^{\bar\alpha_2} |\theta_R\rangle_{\Delta_3,\bar\alpha_2}
\end{aligned}$ \vspace{6mm} \\ \hspace{-4mm} \includegraphics[width=2.5cm]{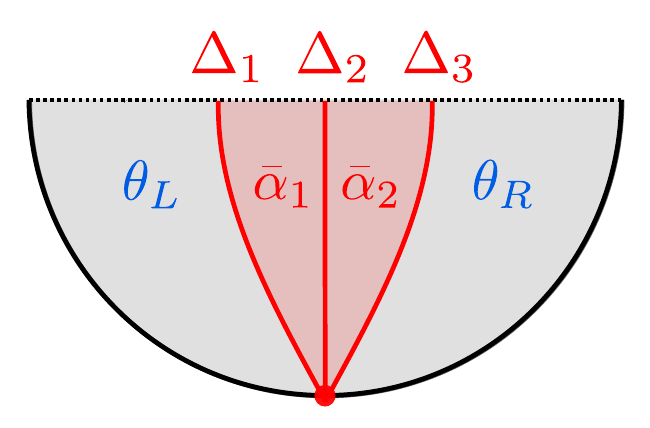}} 
\\   \hline  
$\cdot \cdot \cdot$ & $\cdot \cdot \cdot$ & $\cdot \cdot \cdot$  \\   \hline
$n$-split & \vspace{-16mm} \makecell{ $\begin{aligned} 
&\prod_{i=1}^{n}\left(\int_0^{\pi} d\bar\alpha_i  \rho(\bar\alpha_i) \right) \\
&\langle\theta_L|e^{-\Delta_1 \hat{L}}|\bar\alpha_1\rangle \langle\bar\alpha_1|
 \cdot\cdot\cdot|\bar\alpha_n\rangle \langle\bar\alpha_n|e^{-\Delta_{n}\hat{L}}|\theta_R\rangle     
\end{aligned}$ \vspace{3mm}\\ \includegraphics[width=2.5cm]{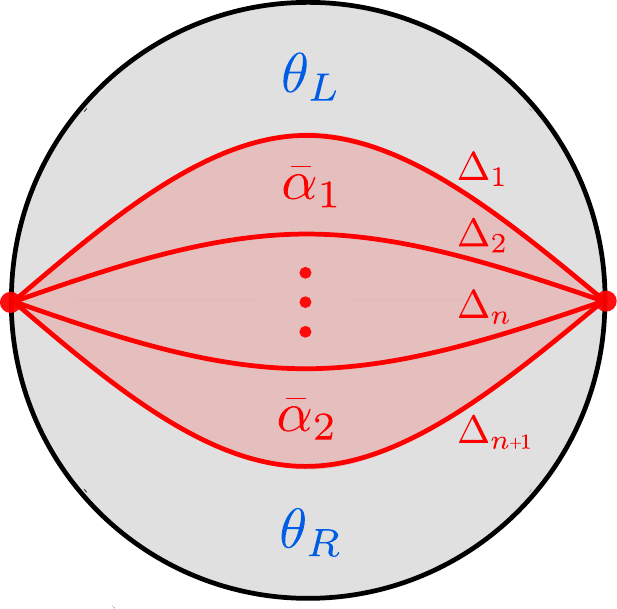}}    & \vspace{-8mm} \makecell{$\begin{aligned}  \\ 
&\prod_{i=1}^{n-1}\left(\int_0^{\pi} d\bar\alpha_i  \rho_{\Delta_{i+1},\bar\alpha_{i+1}}(\bar\alpha_i) \right) \int_0^{\pi} d\bar\alpha_n \rho(\bar\alpha_n) \\
&|{\theta_L}\rangle_{\Delta_1,\bar\alpha_1} \otimes^{\bar\alpha_1} 
|\bar\alpha_1\rangle_{\Delta_{2},\bar\alpha_2} \otimes^{\bar\alpha_2} |\bar\alpha_2\rangle_{\Delta_3,\bar\alpha_3} \otimes^{\bar\alpha_3} \cdot\cdot\cdot
\\
& \otimes^{\bar\alpha_{n-1}} |\bar\alpha_{n-1}\rangle_{\Delta_n,\bar\alpha_n}\otimes^{\bar\alpha_n} |\bar\alpha_n\rangle\otimes^{\bar\alpha_n}
|\theta_R\rangle_{\Delta_{n+1},\bar\alpha_n}   
\end{aligned} $  \vspace{4mm} \\  \hspace{-5.5mm} \includegraphics[width=2.5cm]{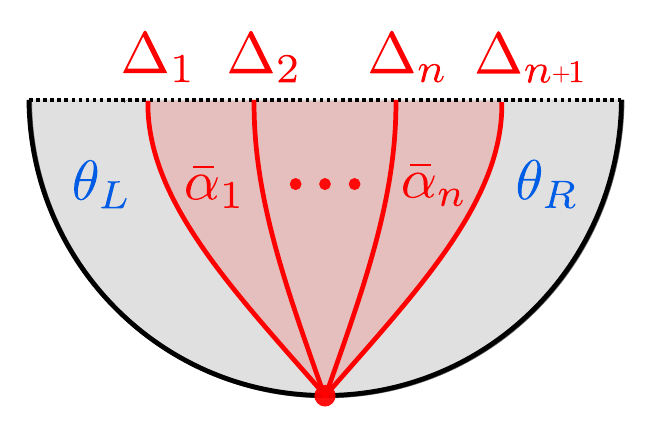}} \\ \hline

\end{tabular}
\captionsetup{justification=raggedright}
\caption{Different representations of the one-particle wormhole Hilbert space. } 
\label{rep_table}
\end{table}

\vspace{5pt}
\subsection{Multi-particle Hilbert space}
\label{multi_parti_HS}
\begin{figure}[htbp]
  \begin{center}
   \includegraphics[width=5cm]{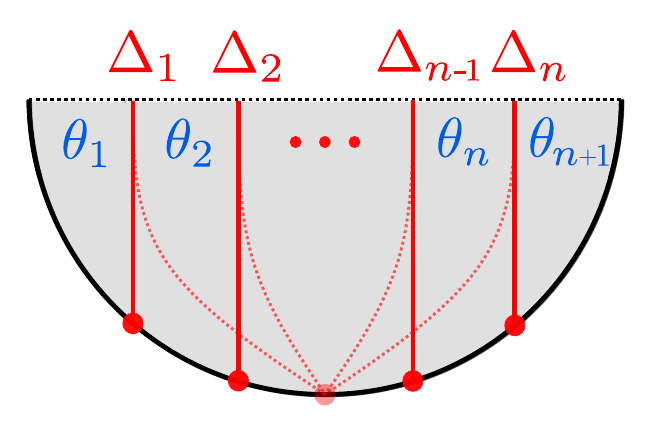}
  \end{center}   
\vspace*{-0.5cm}
\caption{Energy eigenstate in the $n$-particle Hilbert space. The red dotted line is its analogue in $(n-1)$-split  one-particle representation.}
\label{n-parti_energy}
\end{figure}

The different split representations discussed above encode the one-particle Hilbert space in different ways, but in fact they also capture the structure of the multi-particle Hilbert space: when the split states are viewed locally in the bulk  with fixed energies in each region, one cannot distinguish them from states with genuine multi-particle insertions on the boundary! This is shown in Fig.\ref{n-parti_energy}.

Thus, the energy eigenstate in the $n$-particle Hilbert space is simply given by energy basis of ($n-1$)-split representation of one-particle state without the integrals and the density of states in the integrand
\begin{align}
    &|\Delta_1,...,\Delta_n; \theta_1,\theta_2,...,\theta_n,\theta_{n+1}\rangle
    \\
    &=|{\theta_1}\rangle_{\Delta_{1},\theta_2}\otimes^{\theta_2} |\theta_2\rangle_{\Delta_{2},\theta_3}\otimes^{\theta_3} \cdot\cdot\cdot \otimes^{\theta_{n-1}} |\theta_{n-1}\rangle_{\Delta_{n-1},\theta_{n}}\otimes^{\theta_n} |\theta_n\rangle \otimes^{\theta_n}
|\theta_{n+1}\rangle_{\Delta_{n},\theta_n} .  
\end{align}
As before, the untwisted state $|\theta_i\rangle$ can be placed at any position. All such different decompositions can be thought of as coming from a combinatorial problem: one assigns each EOW brane to one of its two adjacent regions, in such a way that each region  contains at most one EOW brane. This results in $n$ twisted states and one untwisted state.

For the length eigenstate of n-particle state, the only difference from the $(n-1)$-split length basis is that now the middle regions can  now have arbitrary separations. Therefore, it can be written as the following factorized form
\begin{align}
    |\Delta_1,...,\Delta_n; n_1,n_2,...n_n,n_{n+1}\rangle= |n_1\rangle \otimes |n_2\rangle \otimes \cdot \cdot \cdot \otimes |n_n\rangle \otimes |n_{n+1}\rangle.
\end{align}

\begin{figure}[htbp]
  \begin{center}
   \includegraphics[width=5.8cm]{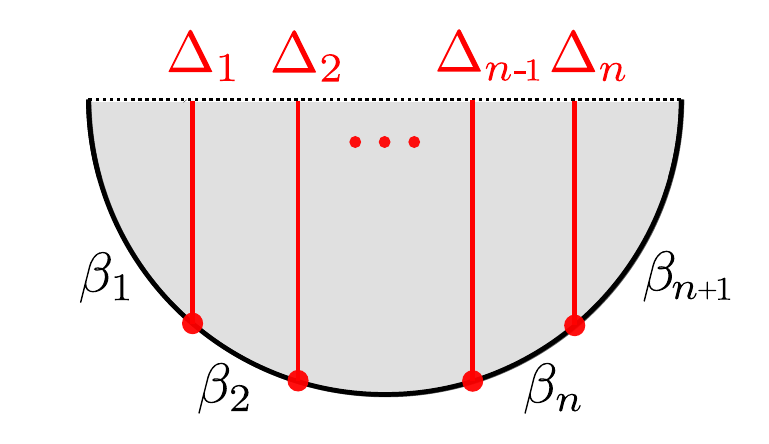}
  \end{center}   
\vspace*{-0.5cm}
\caption{Hartle-Hawking state in $n$-particle Hilbert space}
\label{n-parti_beta}
\end{figure}

The major difference between $n$-particle Hilbert space and ($n-1$)-split one-particle Hilbert space is that the former has additional boundary evolutions. This is explicit when we construct the $n$-particle Hartle-Hawking states, where the asymptotic boundary conditions $\beta_1,..., \beta_{n+1}$ are specified, as shown in Fig.\ref{n-parti_beta}. Referring to the energy eigenstate, the Hartle-Hawking states now take the form

\begin{align}
     &|\Delta_1,...,\Delta_n; \beta_1,...,\beta_{n+1}\rangle
     \nonumber\\
     &=\prod_{i=1}^{n-1}\left(\int_0^{\pi} d\theta_i  \,\rho_{\Delta_{i},\theta_{i+1}}(\theta_i) \right) \int_0^{\pi} d\theta_n \, \rho(\theta_n) \int_0^{\pi} d\theta_{n+1} \, \rho_{\Delta_n,\theta_n}(\theta_{n+1}) \,  e^{-\beta_1\hat{H}_{m_1,\mu_2}}|{\theta_1}\rangle_{\Delta_{1},\theta_2}\otimes^{\theta_2} \cdot\cdot\cdot  \nonumber
     \nonumber\\
     &\quad \otimes^{\theta_{n-1}} e^{-\beta_{n-1}\hat{H}_{m_{n-1},\mu_{n}}} |\theta_{n-1}\rangle_{\Delta_{n-1},\theta_{n}} \otimes^{\theta_n}  e^{-\beta_n\hat{H}_{\mathrm{SD}}}|\theta_n\rangle \otimes^{\theta_n} e^{-\beta_{n+1}\hat{H}_{m_n,\mu_n}}|\theta_{n+1}\rangle_{\Delta_{n},\theta_n}
     \nonumber\\
     &= \prod_{i=1}^{n-1}\left(\int_0^{\pi} d\theta_i  \,\rho_{\Delta_{i},\theta_{i+1}}(\theta_i) \,e^{\beta_i\frac{\cos(\theta_i)}{\hbar}} \right) \int_0^{\pi} d\theta_n  \,\rho(\theta_n) \,e^{\beta_n\frac{\cos(\theta_n)}{\hbar}} \int_0^{\pi} d\theta_{n+1} \, \rho_{\Delta_n,\theta_n}(\theta_{n+1}) \, e^{\beta_{n+1}\frac{\cos(\theta_{n+1})}{\hbar}} \nonumber
     \\
     &\quad \cdot |{\theta_1}\rangle_{\Delta_{1},\theta_2}\otimes^{\theta_2} |\theta_2\rangle_{\Delta_{2},\theta_3}\otimes^{\theta_3} \cdot\cdot\cdot \otimes^{\theta_{n-1}} |\theta_{n-1}\rangle_{\Delta_{n-1},\theta_{n}}\otimes^{\theta_n} |\theta_n\rangle \otimes^{\theta_n}
|\theta_{n+1}\rangle_{\Delta_{n},\theta_n} .
\end{align}
When setting $\beta_2,..,\beta_n=0$, this state becomes $(n-1)$-split one-particle Hartle-Hawking state again.

The inner product of two such states indeed gives the correct uncrossed $n$-point function
\begin{align}
    &\langle\Delta_1,...,\Delta_n; \beta_1^{\prime},...,\beta_{n+1}^{\prime}|\Delta_1,...,\Delta_n; \beta_1,...,\beta_{n+1}\rangle
    \nonumber \\
    &= \prod_{i=1}^{n+1}\left(\int_0^{\pi} d\theta_i  \rho(\theta_i) e^{(\beta_i+\beta_i^{\prime})\frac{\cos(\theta_i)}{\hbar}} \right) 
    \nonumber\\
    &\quad \times \frac{(q^{4\Delta_1};q^2)_{\infty}}{(q^{2\Delta_1}e^{\pm i\theta_1\pm i \theta_2};q^2)_{\infty}} \frac{(q^{4\Delta_2};q^2)_{\infty}}{(q^{2\Delta_2}e^{\pm i\theta_2\pm i \theta_3};q^2)_{\infty}} \cdot\cdot\cdot \frac{(q^{4\Delta_n};q^2)_{\infty}}{(q^{2\Delta_n}e^{\pm i\theta_n\pm i \theta_{n+1}};q^2)_{\infty}}
\end{align}

We note again that in our multi-particle Hilbert space,  the energy eigenstates exhibit a non-trivial  state-dependent structure, while the length eigenstates simply factorize. This is a direct feature that follows from our splitting and gluing procedure in sine-dilaton gravity. The quasi-unitary transformation operator between these two bases has the same form as in (\ref{cob_matrix}), which is non-local between different regions separated by the matter lines. This is in contrast to previous constructions by Lin and Stanford \cite{Lin:2022rbf,Lin:2023trc}, which focus on the non-factorized length (or chord number) basis.\footnote{For a more careful comparison, check footnote \ref{LS_compare}.} It would be interesting to find an isomorphism between our construction with theirs, possibly similar to the one discussed in \cite{xu2025vonneumannalgebrasdoublescaled}. 


With the structure of the multi-particle Hilbert space now understood, we are able to calculate arbitrary correlation functions, including crucially the crossed diagrams, which we discuss in detail in the next section.

\vspace{10pt}
\section{Crossed four-point function and the q-deformed 6j-symbol}
\label{OTOC_sec}
In this section, we calculate the crossed four-point function using the representations of the one-particle Hilbert space established in the last section and extract the result which is known to be a q-deformed 6j-symbol in quantum group. As we will see, different split representations of the one-particle Hilbert space result in different representations of the OTOC. However, we will show that these results are in fact equivalent.

\begin{figure}[htbp]
  \begin{center}
   \includegraphics[width=6cm]{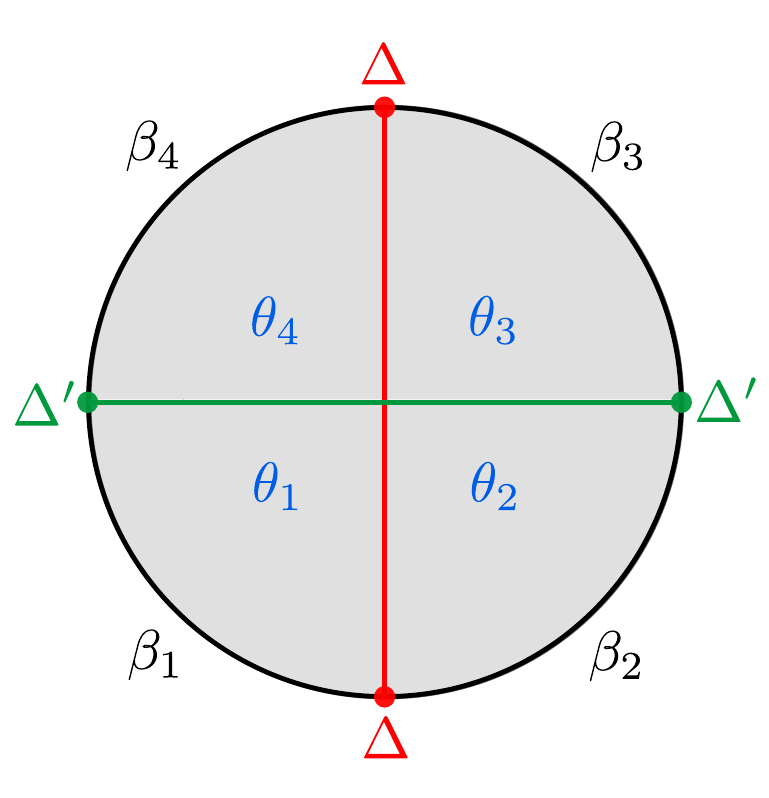}
  \end{center}   
\vspace*{-0.5cm}
\caption{OTOC contour}
\label{OTOC}
\end{figure}

The crossed four-point function is also called the out-of-time correlator (OTOC), which is defined as
\begin{align}
    \mathrm{OTOC}=\mathrm{tr}\left[ \mathcal{O}_{\Delta^{\prime}}e^{-\beta_4 H} \mathcal{O}_{\Delta}e^{-\beta_3 H} \mathcal{O}_{\Delta^{\prime}}e^{-\beta_2 H} \mathcal{O}_{\Delta}e^{-\beta_1 H} \right].
\end{align}
We draw the OTOC contour in the bulk in Fig.\ref{OTOC}.

This OTOC can be thought of as a "two-point" function in the one-particle Hilbert space, where the $\mathcal{O}_{\Delta}$ (vertical) operators are the twist (or defect) operators that generate the one-particle Hilbert space, and the insertion of $\mathcal{O}_{\Delta^{\prime}}$ (horizontal) operators generates the boundary-to-boundary propagator $e^{-\Delta^{\prime}L}$ in the bulk. Thus, the OTOC can be re-expressed using the one-particle Hartle-Hawking state as
\begin{align}
     \mathrm{OTOC}=\langle\Delta; \beta_1,\beta_2|e^{-\Delta^{\prime}\hat{L}}|\Delta;\beta_4,\beta_3\rangle.
\end{align}
As we learned in the previous section, the one-particle Hilbert space admits many different representations. This means that the OTOC can be computed in multiple ways, depending on which representation is used for the Hartle-Hawking state. In the following, we derive several representations of the result. This serves as a non-trivial consistency check -- any physical observable such as the OTOC must yield the same result regardless of the chosen representation. We will also see that reproducing the same quantity from different starting point uncovers new mathematical identities.

\vspace{5pt}
\subsection{OTOC in 0-split representation}

In 0-split representation, one-particle Hartle-Hawking state is given by (\ref{0split_HH})
\begin{align}
    |\Delta;\beta_L,\beta_R\rangle_0=\int_0^{\pi}d\theta_L \, \rho(\theta_L) \, e^{\beta_L \frac{\cos(\theta_L)}{\hbar}} \int_0^{\pi} d\theta_R \, \rho_{\Delta,\theta_L}(\theta_R) \, e^{\beta_R\frac{\cos(\theta_R)}{\hbar}} |\theta_L\rangle\otimes^{\theta_L} |\theta_R\rangle_{\Delta,\theta_L}.
\end{align}
Since this representation has two regions, namely the left and right regions,  we can separate the length operator in $e^{-\Delta^{\prime}\hat{L}}$ into left and right parts, each acting on the corresponding region
\begin{align}
    e^{-\Delta^{\prime}\hat{L}}=e^{-\Delta^{\prime}\hat{L}_L}\otimes e^{-\Delta^{\prime}\hat{L}_R}.
\end{align}
This separation is valid due to the factorization property of the length basis in one-particle Hilbert space, as given in (\ref{0-split_rep}).
\begin{figure}[htbp]
  \begin{center}
   \includegraphics[width=6cm]{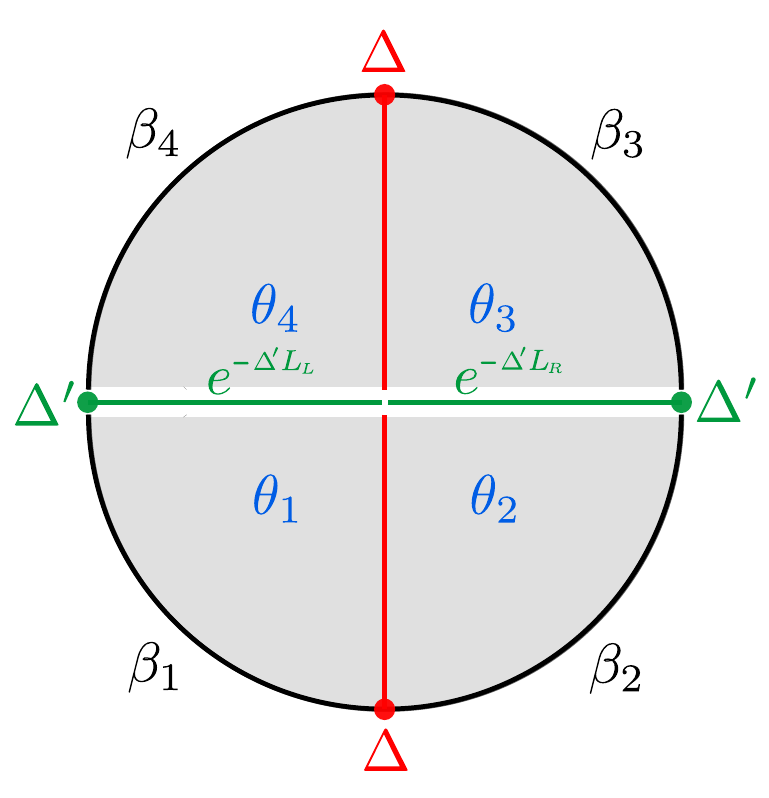}
  \end{center}   
\vspace*{-0.5cm}
\caption{OTOC in 0-split representation}
\label{OTOC_0split}
\end{figure}

Referring to Fig.\ref{OTOC_0split}, the OTOC can be calculated in 0-split representation as 
\begin{align}
   \mathrm{OTOC}&= {_0}\langle\Delta; \beta_1,\beta_2|e^{-\Delta^{\prime}\hat{L}_L}\otimes e^{-\Delta^{\prime}\hat{L}_R}|\Delta;\beta_4,\beta_3\rangle_0
    \nonumber
    \\
    &= \prod_{i=1}^{4}\left( \int_{0}^{\pi}d\theta_i \exp\left( \beta_i\frac{\cos(\theta_i)}{\hbar}\right) \right) \nonumber
    \\
    &\quad \times \rho(\theta_1)\, \rho(\theta_4) \, \rho_{\Delta,\theta_1}(\theta_2) \, \rho_{\Delta,\theta_4}(\theta_3) 
  \,
    \langle\theta_4 | e^{-\Delta^{\prime} \hat{L}_L}|\theta_1\rangle \,{_{\Delta,\theta_4}\langle\theta_3} | e^{-\Delta^{\prime}\hat{L}_R}|\theta_2\rangle_{\Delta,\theta_1} 
    \label{OTOC_begin_0split}
\end{align}
The first correlation function on the last line is given by (\ref{2-pt_correlator2})
\begin{align}
    \langle\theta_4 | e^{-\Delta^{\prime}\hat{L}_L}|\theta_1\rangle=\langle\theta_4 |q^{2\Delta^{\prime}\hat{n}_L}|\theta_1\rangle= \frac{(q^{4\Delta^{\prime}};q^2)_{\infty}}{(q^{2\Delta^{\prime}}e^{\pm i\theta_4\pm i \theta_1};q^2)_{\infty}}.
\end{align}
The second correlation function on the RHS can be calculated by the insertion of an identity operator in length basis
\begin{align}
    &{_{\Delta,\theta_4}\langle\theta_3} | e^{-\Delta^{\prime}\hat{L}_R}|\theta_2\rangle_{\Delta,\theta_1} \nonumber\\&=\sum_{n=0}^{\infty}q^{2\Delta^{\prime}n}  {_{\Delta,\theta_4}\langle\theta_3}|n\rangle \langle n| \theta_2\rangle_{\Delta,\theta_1} , \nonumber
    \\
    &=\sum_{n=0}^{\infty} \frac{q^{2n\Delta^{\prime}}}{(q^2,q^{4\Delta};q^2)_n} ~Q_n(\cos(\theta_3)|q^{2\Delta} e^{i\theta_4}, q^{2\Delta} e^{-i\theta_4};q^2)~ Q_n(\cos(\theta_2)|q^{2\Delta} e^{i\theta_1}, q^{2\Delta} e^{-i\theta_1};q^2)
     \nonumber\\
    &=\frac{\left( q^{2\Delta^{\prime}}e^{i(-\theta_1-\theta_4)},q^{2\Delta} q^{2\Delta^{\prime}}e^{i(\theta_1\pm \theta_3)},q^{2\Delta} q^{2\Delta^{\prime}}e^{i(\theta_4\pm \theta_2)};q^2\right)_{\infty}}{\left(q^{4\Delta}q^{2\Delta^{\prime}}e^{i(\theta_4+\theta_1)},q^{2\Delta^{\prime}}e^{i(\pm\theta_3\pm\theta_2)};q^2\right)_{\infty}}  \nonumber
    \\
    &\quad \times{_8W_7} \left(\frac{q^{4\Delta}q^{2\Delta^{\prime}}e^{i(\theta_4+\theta_1)}}{q^2}, q^{2\Delta^{\prime}}e^{i(\theta_1+\theta_4)},q^{2\Delta}e^{i(\theta_4\pm \theta_3)},q^{2\Delta}e^{i(\theta_1\pm\theta_2)};q^2, q^{2\Delta^{\prime}}e^{i(-\theta_1-\theta_4)} \right) 
     \nonumber\\
    &=\frac{[(q^{2\Delta^{\prime}}e^{i(\pm \theta_4\pm \theta_1)},q^{2\Delta}e^{i(\pm \theta_1\pm \theta_2)},q^{2\Delta}e^{i(\pm \theta_4\pm \theta_3)};q^2)_{\infty}]^{1/2}}{\left(q^{4\Delta};q^2\right)_{\infty}[(q^{2\Delta^{\prime}}e^{i(\pm\theta_3\pm\theta_2)};q^2)_{\infty}]^{1/2}}  R_{\theta_2\theta_4}^{q^2}\begin{bmatrix}
       \theta_1&\Delta\\\theta_3& \Delta^{\prime}
   \end{bmatrix},
   \label{8W7_R}
\end{align}
where we use the Poisson kernel of Al-Salam-Chihara polynomials \cite{askey1985some} to obtain the well-poised hypergeometric series $_8W_7$. In the last line we identify  the 6j-symbol of the quantum group $\mathcal{U}_q(su(1,1))$ as written in the R-matrix form
\begin{align}
& R_{\theta_2\theta_4}^{q^2}\begin{bmatrix}
       \theta_1&\Delta\\\theta_3& \Delta^{\prime}
   \end{bmatrix} \nonumber \nonumber\\&=
    \frac{\left( q^{2\Delta^{\prime}}e^{i(-\theta_4-\theta_1)},q^{2\Delta} q^{2\Delta^{\prime}}e^{i(\theta_1\pm \theta_3)},q^{2\Delta} q^{2\Delta^{\prime}}e^{i(\theta_4\pm \theta_2)};q^2\right)_{\infty}}{\left(q^{4\Delta}q^{2\Delta^{\prime}}e^{i(\theta_4+\theta_1)};q^2\right)_{\infty}} \nonumber
    \\
    & \quad \times \frac{(q^{4\Delta};q^2)_{\infty}}{[(q^{2\Delta^{\prime}}e^{i(\pm \theta_4\pm \theta_1)},q^{2\Delta^{\prime}}e^{i(\pm \theta_3\pm \theta_2)},q^{2\Delta}e^{i(\pm \theta_1\pm \theta_2)},q^{2\Delta}e^{i(\pm \theta_4\pm \theta_3)};q^2)_{\infty}]^{1/2}} \nonumber
    \\
    &\quad \times {_8W_7} \left(\frac{q^{4\Delta}q^{2\Delta^{\prime}}e^{i(\theta_4+\theta_1)}}{q^2}, q^{2\Delta^{\prime}}e^{i(\theta_4+\theta_1)},q^{2\Delta}e^{i(\theta_4\pm \theta_3)},q^{2\Delta}e^{i(\theta_1\pm\theta_2)};q^2, q^{2\Delta^{\prime}}e^{i(-\theta_4-\theta_1)} \right).
    \label{R-matrix}
\end{align}
Together with density of states  of the EOW brane Hilbert space given in (\ref{dos_so}), the OTOC is obtained as
\begin{align}
   & \prod_{i=1}^{4}\left( \int_{0}^{\pi}d\theta_i \exp\left( \beta_i\frac{\cos(\theta_i)}{\hbar}\right) \rho(\theta_i) \right)
    ~ R_{\theta_2\theta_4}^{q^2}\begin{bmatrix}
       \theta_1&\Delta\\\theta_3& \Delta^{\prime}
   \end{bmatrix} \nonumber
   \\
   &\times \left( \frac{(q^{4\Delta^{\prime}};q^2)_{\infty}}{(q^{2\Delta^{\prime}}e^{i(\pm \theta_4\pm \theta_1)};q^2)_{\infty}}   \frac{(q^{4\Delta^{\prime}};q^2)_{\infty}}{(q^{2\Delta^{\prime}}e^{i(\pm \theta_2\pm \theta_3)};q^2)_{\infty}}  \frac{(q^{4\Delta};q^2)_{\infty}}{(q^{2\Delta}e^{i(\pm \theta_1\pm \theta_2)};q^2)_{\infty}}  \frac{(q^{4\Delta};q^2)_{\infty}}{(q^{2\Delta}e^{i(\pm \theta_3\pm \theta_4)};q^2)_{\infty}}\right)^{1/2}.
   \label{OTOC_result}
\end{align}
This result matches exactly to the crossed 4-point function obtained in DSSYK, as given in (5.5) of \cite{Berkooz:2018jqr}, which we have now derived from a bulk perspective.

One slightly dissatisfying aspect of the computation is that, despite the final result being symmetric between the left and right regions, the starting point is manifestly asymmetric, as can be seen from (\ref{OTOC_begin_0split}). This asymmetry is a feature of 0-split representation, in which the EOW brane must be associated with one region. This merely reflects a choice of quantization procedure, rather than any physical asymmetry. Indeed, in the next subsection, we will calculate the OTOC in the 1-split representation, where the left and right regions are treated symmetrically from the outset.

\vspace{5pt}
\subsection{OTOC in 1-split representation}
The one-particle Hartle-Hawking state in 1-split representation is written as (\ref{1split_HH})
\begin{align}
    &|\Delta;\beta_L,\beta_R\rangle_1 
     \nonumber\\
     &=\int_0^{\pi} d\bar\alpha \, \rho(\bar\alpha) \int_0^{\pi}  d\theta_L   \, \rho_{\Delta_L,\bar\alpha}(\theta_L) \, e^{\beta_L\frac{\cos(\theta_L)}{\hbar}} \int_0^{\pi} d\theta_R \,
     \rho_{\Delta_R,\bar\alpha}(\theta_R)  \, e^{\beta_R\frac{\cos(\theta_R)}{\hbar}}
     |{\theta_L}\rangle_{\Delta_L,\bar\alpha} \otimes^{\bar\alpha} |\bar\alpha\rangle \otimes^{\bar\alpha} |\theta_R\rangle_{\Delta_R,\bar\alpha}.
\end{align}
Since this representation now involves the left, middle and right regions, the operator $e^{-\Delta^{\prime}\hat{L}}$ can be separated as
\begin{align}
    e^{-\Delta^{\prime}\hat{L}}=e^{-\Delta^{\prime}\hat{L}_L}\otimes e^{-\Delta^{\prime}\hat{L}_M}\otimes e^{-\Delta^{\prime}\hat{L}_R}.
\end{align}
Again, this separation follows from the factorization of the length basis in 1-split representation.

\begin{figure}[htbp]
  \begin{center}
   \includegraphics[width=6cm]{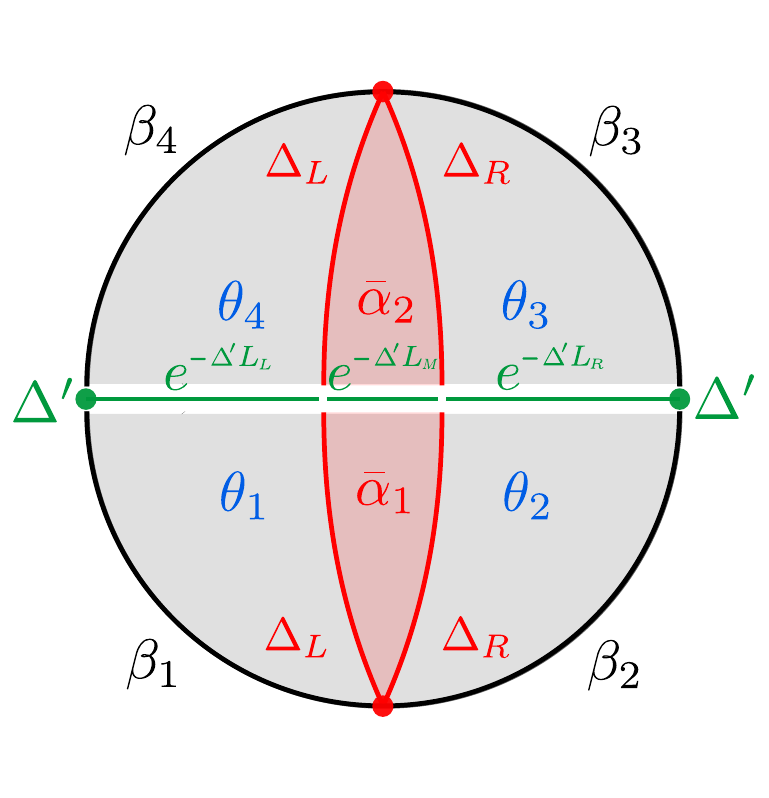}
  \end{center}   
\vspace*{-0.5cm}
\caption{OTOC in 1-split representation}
\label{OTOC_1split}
\end{figure}

Referring to Fig.\ref{OTOC_1split}, the OTOC in the bulk can be written as 
\begin{align}
   \mathrm{OTOC}=& {_1}\langle\Delta; \beta_1,\beta_2|e^{-\Delta^{\prime}\hat{L}_L}\otimes e^{-\Delta^{\prime}\hat{L}_M}\otimes e^{-\Delta^{\prime}\hat{L}_R}|\Delta;\beta_4,\beta_3\rangle_1
    \nonumber
    \\
    =& \int_0^{\pi}\int_0^{\pi}d\bar{\alpha}_1 d\bar{\alpha}_2 \, \rho(\bar{\alpha}_1) \, \rho(\bar{\alpha}_2) \,\prod_{i=1}^{4}\left( \int_{0}^{\pi}d\theta_i  \exp\left( \beta_i\frac{\cos(\theta_i)}{\hbar}\right) \right) \nonumber
    \\
    &\times  \rho_{\Delta_L,\bar\alpha_1}(\theta_1)\,\rho_{\Delta_R,\bar\alpha_1}(\theta_2)\, \rho_{\Delta_R,\bar\alpha_2}(\theta_3) \,\rho_{\Delta_L,\bar\alpha_2}(\theta_4) \nonumber
    \\
    &\times
    {_{\Delta_L,\bar\alpha_2}}\langle\theta_4 | e^{-\Delta^{\prime}\hat{L}_L}|\theta_1\rangle_{\Delta_L,\bar\alpha_1}\, \langle \bar\alpha_2|e^{-\Delta^{\prime}\hat{L}_{M}}|\bar\alpha_1\rangle  \, {_{\Delta_R,\bar\alpha_2}}\langle\theta_3 | e^{-\Delta^{\prime}\hat{L}_R}|\theta_2\rangle_{\Delta_R,\bar\alpha_1}.
    \label{appr2_4prod}
\end{align}
Note that now the left and right regions are completely symmetric. Following the similar calculation in the last subsection by using (\ref{8W7_R}), the OTOC takes the form
\begin{align}
   &\prod_{i=1}^{4}\left( \int_{0}^{\pi}d\theta_i \exp\left( \beta_i\frac{\cos(\theta_i)}{\hbar}\right) \rho(\theta_i) \right) \frac{(q^{4\Delta^{\prime}};q^2)_{\infty}}{[(q^{2\Delta^{\prime}}e^{i(\pm \theta_4\pm \theta_1)},q^{2\Delta^{\prime}}e^{i(\pm \theta_3\pm \theta_2)};q^2)_{\infty}]^{1/2}  } \nonumber
    \\
    &\times \int_0^{\pi}\int_0^{\pi}d\bar{\alpha}_1 d\bar{\alpha}_2 \,\rho(\bar{\alpha}_1) \, \rho(\bar{\alpha}_2) \, R_{\theta_1\bar\alpha_2}^{q^2}\begin{bmatrix}
       \bar\alpha_1&\Delta_L\\\theta_4& \Delta^{\prime}
   \end{bmatrix} \,R_{\theta_2\bar\alpha_2}^{q^2}\begin{bmatrix}
       \bar\alpha_1&\Delta_R\\\theta_3& \Delta^{\prime}
   \end{bmatrix}  \nonumber\\
    &\times \left( \frac{(q^{4\Delta_{L}};q^2)_{\infty}}{(q^{2\Delta_{L}}e^{i(\pm \theta_1\pm \bar\alpha_1)};q^2)_{\infty}}   \frac{(q^{4\Delta_{L}};q^2)_{\infty}}{(q^{2\Delta_{L}}e^{i(\pm \theta_4\pm \bar\alpha_2)};q^2)_{\infty}}  \frac{(q^{4\Delta_R};q^2)_{\infty}}{(q^{2\Delta_R}e^{i(\pm \theta_2\pm \bar\alpha_1)};q^2)_{\infty}}  \frac{(q^{4\Delta_R};q^2)_{\infty}}{(q^{2\Delta_R}e^{i(\pm \theta_3\pm \bar\alpha_2)};q^2)_{\infty}}\right)^{1/2} 
    .
    \label{OTOC_1split_formula}
\end{align}
To our knowledge this is a new expression for the OTOC, involving a convolution of two R-matrices. To make connection to the known expression, 
we use the following interesting identity for the integration over two R-matrices, which we have verified numerically
\begin{align}
    &\int_0^{\pi}\int_0^{\pi}d\bar{\alpha}_1 d\bar{\alpha}_2 \,\rho(\bar{\alpha}_1) \, \rho(\bar{\alpha}_2) \, R_{\theta_1\bar\alpha_2}^{q^2}\begin{bmatrix}
       \bar\alpha_1&\Delta_L\\\theta_4& \Delta^{\prime}
   \end{bmatrix} \,R_{\theta_2\bar\alpha_2}^{q^2}\begin{bmatrix}
       \bar\alpha_1&\Delta_R\\\theta_3& \Delta^{\prime}
   \end{bmatrix}  \nonumber\\
    &\times \left( \frac{(q^{4\Delta_{L}};q^2)_{\infty}}{(q^{2\Delta_{L}}e^{i(\pm \theta_1\pm \bar\alpha_1)};q^2)_{\infty}}   \frac{(q^{4\Delta_{L}};q^2)_{\infty}}{(q^{2\Delta_{L}}e^{i(\pm \theta_4\pm \bar\alpha_2)};q^2)_{\infty}}  \frac{(q^{4\Delta_R};q^2)_{\infty}}{(q^{2\Delta_R}e^{i(\pm \theta_2\pm \bar\alpha_1)};q^2)_{\infty}}  \frac{(q^{4\Delta_R};q^2)_{\infty}}{(q^{2\Delta_R}e^{i(\pm \theta_3\pm \bar\alpha_2)};q^2)_{\infty}}\right)^{1/2} \nonumber
   \\
   &=   R_{\theta_2\theta_4}^{q^2}\begin{bmatrix}
       \theta_1&\Delta_L+\Delta_R\\\theta_3& \Delta^{\prime}\end{bmatrix}
       ~\frac{(q^{4(\Delta_L+\Delta_R)};q^2)_{\infty}}{[(q^{2(\Delta_L+\Delta_R))}e^{i(\pm \theta_1\pm \theta_2)},q^{2(\Delta_L+\Delta_R))}e^{i(\pm \theta_3\pm \theta_4)};q^2)_{\infty}]^{1/2}} .
       \label{R_relation}
\end{align} 
Putting (\ref{R_relation}) back into (\ref{OTOC_1split_formula}) recovers the OTOC result in DSSYK (\ref{OTOC_result})
\begin{align}
   & \prod_{i=1}^{4}\left( \int_{0}^{\pi}d\theta_i \exp\left( \beta_i\frac{\cos(\theta_i)}{\hbar}\right) \rho(\theta_i) \right)
    ~ R_{\theta_2\theta_4}^{q^2}\begin{bmatrix}
       \theta_1&\Delta\\\theta_3& \Delta^{\prime}
   \end{bmatrix} \nonumber
   \\
   &\times \left( \frac{(q^{4\Delta^{\prime}};q^2)_{\infty}}{(q^{2\Delta^{\prime}}e^{i(\pm \theta_4\pm \theta_1)};q^2)_{\infty}}   \frac{(q^{4\Delta^{\prime}};q^2)_{\infty}}{(q^{2\Delta^{\prime}}e^{i(\pm \theta_2\pm \theta_3)};q^2)_{\infty}}  \frac{(q^{4\Delta};q^2)_{\infty}}{(q^{2\Delta}e^{i(\pm \theta_1\pm \theta_2)};q^2)_{\infty}}  \frac{(q^{4\Delta};q^2)_{\infty}}{(q^{2\Delta}e^{i(\pm \theta_3\pm \theta_4)};q^2)_{\infty}}\right)^{1/2}. 
\end{align}

The formula (\ref{R_relation}) is a new identity for the 6j-symbols\footnote{ To our knowledge, this identity does not appear in the existing literature. It may, however, be related to the Biedenharn-Eliott identity of quantum group $\mathcal{U}_q(su(1,1))$ \cite{groenevelt2005wilsonfunctiontransformsrelated}. We thank Misha Isachenkov for pointing this out.}, which is similar to the familiar "pentagon identity", but with the ends of the vertical lines  attached together. We present this relation pictorially
\begin{align}
    \int_0^{\pi}\int_0^{\pi}d\bar\alpha_1 d\bar\alpha_2 \, \rho(\bar\alpha_1) \, \rho(\bar\alpha_2)  \quad \adjincludegraphics[width=4.5cm,valign=c]{figure_pdf/R-rel_L.pdf}=\quad \adjincludegraphics[width=4.5cm,valign=c]{figure_pdf/R-rel_R.pdf}.
\end{align}
By using this new identity recursively, one can show that all other representations of the one-particle Hilbert space in Section~\ref{1-parti_HS}  yield the same OTOC result as given in  (\ref{OTOC_result}).

In order to get a representation of the OTOC in which we are gluing factors associated with each of the four spacetime regions, as in  \cite{Blommaert:2018oro, Iliesiu_2019}, one can also split the horizontal operator $e^{-\Delta^{\prime}\hat{L}}$ by the same technique of inserting an identity operator. This results in a "pizza slicing" representation\footnote{MR thanks Micha Berkooz, Misha Isachenkov and Simon Ross for relevant conversations and for the descriptive terminology.} of the OTOC, which we present in Appendix \ref{pizza_OTOC}.  Presumably each factor in this representation corresponds to matrix elements of a quantum group representation. Clarifying the algebraic structure underlying this construction is an interesting future direction.

\vspace{10pt}
\section{General n-point functions and the Feynman rules}
\label{n-pt_sec}
In this section, we present several examples of correlation function calculations in sine-dilaton gravity, using the multi-particle Hilbert space constructed in Section \ref{multi_parti_HS}. We then present the bulk Feynman rules for computing general correlation functions, which extend somewhat the Feynman rules derived from the chord diagrams on the disk in DSSYK to also include the bulk correlators and the double trumpet geometry.

\vspace{5pt}
\subsection{n-point functions on the disk}
\begin{figure}[htbp]
  \begin{center}
   \includegraphics[width=5cm]{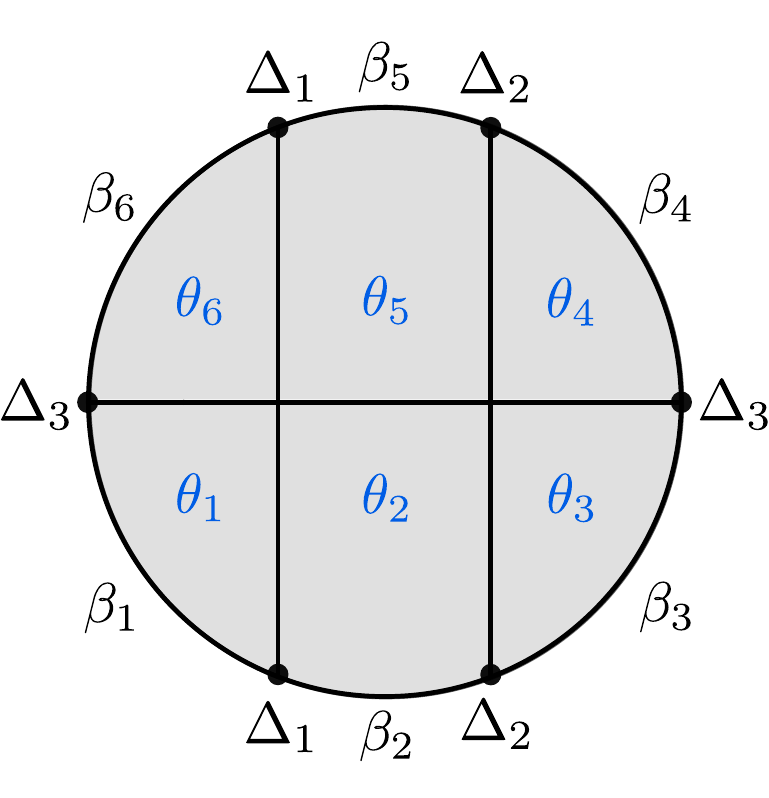}
  \end{center}   
\vspace*{-0.5cm}
\caption{Crossed 6-point function}
\label{6pt}
\end{figure}

\vspace{5pt}
\paragraph{6-point functions} We present two examples of 6-point functions.  The first one is the crossed 6-point function as shown in Fig.\ref{6pt}.
Referring to the Hartle-Hawking state in two-particle Hilbert space, this 6-point function (6pf) can be calculated as
\begin{align}
\mathrm{6pf}=&\langle\Delta_1,\Delta_2;\beta_6,\beta_5,\beta_4|e^{-\Delta_3 \hat{L}_1}\otimes e^{-\Delta_3 \hat{L}_2} \otimes e^{-\Delta_3 \hat{L}_3}|\Delta_1,\Delta_2;\beta_1,\beta_2,\beta_3\rangle \nonumber
    \\
    =&\prod_{i=1}^{6}\left(\int_0^{\pi} d\theta_i  \, \exp\left({\beta_i\frac{\cos(\theta_i)}{\hbar}}\right) \right) 
    \rho_{\Delta_1,\theta_2}(\theta_1)  \rho_{\Delta_2,\theta_2}(\theta_3) \rho_{\Delta_5,\theta_2}(\theta_4) \rho_{\Delta_5,\theta_1}(\theta_6)
     \rho(\theta_2) \rho(\theta_5)
    \nonumber
     \\
     & \times {_{\Delta_1,\theta_5}}\langle\theta_6|e^{-\Delta_3 \hat{L}_1} |{\theta_1}\rangle_{\Delta_{1},\theta_2}\,   \, {_{\Delta_2,\theta_5}}\langle \theta_4| e^{-\Delta_3 \hat{L}_3}|\theta_{3}\rangle_{\Delta_{2},\theta_{2}} \, \langle \theta_5|e^{-\Delta_3 \hat{L}_2}|\theta_2\rangle 
     \label{6pt_1}
     \\
     =& \prod_{i=1}^{6}\left( \int_{0}^{\pi}d\theta_i \exp\left( \beta_i\frac{\cos(\theta_i)}{\hbar}\right) \rho(\theta_i) \right)
    ~ R_{\theta_2\theta_6}^{q^2}\begin{bmatrix}
       \theta_1&\Delta_1\\\theta_5& \Delta_3
   \end{bmatrix} \,  R_{\theta_3\theta_5}^{q^2}\begin{bmatrix}
       \theta_2&\Delta_2\\\theta_4& \Delta_3
   \end{bmatrix}  \nonumber
   \\
   &\times \Bigg( \frac{(q^{4\Delta_{1}};q^2)_{\infty}}{(q^{2\Delta_{1}}e^{i(\pm \theta_1\pm \theta_2)};q^2)_{\infty}}   \frac{(q^{4\Delta_{2}};q^2)_{\infty}}{(q^{2\Delta\Delta_{2}}e^{i(\pm \theta_2\pm \theta_3)};q^2)_{\infty}}  \frac{(q^{4\Delta_3};q^2)_{\infty}}{(q^{2\Delta_3}e^{i(\pm \theta_3\pm \theta_4)};q^2)_{\infty}} \nonumber 
   \\
  & \quad \quad \quad 
   \frac{(q^{4\Delta_2};q^2)_{\infty}}{(q^{2\Delta_2}e^{i(\pm \theta_4\pm \theta_5)};q^2)_{\infty}} \frac{(q^{4\Delta_1};q^2)_{\infty}}{(q^{2\Delta_1}e^{i(\pm \theta_5\pm \theta_6)};q^2)_{\infty}} 
   \frac{(q^{4\Delta_3};q^2)_{\infty}}{(q^{2\Delta_3}e^{i(\pm \theta_6\pm \theta_1)};q^2)_{\infty}} 
   \Bigg)^{1/2}. 
\end{align}
Note that the bulk two-point function $\langle \theta_5|e^{-\Delta_3 \hat{L}_2}|\theta_2\rangle$ is canceled by the factors from other two-point functions in (\ref{6pt_1}), in agreement with \cite{Aguilar-Gutierrez:2025mxf}. The result involves the convolution of two R-matrices. However, one cannot use (\ref{R_relation}) to reduce it to a single R-matrix due to the extra $\exp\left({\beta_i\frac{\cos(\theta_i)}{\hbar}}\right)$ factors.

\begin{figure}[htbp]
  \begin{center}
   \includegraphics[width=5cm]{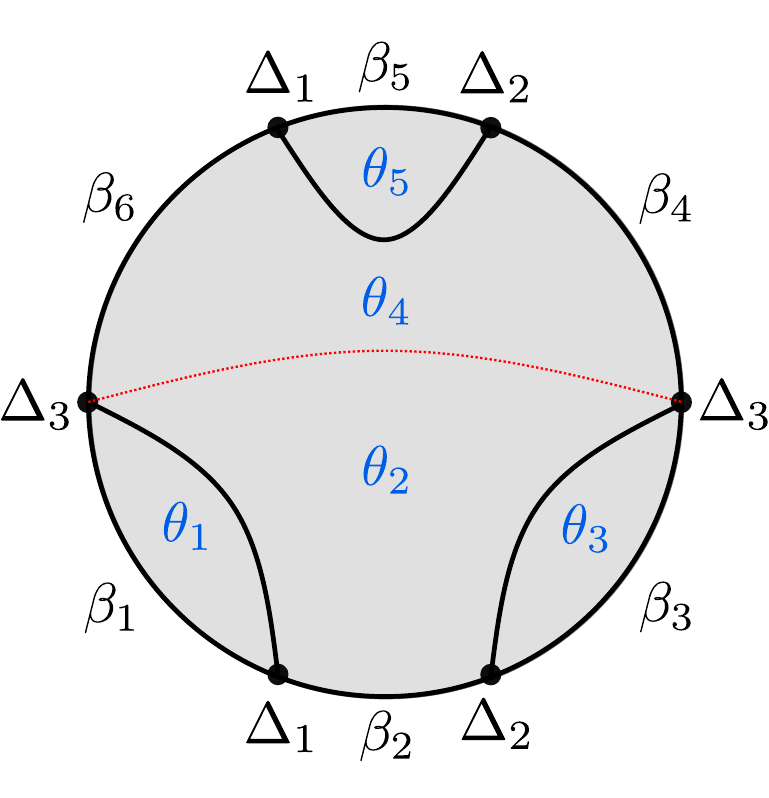}
  \end{center}   
\vspace*{-0.5cm}
\caption{Uncrossed 6-point function}
\label{6pt_2}
\end{figure}

The second one is shown in Fig.\ref{6pt_2}.
 To calculate this diagram, we insert an identity operator in length basis to cut the disk along the red dotted line shown in the figure. The region above the red line can be calculated in the fully-open channel, and the lower part can be calculated in semi-open channel using the two-particle Hartle-Hawking state. This 6-point function can thus be computed as follows
 \begin{align}
     \mathrm{6pf}&=\sum_{m=0}^{\infty} \langle n=0|e^{-\beta_5 \hat{H}_{\mathrm{SD}}}e^{-\Delta_3 \hat{L}}e^{-(\beta_4+\beta_6)\hat{H}_{\mathrm{SD}}}|m\rangle
     \langle \Delta_1,\Delta_2; n_L=0,m,n_R=0| \Delta_1,\Delta_2; \beta_1,\beta_2,\beta_3\rangle
     \\
     &=\sum_{m=0}^{\infty} \Bigg(  \int_0^{\pi} d\theta_5 d\theta_4 \, \rho(\theta_5) \, \rho(\theta_4) \, \exp\left( \beta_5 \frac{\cos(\theta_5)}{\hbar}\right) \exp\left( (\beta_4+\beta_6) \frac{\cos(\theta_4)}{\hbar}\right) \langle \theta_5|e^{-\Delta_4 \hat{L}}|\theta_4\rangle \langle \theta_4|m\rangle  \nonumber 
     \\
     &\quad \quad \quad \quad \times \left(\prod_{i=1}^{3} \int_0^{\pi}d\theta_i  \,\exp\left( \beta_i \frac{\cos(\theta_i)}{\hbar} \right)\right) \rho_{\Delta_1,\theta_2}(\theta_1) \,  \rho(\theta_2)\,\rho_{\Delta_2,\theta_2}(\theta_1) \nonumber
     \\
     & \quad \quad \quad \quad \times
     \langle n_L=0|\theta_1\rangle_{\Delta_1,\theta_2} \, \langle n_R=0|\theta_3\rangle_{\Delta_2,\theta_2} \,
     \langle m|\theta_2\rangle
     \Bigg)
     \\
     &=\prod_{i=1}^{5}\left( \int_{0}^{\pi}d\theta_i \exp\left( \beta_i\frac{\cos(\theta_i)}{\hbar}\right) \rho(\theta_i) \right) \exp\left( \beta_6\frac{\cos(\theta_4)}{\hbar}\right) \nonumber
     \\
     &\quad  \times
      \frac{(q^{4\Delta_1};q^2)_{\infty}}{(q^{2\Delta_1}e^{i(\pm \theta_1\pm \theta_2)};q^2)_{\infty}} \frac{(q^{4\Delta_2};q^2)_{\infty}}{(q^{2\Delta_2}e^{i(\pm \theta_2\pm \theta_3)};q^2)_{\infty}} 
   \frac{(q^{4\Delta_3};q^2)_{\infty}}{(q^{2\Delta_3}e^{i(\pm \theta_4\pm \theta_5)};q^2)_{\infty}} .
 \end{align}

\begin{figure}[htbp]
  \begin{center}
   \includegraphics[width=5cm]{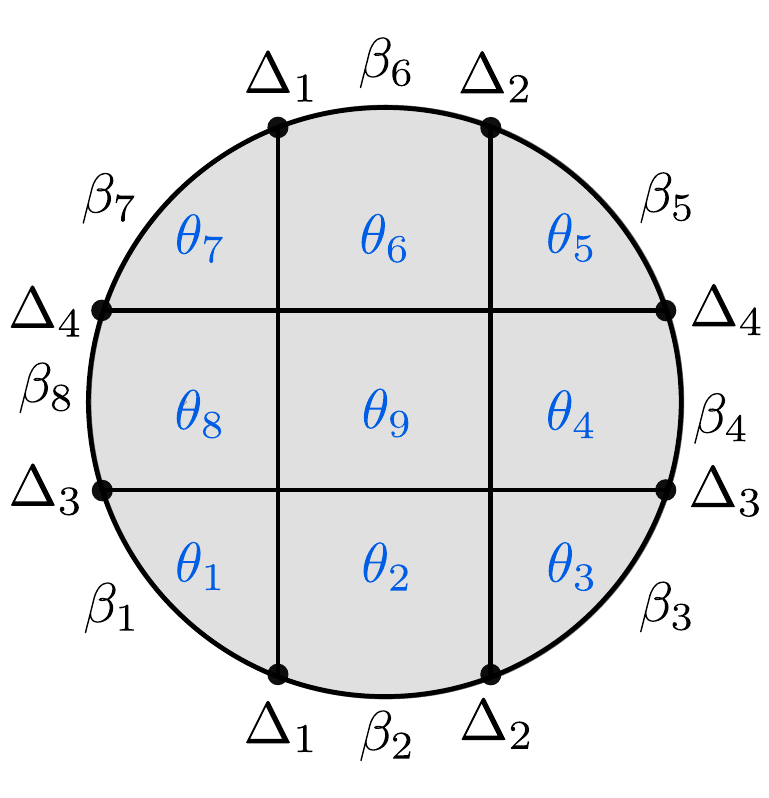}
  \end{center}   
\vspace*{-0.5cm}
\caption{crossed 8-point function}
\label{8pt}
\end{figure}

\vspace{5pt}
\paragraph{Crossed 8-point function} The crossed 8-point function is given in Fig.\ref{8pt}.
This correlation function differs from the previous crossed 6-point function in that it involves two vertical insertions, which are separated by boundary evolutions with lengths $\beta_4$ and $\beta_8$. These intermediate evolutions can be captured by inserting an identity operator in the two-particle Hilbert space expressed in energy basis.\footnote{Here we place the untwisted state in the middle of the tensor product.} Thus, this 8-point function (8pf) can be written as
\begin{align}
\mathrm{8pf}=& \int_0^{\pi} \int_0^{\pi} \int_0^{\pi} d\theta_8 d\theta_9 d\theta_4  \, \exp\left({\beta_8\frac{\cos(\theta_8)}{\hbar}}\right) \exp\left({\beta_4\frac{\cos(\theta_4)}{\hbar}}\right)  
    \rho_{\Delta_1,\theta_9}(\theta_8)  \rho_{\Delta_2,\theta_9}(\theta_4) \rho(\theta_9)  \nonumber
\\
&\times \langle\Delta_1,\Delta_2;\beta_6,\beta_5,\beta_4|e^{-\Delta_3 \hat{L}_1}\otimes e^{-\Delta_3 \hat{L}_2} \otimes e^{-\Delta_3 \hat{L}_3}|\Delta_1,\Delta_2;\theta_8,\theta_9,\theta_4\rangle \nonumber
    \\
    &\times \langle \Delta_1,\Delta_2;\theta_8,\theta_9,\theta_4|e^{-\Delta_4 \hat{L}_1}\otimes e^{-\Delta_4 \hat{L}_2} \otimes e^{-\Delta_4 \hat{L}_3}|\Delta_1,\Delta_2;\beta_1,\beta_2,\beta_3\rangle \nonumber
    \\
    =&\prod_{i=1}^{8}\left( \int_{0}^{\pi}d\theta_i \exp\left( \beta_i\frac{\cos(\theta_i)}{\hbar}\right) \rho(\theta_i) \right)  \int_{0}^{\pi}d\theta_9  \rho(\theta_9) \nonumber
    \\
    &
     R_{\theta_2\theta_8}^{q^2}\begin{bmatrix}
       \theta_1&\Delta_1\\\theta_9& \Delta_3
   \end{bmatrix} \,  R_{\theta_3\theta_9}^{q^2}\begin{bmatrix}
       \theta_2&\Delta_2\\\theta_4& \Delta_3
   \end{bmatrix} \,  R_{\theta_4\theta_6}^{q^2}\begin{bmatrix}
       \theta_9&\Delta_2\\\theta_5& \Delta_4
   \end{bmatrix} \, R_{\theta_9\theta_7}^{q^2}\begin{bmatrix}
       \theta_8&\Delta_1\\\theta_6& \Delta_4
   \end{bmatrix} \nonumber
   \\
   &\times \Bigg( \frac{(q^{4\Delta_{1}};q^2)_{\infty}}{(q^{2\Delta_{1}}e^{i(\pm \theta_1\pm \theta_2)};q^2)_{\infty}}   \frac{(q^{4\Delta_{2}};q^2)_{\infty}}{(q^{2\Delta\Delta_{2}}e^{i(\pm \theta_2\pm \theta_3)};q^2)_{\infty}}  \frac{(q^{4\Delta_3};q^2)_{\infty}}{(q^{2\Delta_3}e^{i(\pm \theta_3\pm \theta_4)};q^2)_{\infty}} \frac{(q^{4\Delta_{4}};q^2)_{\infty}}{(q^{2\Delta_{4}}e^{i(\pm \theta_4\pm \theta_5)};q^2)_{\infty}} \nonumber 
   \\
  & \quad \quad
   \frac{(q^{4\Delta_2};q^2)_{\infty}}{(q^{2\Delta_2}e^{i(\pm \theta_5\pm \theta_6)};q^2)_{\infty}} \frac{(q^{4\Delta_1};q^2)_{\infty}}{(q^{2\Delta_1}e^{i(\pm \theta_6\pm \theta_7)};q^2)_{\infty}} 
   \frac{(q^{4\Delta_4};q^2)_{\infty}}{(q^{2\Delta_4}e^{i(\pm \theta_7\pm \theta_8)};q^2)_{\infty}} \frac{(q^{4\Delta_{3}};q^2)_{\infty}}{(q^{2\Delta_{3}}e^{i(\pm \theta_8\pm \theta_1)};q^2)_{\infty}} 
   \Bigg)^{1/2}. 
\end{align}
{The reason for presenting this 8-point function calculation is to show that, for the middle region $\theta_9$, which is completely disconnected from the asymptotic boundary, the bulk two-point functions around this regions are conspicuously canceled out in the final expression. This is consistent from a bulk-to-boundary map point of view, as the detached bulk correlation functions cannot be encoded to the boundary in a gauge invariant way.

\vspace{5pt}
\subsection{Two-point functions on the double trumpet}
The bulk derivation presented here allows us to go beyond the results derivable from chord diagrams, to include matter correlators not only on the disk but also on the double trumpet\footnote{The path integral for sine-dilaton gravity on general higher genus surfaces, even in the absence of matter, is currently not well-understood. We expect that adding matter should follow the general procedure we outline here.}. Here we present two simple examples of such a calculation.

The wormhole Hilbert space allows us to calculate the two-point function on double trumpet (2pf-DT), as shown in Fig.\ref{DT} (a).
\begin{figure}[htbp]
    \centering
    \subfigure[]{\includegraphics[width=0.35\textwidth]{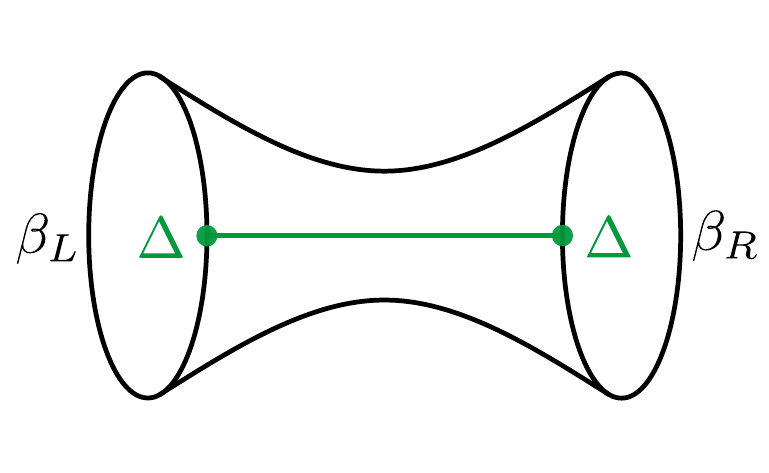}} 
    \subfigure[]{\includegraphics[width=0.35\textwidth]{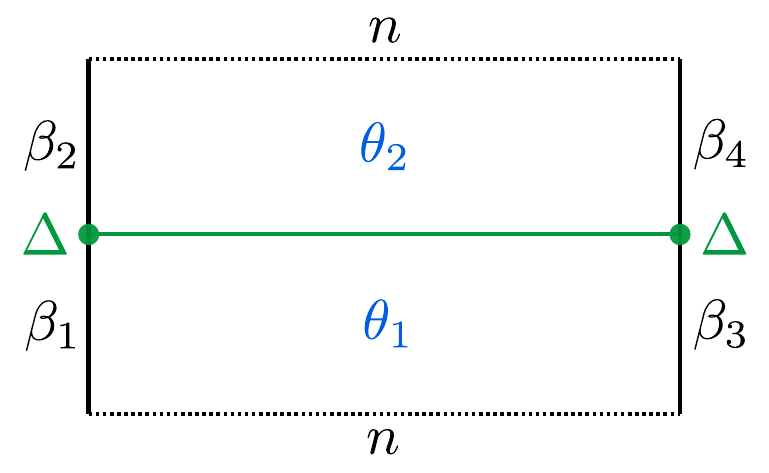}} 
    \caption{ Two-point function on double trumpet }
    \label{DT}
\end{figure}
By inserting an identity operator in the length basis, the double trumpet can be cut open as in Fig.\ref{DT} (b), with $\beta_L=\beta_1+\beta_2$ and $\beta_R=\beta_3+\beta_4$. The two-point function can be computed as
\begin{align}
    \text{2pt-DT}=&\sum_{n=0}^{\infty} \langle n| e^{-(\beta_1+\beta_3)\hat{H}_{\mathrm{SD}}}e^{-\Delta \hat{L}}   e^{-(\beta_2+\beta_4)\hat{H}_{\mathrm{SD}}}    |n\rangle
    \\
    =&\prod_{i=1}^{2}\left(\int_0^{\pi}d\theta_i \, \rho(\theta_i) \exp\left({(\beta_i+\beta_{i+2})\frac{\cos(\theta_i)}{\hbar}}\right)   \right)  \langle\theta_2|e^{-\Delta \hat{L}}|\theta_1\rangle  \left(  \sum_{n=0}^{\infty} \langle n|\theta_2\rangle\langle\theta_1| n \rangle \right)
    \\
    =&\frac{(q^{4\Delta};q^2)_{\infty}}{(q^{2\Delta};q^2)_{\infty}^2}\int_0^{\pi}d\theta_1 \rho(\theta_1) \exp\left({(\beta_L+\beta_{R})\frac{\cos(\theta_1)}{\hbar}}\right) \,   \frac{1}{(q^{2\Delta}e^{\pm 2i\theta_1};q^2)_{\infty}}
\end{align}
This is the same result as given in \cite{Okuyama:2023yat}.
Note that this two-point function diverges as taking $\Delta \rightarrow 0$. This divergence arises because, in this limit, the matter lines in the bulk can wind around the trumpet indefinitely, leading to a zero-mode divergence. These windings should be quotiented out in the calculation of the pure double trumpet partition function, whose result is given in \cite{Blommaert:2025avl}.

We can go further by adding a matter loop, involving a single trace operator with scaling dimension $\Delta^{\prime}$, in the double trumpet, as shown in Fig.\ref{DT_loop} (a).
\begin{figure}[htbp]
    \centering
    \subfigure[]{\includegraphics[width=0.35\textwidth]{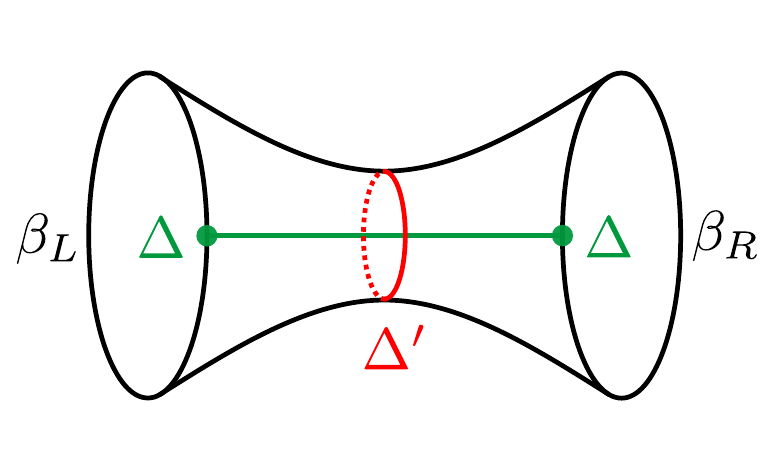}} 
    \subfigure[]{\includegraphics[width=0.35\textwidth]{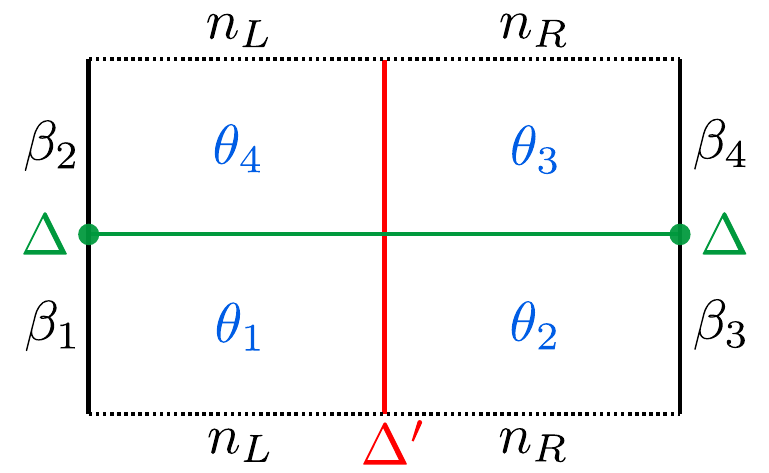}} 
    \caption{ Two-point function on double trumpet with a single trace operator of scaling dimension $\Delta^{\prime}$ in the  matter loop.}
    \label{DT_loop}
\end{figure}
Cutting the double trumpet open as in Fig.\ref{DT_loop} (b) and using the one-particle wormhole Hilbert space, this two-point function with a matter loop (2pf-DT-loop) can be obtained as
\begin{align}
    &\text{2pf-DT-loop}
    \\
    &= \prod_{i=1}^{4}\left(\int_0^{\pi}d\theta_i \, \exp\left( \beta_i\frac{\cos(\theta_i)}{\hbar} \right) \right) 
    \,\rho(\theta_1) \, \rho(\theta_4) \, \rho_{\Delta^{\prime},\theta_1}(\theta_2) \, \rho_{\Delta^{\prime},\theta_4}(\theta_3) \nonumber
    \\
    & \quad\times \sum_{n_L}^{\infty}\left(
    \langle n_L|\theta_1\rangle \langle\theta_1 |e^{-\Delta \hat{L}_L}|\theta_4\rangle \langle \theta_4|n_L\rangle \right) \sum_{n_R}^{\infty}\left(    \langle n_R|\theta_2\rangle_{\Delta^{\prime},\theta_1}  \,{_{\Delta^{\prime},\theta_1}}\langle\theta_2 |e^{-\Delta \hat{L}_R}|\theta_3\rangle_{\Delta^{\prime},\theta_4}  \, {_{\Delta^{\prime},\theta_4}}\langle \theta_3|n_R\rangle \right)
    \\
    &=  \frac{(q^{4\Delta};q^2)_{\infty}}{(q^{2\Delta};q^2)_{\infty}^2}\int_{0}^{\pi} \int_{0}^{\pi}d\theta_1 d\theta_2 \, \rho(\theta_1) \, \rho(\theta_2) \, \exp\left( \beta_L\frac{\cos(\theta_1)}{\hbar}\right)  \exp\left( \beta_R\frac{\cos(\theta_2)}{\hbar}\right)  \nonumber
    \\
    &\quad  \times  R_{\theta_2\theta_1}^{q^2}\begin{bmatrix}
       \theta_1&\Delta^{\prime}\\\theta_2& \Delta
   \end{bmatrix} 
   \left( \frac{1}{(q^{2\Delta}e^{\pm 2i\theta_1};q^2)_{\infty}}   \frac{1}{(q^{2\Delta}e^{\pm 2i \theta_2};q^2)_{\infty}} \right)^{1/2}.
\end{align}
Both results have been numerically verified to be finite for generic values of $\Delta,\Delta^{\prime}\neq 0$ and $\beta$, including the case $\beta=0$. This finiteness is consistent with the result of matter loops on the double trumpet found in DSSYK \cite{Okuyama:2023byh}, where the Hagedorn divergence noted in \cite{Jafferis:2022wez} was shown to be regularized in DSSYK. The sine-dilaton exhibits the same regulated  behaviour.\footnote{Note that \cite{Blommaert:2025avl} found a divergence in the EOW brane cylinder amplitude in sine-dilaton gravity. However, this divergence is canceled once two cylinders are glued together to form a double trumpet, due to  an additional contribution from the gluing measure - denoted as "$b$" in that work.} \footnote{In JT gravity, two-point functions on the double trumpet, with and without the matter loop, were derived in \cite{Jafferis:2022wez}. Our results agree with \cite{Jafferis:2022wez} upon taking the JT limit defined in Section~\ref{JT limit}.}

\vspace{5pt}
\subsection{Feynman rules}
\label{fey_rule}
With all the previous examples on hand, we now formulate a set of Feynman rules for computing general correlation functions, including those defined on the double trumpet. These rules are similar to those given in DSSYK \cite{Berkooz:2018jqr}, but are generalized to also include bulk correlators.


\begin{itemize}
    \item \textbf{Boundary three-point vertex:} Each boundary vertex contributes
    \begin{align}
        \adjincludegraphics[width=3cm,valign=c]{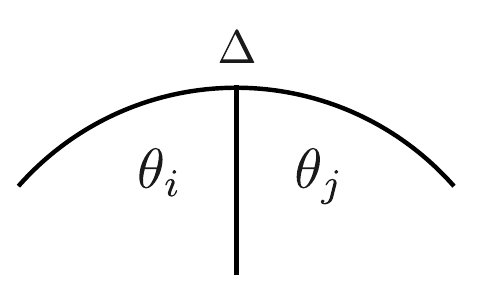}=\left( \frac{(q^{4\Delta};q^2)_{\infty}}{(q^{2\Delta}e^{i(\pm \theta_i\pm \theta_j)};q^2)_{\infty}} \right)^{1/2}.
    \end{align}
    \item \textbf{Boundary evolution:} Each boundary evolution gives
     \begin{align}
        \adjincludegraphics[width=3cm,valign=c]{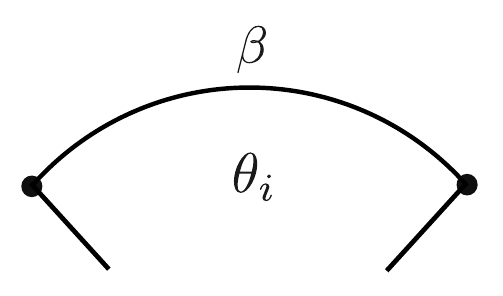}=\exp\left( \beta_i \frac{\cos(\theta_i)}{\hbar}\right).
    \end{align}
   This also applies to the case where the two ends of the boundary evolution are joined together, as in the double trumpet geometry.
    \item \textbf{Bulk four-point vertex:} Each intersection in the bulk gives a 6j-symbol of the quantum group $\mathcal{U}_q(su(1,1))$ in the R-matrix form
     \begin{align}
        \adjincludegraphics[width=4cm,valign=c]{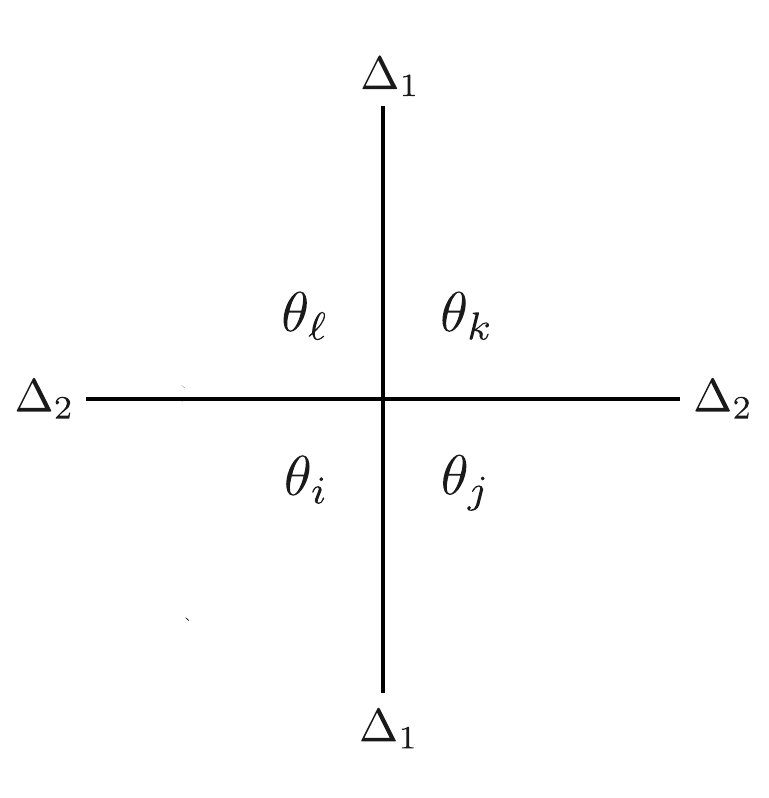}=  R_{\theta_j\theta_{\ell}}^{q^2}\begin{bmatrix}
       \theta_i&\Delta_2\\\theta_k& \Delta_1
   \end{bmatrix}.
    \end{align}
    \item \textbf{Bulk two-point functions:} There is no contribution from the two-point functions localized in the bulk. For example,
    \begin{align}
         \adjincludegraphics[width=4cm,valign=c]{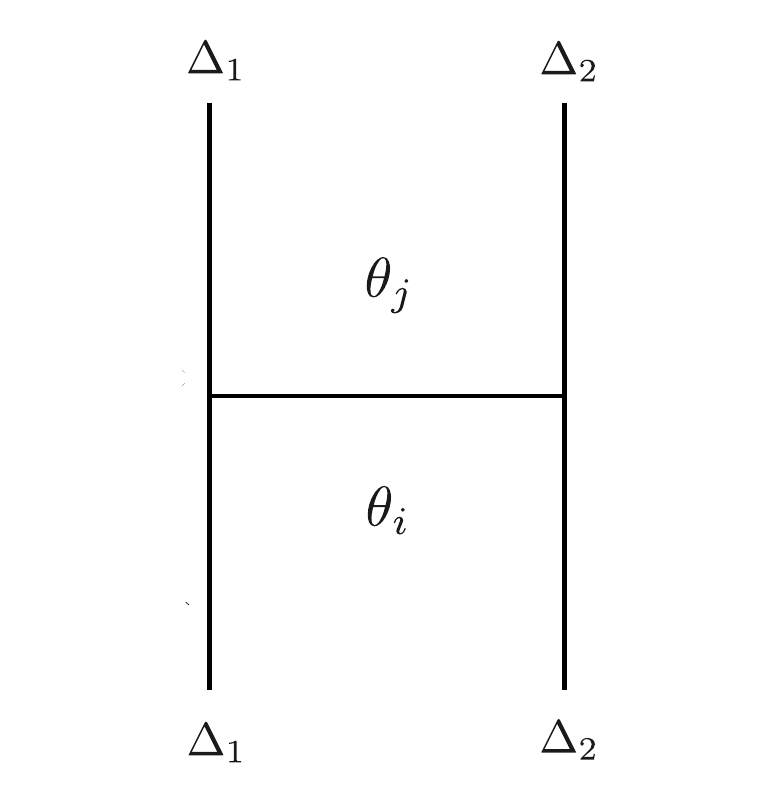}= 1.
    \end{align}
    This can be seen from previous examples, where such bulk two-point functions are always guaranteed  to cancel out.
    \item \textbf{Energy integration:} Finally, we integrate over all the energy parameters using
    \begin{align}
        \int_{0}^{\pi}d\theta_i \, \rho(\theta_i),
    \end{align}
    with the density of states given by
    \begin{align}
         \rho(\theta_i)=\frac{(q^2;q^2)_{\infty}}{2\pi}(e^{\pm2i\theta_i};q^2 )_{\infty}.
    \end{align}
\end{itemize}

\section{JT gravity limit}
\label{JT limit}

Our splitting and gluing process is not directly available in JT gravity, as it uses essentially the description of EOW branes by {\it two} parameters, only one of which is visible in the usual EOW brane formulations of JT gravity \cite{Gao:2021uro}. Nevertheless, the resulting formulas do have a good JT limit, which we now describe briefly.

The JT limit  can be taken as described in \cite{Berkooz:2018jqr}
\begin{align}
 \theta_i=(1-q^2)y_i, \quad q\rightarrow 1^{-},
\end{align}
where $y_i$ is the rescaled energy parameter near the edge of the spectrum.\footnote{In JT gravity, the relation between $y_i$ and the energy $E_i$ is: $E_i=y_i^2$.}

In this limit, most quantities in sine-dilaton gravity diverge, and they map to their JT gravity counterparts as follows
\begin{itemize}
    \item The density of states takes the form
    \begin{align}
   \int_0^{\pi}d\theta_i \,\rho(\theta_i) \rightarrow (q^2;q^2)_{\infty}^{3}(1-q^2)^3  \int_0^{\infty} dy_i \,\rho_{\mathrm{JT}}(y_i). \end{align}
   The JT density of states $\rho_{\mathrm{JT}}(y_i)$ is given by
   \begin{align}
       \rho_{\mathrm{JT}}(y_i)=\frac{1}{2\pi \Gamma(\pm 2iy_i)},
   \end{align}
   where the $\pm$  in the argument denotes a product of two gamma functions with opposite signs: $\Gamma(\pm y)=\Gamma(y)\Gamma(-y)$.
   \item The propagator changes as
   \begin{align}
       \frac{(q^{4\Delta};q^2)_{\infty}}{(q^{2\Delta}e^{i(\pm \theta_i\pm \theta_j)};q^2)_{\infty}} \rightarrow\frac{(1-q^2)^{2\Delta}}{(q^2;q^2)_{\infty}^{3}(1-q^2)^3} \frac{\Gamma(\Delta\pm iy_i \pm iy_j)}{\Gamma(2\Delta)} .
   \end{align}
   \item The R-matrix reduces to 6j-symbol of the $\mathfrak{sl}(2,\mathbb{R})$ algebra in JT gravity as
   \begin{align}
       R_{\theta_2\theta_4}^{q^2}\begin{bmatrix}
       \theta_1&\Delta\\\theta_3& \Delta^{\prime}
   \end{bmatrix} \rightarrow\frac{1}{(q^2;q^2)_{\infty}^3(1-q^2)^3}\,\begin{Bmatrix}
       \Delta^{\prime}&y_3 & y_2\\\Delta& y_1 & y_4
   \end{Bmatrix} .
   \end{align}
   \item Some other q-Pochhammer symbols takes the form
   \begin{align}
       &(q^{4\Delta},q^2)_{\infty}\rightarrow\frac{(q^2;q^2)_{\infty}(1-q^2)^{1-2\Delta}}{\Gamma(2\Delta)}.
       \\
       &\frac{1}{(q^{2\Delta}e^{i(\pm \theta_i\pm \theta_j)};q^2)_{\infty}}\rightarrow\frac{\Gamma(\Delta\pm iy_i \pm iy_j)(1-q^2)^{4\Delta}}{(q^2;q^2)_{\infty}^4 (1-q^2)^4}.
   \end{align}
\end{itemize}

Using  the above map, the Askey-Wilson integral we used for gluing two EOW branes into a single geodesic for reproducing the two-point function in (\ref{AW-integral}) becomes
\begin{align}
    \int_0^{\infty}ds\, \frac{1}{2\pi \Gamma(\pm 2is)}  \frac{\Gamma(\Delta_L\pm iy_L \pm is)}{\Gamma(2\Delta_L)} \frac{\Gamma(\Delta_R\pm iy_R \pm is)}{\Gamma(2\Delta_R)}=\frac{\Gamma((\Delta_L+\Delta_R)\pm iy_L \pm iy_R)}{\Gamma(2(\Delta_L+\Delta_R))}.
\end{align}
This identity confirms the validity of applying the same procedure of gluing two matter lines in JT gravity. It was previously found in \cite{Blommaert:2020yeo}, where the authors used a similar splitting trick in the fully-open channel.

Our new identity for the 6j-symbols (\ref{R_relation}) in this limit becomes
\begin{align}
    &\int_0^{\infty} \int_0^{\infty} ds_1 ds_2 \, \frac{1}{2\pi \Gamma(\pm 2i s_1)} \frac{1}{2\pi \Gamma(\pm 2i s_2)} \frac{1}{\Gamma(2\Delta_L) \Gamma(2\Delta_R)} \begin{Bmatrix}
       \Delta^{\prime}&y_4 & y_1\\\Delta_L& s_1 & s_2
   \end{Bmatrix} \begin{Bmatrix}
       \Delta^{\prime}&y_3 & y_2\\\Delta_R& s_1 & s_2
   \end{Bmatrix} \nonumber
   \\
   &\times   \Big(\Gamma(\Delta_L\pm is_1\pm iy_1) \Gamma(\Delta_R\pm is_1\pm iy_2) \Gamma(\Delta_R\pm is_2\pm iy_3) \Gamma(\Delta_L\pm is_2\pm iy_4) \Big)^{1/2}\nonumber
   \\
   &= \frac{1}{\Gamma(2(\Delta_L+\Delta_R))}
   \begin{Bmatrix}
       \Delta^{\prime}&y_3 & y_2\\\Delta_L+\Delta_R& y_1 & y_4
   \end{Bmatrix}\nonumber
   \\
   &\quad \times \Big(\Gamma((\Delta_L+\Delta_R)\pm iy_1\pm iy_2) \Gamma((\Delta_L+\Delta_R)\pm iy_3\pm iy_4) \Big)^{1/2}.
   \label{6jRel_sl2r}
\end{align}
Note that the divergent prefactors on both sides have canceled out. This is a nontrivial analytic check of the correctness of our identity. Using the JT gravity bulk Feynman rules \cite{Mertens:2017mtv}, this new identity again implies a 1-split representation of OTOC as in Fig.\ref{OTOC_1split}.

\section{Conclusions and outlook}

We have analyzed the implementation of splitting and gluing of spacetime regions with EOW branes in sine-dilaton gravity, and shown how this procedure leads to bulk computations of general correlation functions. In the semi-open channel, this splitting and gluing procedure gives rise to the structure of the multi-particle wormhole Hilbert space: the length basis factorizes, but the energy basis exhibits a non-local, state-dependent structure arising from the EOW brane quantization.  We end the paper with further comments and future directions.

\subsection{Concrete open questions}
\begin{itemize}
    \item Is there a direct bulk derivation of the conjectured dictionary (\ref{dic_mass})? Since the boundary of sine-dilaton is essentially non-geometric and singular \cite{Belaey:2025kiu},  the standard way of deriving the holographic dictionary that relates the mass of a bulk matter field and the scaling dimension of its dual operator at the boundary in $\mathrm{AdS}_2$ spacetime does not straightforwardly apply. 
    \item In the EOW brane quantization, an additional complex conjugation was applied by hand in order to obtain consistent results. Is there a canonical principle to determine when such a conjugation should be applied?
    \item What is the quantum group derivation of our new identity (\ref{R_relation})? A possible starting point may be the  Biedenharn-Eliott identity for $\mathcal{U}_q(su(1,1))$ in \cite{groenevelt2005wilsonfunctiontransformsrelated}. Alternatively, one could also start from its JT limit version (\ref{6jRel_sl2r}) for the $\mathfrak{sl}(2,\mathbb{R})$ algebra.
    \item Our multi-particle Hilbert space enables the directly bulk computation of most correlation functions, except for those involving the Yang-Baxter move symmetry. For example:
    \begin{align}
         \adjincludegraphics[width=4.5cm,valign=c]{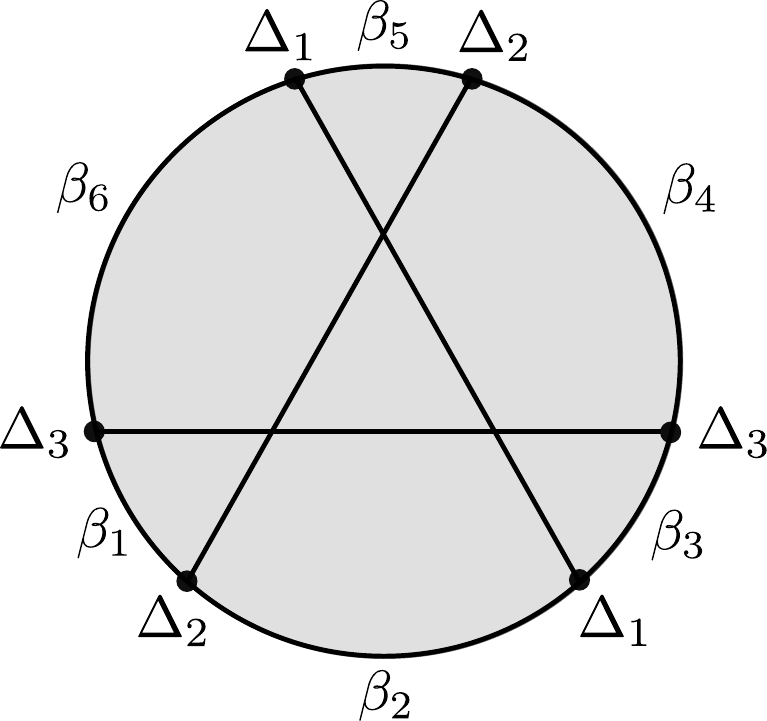}.
    \end{align}
    This diagram is expected to follow the proposed Feynman rules, but a direct bulk calculation would require handling the crossing of matter lines in $n$-particle states. While such crossings should of course just reproduce a 6j-symbol, making this explicit within our splitting and gluing procedure might require extra work.\footnote{A concrete obstruction is that, in this case, three vertices form a triangular region in the center of the bulk. Once one line in the bulk is treated as a vertical EOW brane,  the calculation of two vertices on this line automatically fixes the remaining two lines to be horizontal operators. As a result, there is inevitably at least one vertex at which two lines are both horizontal. For such a vertex, our splitting–gluing procedure does not provide a well-defined prescription, and therefore cannot be directly applied to compute the corresponding contribution.}
\end{itemize}

\subsection{Future directions}
\paragraph{Quantum group origin}
The natural language for our splitting and gluing process is likely the representation theory of the quantum group $\mathcal{U}_q(su(1,1))$ \cite{Fronsdal:1991gf, Berkooz:2022mfk}. Indeed, the structure of the co-product of the dual Hopf algebra seems related to our gluing process. In \cite{Belaey:2025ijg}, the co-product was given a gravitational interpretation as the product of wavefunctions of two factorized bulk subregions. The contracted matrix elements in the co-product are then interpreted as labels of "edge states" living at the entangling surface between two bulk subregions, denoted by "$s$" in \cite{Belaey:2025ijg}. Notably, the label $s$ is periodic due to the discreteness of the spatial coordinate. Our gluing condition, which involves integrating over periodic brane parameter $\bar\alpha$, is very similar to this product, suggesting a possible identification between the edge state label s and the brane parameter $\bar\alpha$. It is interesting to make this connection more precise, and in particular to understand our new identity (\ref{R_relation}) in this language. If we associate each spacetime region with matrix elements in a specific representation of the quantum group, we could understand the emergence of group theoretical quantities like the R-matrix more directly.

\textbf{Note added}: The quantum group used in \cite{Belaey:2025ijg} is $\mathcal{U}_q(sl(2,\mathbb{R}))$. However, after our paper appeared, a later work \cite{vanderHeijden:2025zkr} reproduced exactly the same multi-particle Hilbert space structure as ours using the coproduct of the quantum group $\mathcal{U}_q(su(1,1))$. This suggests that the quantum group $\mathcal{U}_q(su(1,1))$ is more directly relevant to our construction.

\paragraph{Bulk reconstruction}Another fascinating future direction is a better understanding of purely bulk regions that are disconnected from the boundary, using our multi-particle Hilbert space. In some way the state-dependence structure we find is to be expected from previous studies of bulk reconstruction and quantum error correction. However, we expect a new layer of non-locality or non-commutativity to emerge due to the quantum group structure. By comparing to the JT limit, it may be possible to disentangle these two aspects of bulk reconstruction. 

\paragraph{Closed universe} Pure bulk regions  we separate out using our splitting and gluing procedure can also correspond to closed cosmologies bounded by two EOW branes, such as the $\theta_9$ region in Fig.\ref{8pt}. This is a controllable arena to study the role of an observer in constructing the bulk Hilbert space of the closed universe \cite{Harlow:2025pvj, Abdalla:2025gzn}, which is  trivial in the absence of an observer. In fact, it is not difficult to generalize our framework along the lines of the thin-shell construction in \cite{Antonini:2024mci} and analyze such spacetime very precisely using our language, for example via the swap test. 
In particular, each matter line in the bulk is labeled by two EOW brane parameters, and the state-dependent structure of the multi-particle Hilbert space implies that the matter line naturally prepares an ensemble of  Hamiltonians for the adjacent bulk regions,  labeled by its second brane parameter. An interesting perspective is that one can think of the gluing process as preparing the matter line as a coherent bulk observer, which after gluing is only labeled by the first brane parameter. The gluing condition over the second brane parameter then plays a role analogous to an ensemble average over nearby Hamiltonians. In this sense, the presence of an observer naturally brings in an ensemble average into the bulk when one prepares a Hartle-Hakwing state involving observers.  For recent discussions of closed universes and observers, see \cite{Engelhardt:2025azi, Antonini:2025ioh, Chen:2025fwp, Wei:2025guh}.

One of the main motivations to study sine-dilaton gravity is its role as a new paradigm of holographic duality (see e.g. \cite{Belaey:2025kiu}) and as a novel framework for realizing UV complete behavior in gravity. We hope our observations here will be useful for shedding light on the essential mechanism responsible for these surprising aspects of 
sine-dilaton gravity.

\section*{Acknowledgments}

The authors thank Sergio Aguilar, Micha Berkooz, Andreas Blommaert, Hong Zhe (Vincent) Chen,  Misha  Isachenkov, Thomas Mertens, Abhisek Sahu and Jiuci Xu for interesting discussions. The work is supported by a Discovery grant from NSERC.

\appendix
\section{Pizza slicing representation of the OTOC}
\label{pizza_OTOC}
In the 1-split representation of the OTOC, one can further split the horizontal line by inserting an identity operator as $e^{-\Delta^{\prime}\hat{L}}=e^{-\Delta_1^{\prime}\hat{L}}\mathds{1}e^{-\Delta_2^{\prime}\hat{L}}$ with the identity operator taking the form
\begin{align}
     \mathds{1}=\int_0^{\pi} d\bar\alpha_5 \, \rho(\bar\alpha_5)\, &\Bigg( \int_0^{\pi} d\bar\alpha_3 \, \rho_{\Delta_1,\bar\alpha_5}(\bar\alpha_3)\,|\bar\alpha_3\rangle_{\Delta_1,\bar\alpha_5}\, {_{\Delta_1,\bar\alpha_5}}\langle  \bar\alpha_3|\Bigg) \otimes\Bigg( |\bar\alpha_5\rangle \langle\bar\alpha_5| \Bigg)
     \nonumber\\
     &  \otimes \Bigg( \int_0^{\pi} d\bar\alpha_4 \, \rho_{\Delta_2,\bar\alpha_5}(\bar\alpha_4)\,|\bar\alpha_4\rangle_{\Delta_2,\bar\alpha_5}\, {_{\Delta_2,\bar\alpha_5}}\langle  \bar\alpha_4| \Bigg).
\end{align}
With this insertion, the OTOC is represented as a "pizza slicing" as in Fig.\ref{Pizza}.
\begin{figure}[htbp]
  \begin{center}
   \includegraphics[width=5cm]{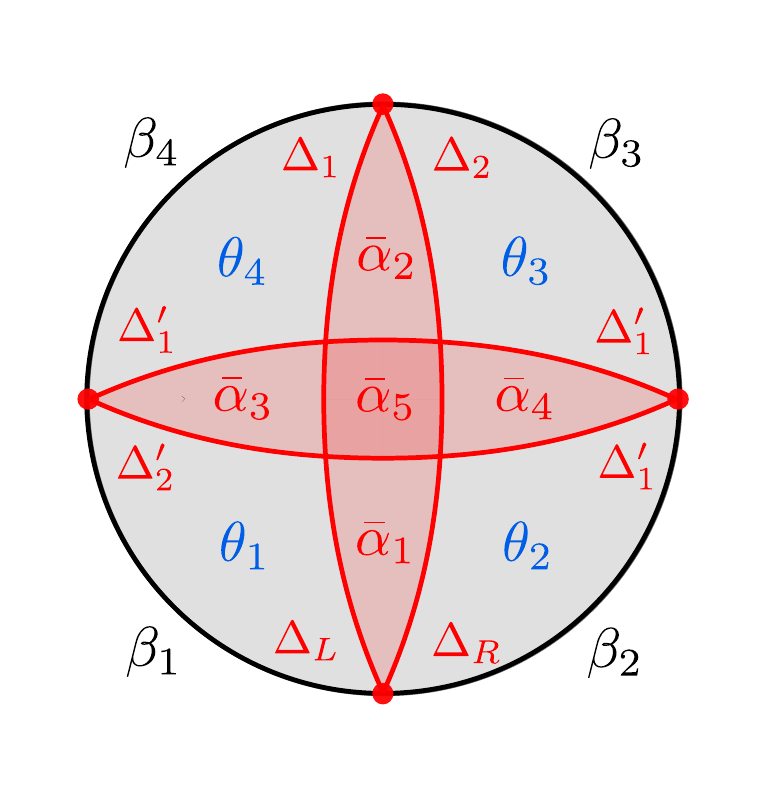}
  \end{center}   
\vspace*{-0.5cm}
\caption{OTOC in pizza slicing representation}
\label{Pizza}
\end{figure}

Referring to (\ref{appr2_4prod}), the OTOC now takes the form
\begin{align}
 \mathrm{OTOC}=& \prod_{i=1}^{5}\left( \int_0^{\pi}d\bar{\alpha}_i \,\rho(\bar\alpha_i) \right) \,\prod_{i=1}^{4}\left( \int_{0}^{\pi}d\theta_i  \exp\left( \beta_i\frac{\cos(\theta_i)}{\hbar}\right) \right) \nonumber
    \\
    &\times  \rho_{\Delta_1,\bar\alpha_1}(\theta_1)\,\rho_{\Delta_2,\bar\alpha_1}(\theta_2)\, \rho_{\Delta_2,\bar\alpha_2}(\theta_3) \,\rho_{\Delta_1,\bar\alpha_2}(\theta_4) \,  \rho_{\Delta_1,\bar\alpha_5}(\bar\alpha_3)\,\rho_{\Delta_2,\bar\alpha_5}(\bar\alpha_4)\, \rho(\bar\alpha_5) \nonumber
    \\
    &\times
    {_{\Delta_1,\bar\alpha_2}}\langle\theta_4 | e^{-\Delta^{\prime}_1\hat{L}_L} |\bar\alpha_3\rangle_{\Delta_1,\bar\alpha_5}\, {_{\Delta_1,\bar\alpha_5}}\langle  \bar\alpha_3| e^{-\Delta^{\prime}_2\hat{L}_L}|\theta_1\rangle_{\Delta_1,\bar\alpha_1} \nonumber
    \\
    &\times \langle \bar\alpha_2|e^{-\Delta^{\prime}_1\hat{L}_{M}} |\bar\alpha_5\rangle \langle\bar\alpha_5|  e^{-\Delta^{\prime}_2\hat{L}_{M}}|\bar\alpha_1\rangle  \nonumber 
    \\
    & \times {_{\Delta_2,\bar\alpha_2}}\langle\theta_3 | e^{-\Delta^{\prime}_1\hat{L}_R} |\bar\alpha_4\rangle_{\Delta_2,\bar\alpha_5}\, {_{\Delta_2,\bar\alpha_5}}\langle  \bar\alpha_4| e^{-\Delta^{\prime}_2\hat{L}_R}|\theta_2\rangle_{\Delta_2,\bar\alpha_1}.
\end{align}
Using (\ref{8W7_R}), we can organize the expression to 
\begin{align}
   & \quad \prod_{i=1}^{5}\left( \int_0^{\pi}d\bar{\alpha}_i \,\rho(\bar\alpha_i) \right) \,\prod_{i=1}^{4}\left( \int_{0}^{\pi}d\theta_i  \exp\left( \beta_i\frac{\cos(\theta_i)}{\hbar}\right) \right) \nonumber
    \\
    &\quad \times  \left( \frac{(q^{4\Delta_1};q^2)_{\infty}}{(q^{2\Delta}e^{i(\pm \theta_4\pm \bar\alpha_2)};q^2)_{\infty}}
    
    \frac{(q^{4\Delta_2};q^2)_{\infty}}{(q^{2\Delta}e^{i(\pm \theta_3\pm \bar\alpha_2)};q^2)_{\infty}}
    
    \frac{(q^{4\Delta_1};q^2)_{\infty}}{(q^{2\Delta}e^{i(\pm \theta_1\pm \bar\alpha_1)};q^2)_{\infty}}  
    
    \frac{(q^{4\Delta_2};q^2)_{\infty}}{(q^{2\Delta}e^{i(\pm \theta_2\pm \bar\alpha_1)};q^2)_{\infty}} \right)^{1/2} \nonumber
    \\

    & \quad \times \left( \frac{(q^{4\Delta_1^{\prime}};q^2)_{\infty}}{(q^{2\Delta}e^{i(\pm \theta_4\pm \bar\alpha_3)};q^2)_{\infty}}
    
    \frac{(q^{4\Delta_2^{\prime}};q^2)_{\infty}}{(q^{2\Delta}e^{i(\pm \theta_1\pm \bar\alpha_3)};q^2)_{\infty}}
    
    \frac{(q^{4\Delta_1^{\prime}};q^2)_{\infty}}{(q^{2\Delta}e^{i(\pm \theta_3\pm \bar\alpha_4)};q^2)_{\infty}}  
    
    \frac{(q^{4\Delta_2^{\prime}};q^2)_{\infty}}{(q^{2\Delta}e^{i(\pm \theta_2\pm \bar\alpha_4)};q^2)_{\infty}} \right)^{1/2}
     \nonumber
    \\
    &\quad\times R_{\bar\alpha_5\theta_{4}}^{q^2}\begin{bmatrix}
       \bar\alpha_3&\Delta_1^{\prime}\\\bar\alpha_2& \Delta_1
   \end{bmatrix} \, R_{\bar\alpha_4\bar\alpha_{2}}^{q^2}\begin{bmatrix}
       \bar\alpha_5&\Delta_1^{\prime}\\\theta_3& \Delta_2
   \end{bmatrix}\, R_{\bar\alpha_1\bar\alpha_{3}}^{q^2}\begin{bmatrix}
       \theta_1&\Delta_2^{\prime}\\\bar\alpha_5& \Delta_1
   \end{bmatrix}\, R_{\theta_2\bar\alpha_{5}}^{q^2}\begin{bmatrix}
       \bar\alpha_1&\Delta_2^{\prime}\\\bar\alpha_4& \Delta_2
   \end{bmatrix}.
\end{align}
This form once again follows from the Feynman rules in Section \ref{fey_rule}.

The utility of the pizza slicing representation is  that  each slice now can be computed individually and then glued together to obtain the full OTOC. Each slice is calculated as 
\begin{align}
     \adjincludegraphics[width=3.5cm,valign=c]{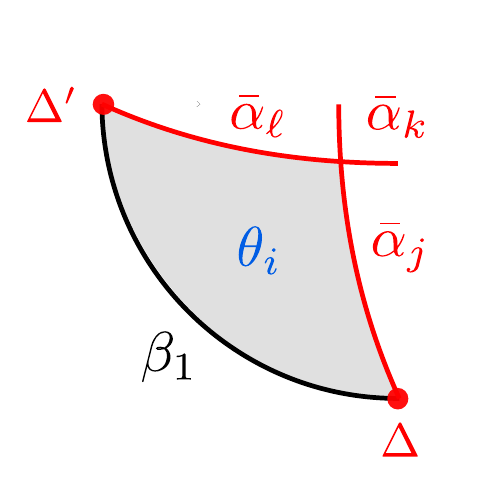}=& \int_{0}^{\pi}d\theta_i  \exp\left( \beta_i\frac{\cos(\theta_i)}{\hbar}\right)  \left( \frac{(q^{4\Delta};q^2)_{\infty}}{(q^{2\Delta}e^{i(\pm \theta_i\pm \bar\alpha_j)};q^2)_{\infty}} \frac{(q^{4\Delta^{\prime}};q^2)_{\infty}}{(q^{2\Delta}e^{i(\pm \theta_i\pm \bar\alpha_{\ell})};q^2)_{\infty}} \right)^{1/2} \nonumber
     \\
     &\times R_{\bar\alpha_j\bar\alpha_{\ell}}^{q^2}\begin{bmatrix}
       \theta_i&\Delta^{\prime}\\\bar\alpha_k& \Delta
   \end{bmatrix}.
\end{align}
  The $\bar\alpha$-indices of each pizza slice are analogous to the labels of the bulk "edge states" discussed in \cite{Belaey:2025ijg}. The OTOC is then obtained by gluing four such slices together
\begin{align}
    \prod_{i=1}^{5}\left( \int_0^{\pi}d\bar{\alpha}_i \,\rho(\bar\alpha_i) \right) \adjincludegraphics[width=5cm,valign=c]{figure_pdf/pizza.pdf}=
    \, \,\, \adjincludegraphics[width=5cm,valign=c]{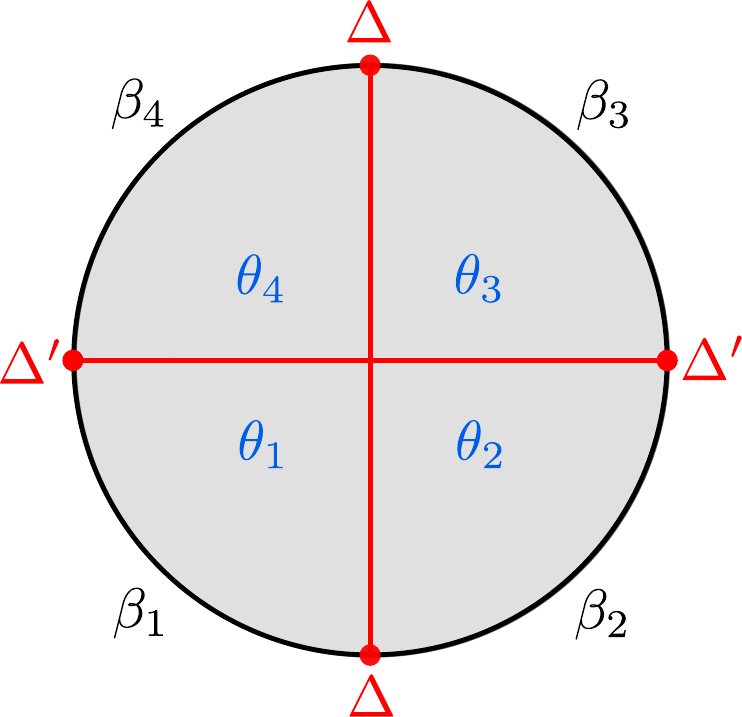},
\end{align}
where the four R-matrices on the LHS are reduced to a single R-matrix by using the identity (\ref{R_relation}).


\vspace{10pt}

\bibliography{sample}

\end{document}